# Prokaryotic genome editing based on the subtype I-B-*Svi* CRISPR-Cas system

## Authors:

Wang-Yu Tong*, De-Xiang Yong, Xin Xu, Cai-Hua Qiu, Yan Zhang, Xing-Wang Yang, Ting-Ting Xia, Qing-Yang Liu, Su-Li Cao, Yan Sun and Xue Li

## Affiliations:

*Integrated Biotechnology Laboratory, School of Life Sciences, Anhui University, 111 Jiulong Road, Hefei 230601, China*

** Corresponding author: tongwy@ahu.edu.cn*

*Tel.: +86-551-63861282*

*Fax: +86-551-63861282*

## Abstract

Type I CRISPR-Cas systems are the most common among six types of CRISPR-Cas systems, however, non-self-targeting genome editing based on a single Cas3 of type I CRISPR-Cas systems has not been reported. Here, we present the subtype I-B-*Svi* CRISPR-Cas system (with three confirmed CRISPRs and a *cas* gene cluster) and genome editing based on this system found in *Streptomyces virginiae* IBL14. Importantly, like the animal-derived bacterial protein *Sp*Cas9 (1368 amino-acids), the single, compact, non-animal-derived bacterial protein *Svi*Cas3 (771 amino-acids) can also direct template-based microbial genome editing through the target cell's own homology-directed repair system, which breaks the view that the genome editing based on type I CRISPR-Cas systems requires a full Cascade. Notably, no off-target changes or indel-formation were detected in the analysis of potential off-target sites. This discovery broadens our understanding of the diversity of type I CRISPR-Cas systems and will facilitate new developments in genome editing tools.

## Introduction

The CRISPR-Cas system is an RNA-guided adaptive microbial immune system widely found in archaea (202 of 232 archea) and bacteria (3059 of 6782 bacteria) genomes (http://crispr.i2bc.paris-saclay.fr/). The names of CRISPR and Cas are two acronyms respectively for clustered regularly interspaced short palindromic repeats and CRISPR-associated system. Almost all of the known CRISPR-Cas systems are classified into two classes that are subdivided into six types. Class 1 systems (includ ing types I, III and IV) existing in bacteria or archaea and class 2 systems (including types II, V and VI) almost totally presenting in bacteria are assigned by the number of Cas proteins in their effector complexes, i.e., with multi-Cases in class 1 effector modules and a single-Cas in class 2 effector complexes, respectively. The types I, II, III, IV, V and VI are mainly characterized by their signature proteins, i.e., Cas3, Cas9, Cas10, undetermined (Csf1?), Cas12 (e.g., Cas12a / previous Cpf1) and Cas13 (e.g., Cas13a / previous C2c2), in proper order [1-4]. By far only proteins Cas9 (e.g., *Sp*Cas9 in *Streptococcus pyogene*) [5,6], Cas12a (in *Acidaminococcus sp*.BV3L6) [7-9] in class 2 CRISPR-Cas systems and CasF / Cas12j in the genomes of huge bacteriophages [10] have been utilized for genome editing, despite the fact that a single Cas13a (C2c2 in *Leptotrichia shahii*) also belonging to class 2 CRISPR-Cas systems could be programmed to knock down specific mRNAs [11,12].

Among the six types of CRISPR-Cas systems, type I systems are the most prevalent (over 60 % of the total systems found in the sequenced genomes of bacteria and archaea) [13] (http://crispr.i2bc.paris-saclay.fr/). Type I Cas systems are subdivided into diverse subtypes according to their architectures of effector complexes [1,14]. Many articles discussing microbial immunity directed by type I CRISPR-Cas systems have been published, mainly relating to the roles and features of spacer (S), repeat (R), CRISPR-RNA (crRNA), R-loop, Cas and CRISPR-associated complex for antiviral defense (Cascade) in subtype I-A [15], I-B [16], I-C [17], I-D [18], I-E [19,20], I-F [21] and I-G [13,15] CRISPR-Cas systems. Although much progress has been made in deciphering structures, roles and mechanism of type I CRISPR-Cas-system-directed immunity to exogenous DNA, a lot of details about the diversities, characteristics and molecular mechanisms of adaptive immunity directed by type I CRISPR-Cas systems remain to be investigated.

Recently, a few of gene editing tools have been developed on the basis of the Type I CRISPR-CAS system [22-28], but these have a complete Cascade and implement the gene editing mainly through the non-homologous end joining (NHEJ) system of the cell itself. Importantly, non-self-targeting gene editing tools ground on a single Cas3 of type I CRISPR-Cas systems have not been reported to date [24,28]. This paper aims to introduce the subtype I-B-*Svi* (*Streptomyces virginiae* IBL14) [29] CRISPR-Cas system and the genome editing based on this system. We analyzed the characteristics and features of the subtype I-B-*Svi* CRISPR-Cas system with the aid of bioinformatics resources and developed several sets of gene editing tools composed of gene editing plasmid (named plasmid-t/g-*target gene abbreviation*) and/or Cas expression plasmid (named plasmid-*cas gene abbreviation*) based on the system.

## Results

## The three confirmed CRISPRs

The genome of *S. virginiae* IBL14 contains three confirmed *Svi*CRISPRs, i.e., *Svi*CRISPRs 1, 2 and 3 (Table 1). All 26 spacers in the three *Svi*CRISPRs are confined to a length range of 34 nt to 44 nt, a GC content range of 51% to 86% and a Tm range of 66 $^{o}$C to 84 $^{o}$C. A total of 29 repeats in the three *Svi*CRISPRs are composed of 30 nucleotides (nt) and share a GC content range of 57% to 70% (lower than the average GC content of 72.98% in the *S. virginiae* IBL14 genome) and a melting temperature (Tm) range of 67 $^{o}$C to 74$^{o}$C. *Svi*CRISPRs 1, 2 and 3 have 13, 5 and 11 repeats, respectively. All 13 repeats in *Svi*CRISPR 1 have the same base sequences, but the base sequences in the 5 repeats of *Svi*CRISPR 2 are not the same. Two and four out of the 11 repeats in *Svi*CRISPR 3 have the same base sequences, respectively. In addition, there is one repeat in *Svi*CRISPR 2 that shares the same base sequences with two repeats in *Svi*CRISPR 3. In comparison to the repeat architecture (gcttcAACCCcacaaGGGTTcgtctgaaac, a total of 30 nt) of the subtype I-B CRISPR-Cas system in *Haloferax volcanii* [16], 26 out of the 29 repeats have a pentameric palindrome composed of two inverted repeats interspaced by 3 nt with the following stoichiometry: gt-5 nt-CCCCT-3 nt-AGGGG-10 nt or gt-8 nt-CCCCT-3 nt-AGGGG-7 nt, implying that in the 30 nt repeats, a 5 nt-palindrome with a higher GC content is probably important for subtype I-B CRISPR-Cas-system-directed adaptive immunity to invader DNA. All 29 repeats begin with the base “g” in the 5’-terminal and 27 out of the 29 repeats end with the two bases “ac” in the 3’-terminal, similar to the repeats in the I-B CRISPR-Cas system in *Haloferax volcanii* [16]. It is worth noting that the same sequence motif "CCC/GGG" in the repeats of the two subtype I-B CRISPR-Cas systems also exists in the repeats of the subtype I-E (gtgtTCCCCGCgccaGCGGGGAtaaacc-28 nt, in *E. coli*) [19] and type VI (ccaCCCCaatatcgaaGGGGactaaaac-28 nt, in *L. shahii*) [11] CRISPR-Cas systems, suggesting that in adaptive immunity to invasive genetic factors (DNA or RNA) directed by CRISPR-Cas systems, the complementary motif is probably necessary for maintaining the stability of crRNA secondary structure.

Table 1 Three confirmed CRISPRs in *S. virginiae* IBL14 genome*

| ***Svi-*** **CRISPR** | **Direct repeat (R)** | **Spacer (S)** |
|---|---|---|
| ***1*** | ***30 bp, 67%GC, Tm 71°C*** | ***35-37 bp, 51-86%GC, Tm 66-84°C*** |
| | gttgcga**cccct**cgt**agggg**cgatgaggac | ggtccgcaccctggtggagaggctcccgtacgtgct |
| | gttgcga**cccct**cgt**agggg**cgatgaggac | aggtcgccgaggcggagcgggacaccgccgtaaaa |
| | gttgcga**cccct**cgt**agggg**cgatgaggac | cccgtctacggcgctcgcgccgacatcctgcacgccgc |
| | gttgcga**cccct**cgt**agggg**cgatgaggac | ctggtacctggcccgctgggccgagtggatcgcctg |
| | gttgcga**cccct**cgt**agggg**cgatgaggac | ttcaccatcaacgccaccgcatgcgggaccctcgc |
| | gttgcga**cccct**cgt**agggg**cgatgaggac | cgagccgggcggtggctgggatgtccaccgtcgcgc |
| | gttgcga**cccct**cgt**agggg**cgatgaggac | *cccgccgtccggcgggccg*ggtgccttccgtcagcggg-*gvgl*005738 |
| | gttgcga**cccct**cgt**agggg**cgatgaggac | ccaccgccgcccgcacgagcgggcggagtcgagtct |
| | gttgcga**cccct**cgt**agggg**cgatgaggac | ctgacccttagatcgccaacacgcaaccccaatccca |
| | gttgcga**cccct**cgt**agggg**cgatgaggac | tcgggtgacgccctgggcggcggctgcgtcaagcatcgcgt |
| | gttgcga**cccct**cgt**agggg**cgatgaggac | ccttgatcgaccattcctgactacaacaggaggac |
| | gttgcga**cccct**cgt**agggg**cgatgaggac | cgggccgcgcagcggccggcctcggcaggcactgc |
| | gttgcga**cccct**cgt**agggg**cgatgaggac | The 13 R sequences in the CRISPR 1 are the same. |
| *2* | ***30 bp, 57-70%GC, Tm 67-71ºC***<br>gtcctcatcg**cccct**gcg**agggg**tcgcaac<br>gtcttcatcg**cccct**gcg**agggg**tcgcaac<br>gtcttcatcg**cccctgcagggg**gtcacaac<br>gtcctcatc**gaccct**tgg**agggtc**cgcaac<br>gttctaatcg**cccct**ttg**agggg**tcgcaac | ***35-37 bp, 66-83%GC, Tm 75-83ºC***<br>gtctccggc*tcggccagctcgatcagcc*cgtgaccgg-*gvgl*005245<br>ccgaggcg*accggcggccaggccgatc*cgctcgccg-*gvgl*002878<br>atggatccggtacggctgatcagcgcggcggtctg<br>aggcgaagggcaagtttgaggaggcccaggccgccg<br>The first R sequence in the CRISPR 2 is the same as the ninth and eleventh R sequences in CRISPR 3. |
| *3* | ***30 bp, 57-70%GC, Tm 67-74ºC***<br>gtcctcatcg**cccct**tcg**agggg**tcgcaac<br>gtcctcatcg**cccct**tcg**agggg**tcgcaac<br>gtcctcatcg**cccct**tcg**agggg**tcgcaac<br>gtcctcatcg**cccct**tcg**agggg**tcgcaac<br>gtgctcatcg**cccct**gcg**agggg**tcgcaac<br>gtcctcatcg**cccct**gtg**agggg**tcgcaat<br>gtcctcat**cgaccc**tatgaag**gggtcg**caa<br>gtcctcat**cga**t**cct**ctg**agg**gg**tcg**caac<br>gtcctcatcg**cccct**gcg**agggg**tcgcaac | ***36-44 bp, 62-81%GC, Tm 75-84ºC***<br>tgcggagcg*ggaggccggccagctcggc*ccagtcggc-*gvgl*001077<br>ctcccagacgaggacggcgtactgcccgacgaccag<br>aggaggcgggtatggatg*gtgccgccgccctcgacc*-*gvgl*006237<br>gcctcgggcacctcgtcctccaggcgcagccccgca<br>accagatcgcgttcgacgacaccgtcaccccggca<br>ctgttgcgctcatggcgcgcagccttgcgggctccgg<br>cgtggcgaggctggcagcggggcagcggggcagcgggtgacgtt<br>atgaacaggctgcggccctcgccgtcctggttggtc<br>gtctggacgcccgtgccccagcggccgatctcgcca |

| | |
|---|---|
| gtcttcatcg**cccct**gac**agggg**tcgcaac | ataaccagcagcgctgtcgatcgcgggtccaacccgt |
| gtcctcatcg**cccct**gcg**agggg**tcgcaac | The sequences of the first four Rs in the CRISPR 3 are the same and were utilized in this study |

* Purple, orange and red letters indicating the same base sequences in the repeats, respectively; Boldface letters with blue color indicating complementary sequences; Italic letters with green color indicating that the fragment sequences in the spacers can also be found in other sites of the *S. virginiae* IBL14 genome; *gvgl*00xxxx representing the temporary gene number in the *S. virginiae* IBL14 genome.

It was reported that approximately 0.4% of spacers (100 of 23550 spacers) were self-targeting spacers (fully matching a portion of endogenous genomic sequences that is not part of a CRISPR array, termed self-proto-spacer) and that such self-targeting was a mode of autoimmunity rather than gene regulation [30]. Although there are no full self-targeting spacers in the three *Svi*CRISPR arrays (Table 1), partial sequences (18 nt to 19 nt) of five spacers match a portion of five encoding gene sequences in the strain IBL14 genome (the five genes *gvgl*005738, *gvgl*005245, *gvgl*002878, *gvgl*001077 and *gvgl*006237 encode squalene-hopene cyclase/EC5.4.99.17, thymidine kinase/EC 2.7.1.21, ATP-binding cassette/ABC transporter integral membrane protein, TetR family transcriptional regulator and putative dehydrogenase, in proper sequence) (http://www.kegg.jp/; http://www.ncbi.nlm.nih.gov; http://www.uniprot.org/help/uniprotkb). It is worth noting that the nucleotide sequences of *Svi*CRISPR 3 fall within the sequences of the coding gene *gvgl*007771 (accession number in the NCBI database: KY243079) that shares the highest identity (77.34%) with the DUF721 domain-containing protein of *Streptomyces subrutilus* (WP_069924566.1). The gene *gvgl*007771 encodes an unknown functional protein, GVGL007771, which is widely present in bacteria (http://www.ncbi.nlm.nih.gov).

**The subtype I-B-*Svi* Cas system**

The architecture of the CRISPR-Cas system in the *S. virginiae* IBL14 genome was delineated as shown in Figure 1A (*gvgl*007770-*cas*7-*cas*5-*cas*3-*cas*4-*cas*1-*cas*2 in transcriptional direction). In contrast to the typical subtype I-B Cas system containing eight *cas* genes with the arrangement *cas*6-*cas*8b-*cas*7-*cas*5-*cas*3-*cas*4-*cas*1-*cas*2 in the transcriptional direction found in *Lactobacillus crispatus* VMC3, *Thermotoga neapolitana* DSM 4359, *Haloarcula marismortui* str. ATCC 43049 and *Clostridium kluyveri* CKL_2758-CKL_2751 [4,14,31], the *Svi*Cas system lacks the common gene *cas*6 and the signature gene *cas*8 used for the subtype classification in type I Cas systems, implying that Gvgl007770 may play the roles of Cas6 or Cas8 or other Cas proteins in the *Svi*Cas system. Based on the architectural similarity of *cas* genes between the *Svi*Cas system and subtype I-B Cas systems [4,14], the *Svi*Cas system was temporarily named the subtype I-B-*Svi* Cas system.

In the sequences of the seven *cas* genes, there are two sets of genes (*cas*7 with *cas*5 and *cas*3 with *cas*4) with the four overlapping bases "gtga", among which "gtg" (valine) serves as a start codon in *Svi*Cas5 and *Svi*Cas4 translation and "tga" as a stop codon in *Svi*Cas7 and *Svi*Cas3 translation, and four superficially redundant sequences, i.e., 29 nt between *gvgl*007770 and *cas*7, 161 nt between *cas* 5 and *cas* 3, 1 nt between *cas* 4 and *cas* 1 and 7 nt between *cas* 1 and *cas* 2. As expected, the number of bases in the overlapping and redundant sequences is not a multiple of three, to prevent misreading of the coding sequences. In addition, the gaps between *Svi*CRISPR 3 and *Svi*CRISPR 2, *Svi*CRISPR 2 and *gvgl*00770 as well as *cas*2 and *Svi*CRISPR 1 in the system are 645 (226+419) nt, 1240 nt and 139 nt, respectively.

**A**

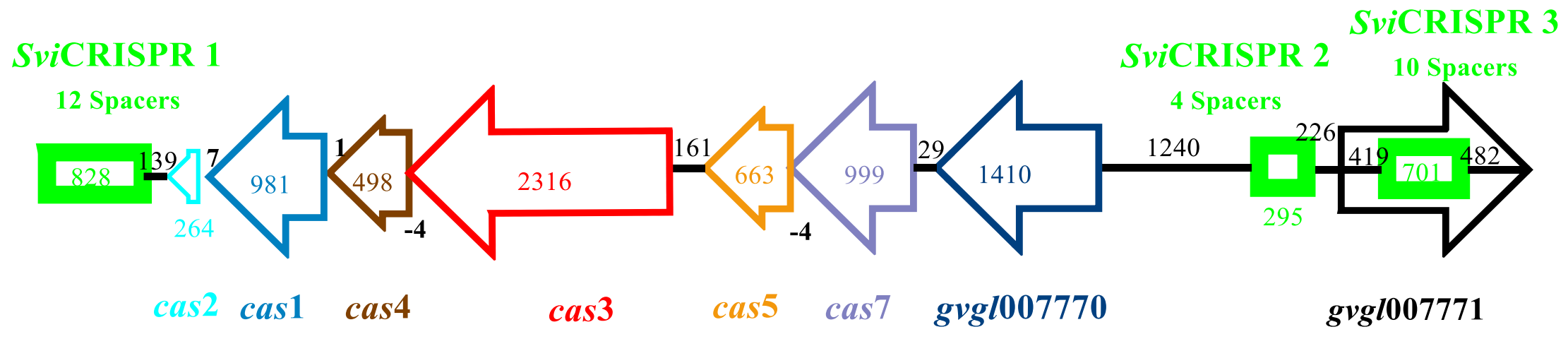

*Svi*CRISPR1: 96665-97492=828 bp; ***cas*2**: 97632-97895=264 bp; ***cas*1**: 97903-98883=981 bp; ***cas*4**: 98885-99382=498 bp; ***cas*3**: 99379-101694=2316 bp; ***cas*5**:101856-102518=663 bp; ***cas*7**: 102515-103513=999 bp; ***gvgl*007770/*cas*6**: 103543:104952=1410 bp; *Svi*CRISPR2: 106193-106487=295 bp; *gvgl*L007771: 106714:108315=1602 bp; *Svi*CRISPR3: 107133-107833=701 bp; arrow: transcriptional direction.

**B**

1 10 20 30 40 50 60 70

**VGRLDAVEDVFGGRFWPVVELAGLTHDAGKIPEGFQRMLAGYSRAWGERHEVASLGFLPALIGDPDVLLW**

c c c c c c c e e e e c c c c c c c e e e e e c c c c c c c c h h h h h h h h h h h h h h h c c h h h h h h h c c c c c c c c c c e e e e e

**VATAVATHHRPLTGQNGRDLQTLYSGVTITELAHRFGPFDPRAVPALEAWLRASAIRVGLPAAAVPDDGT**

e e e e e e c c c c c c c c c c c c c c e e e e e c c c e e e h h h h c c c c c c c c c c h h h h h h h h h h h h h c c c c c c c c c c c c c

**LTDTGVVAGAHQLLEEILDRWADRVRPEVGLAAVLLQGAVTLADHLSSAHQALPTVQPLGAGFRSRLEKE**

e c c c c h h h h h h h h h h h h h h h h h h h c c c c h h h h h h h h h h h h h h h h h h h h h h h c c c c c c c c c c h h h h h h h h

**FAERGRTLRAHQLEAATVTGHLLLRGPTGSGKTEAALLWAASQVEALKAEGRGVPRVFFTLPYLASINAM**

h h h h h h h h h h h h h h h h h h h h c e e e e c c c c c c c h h h h h h h h h h h h h h h h h h c c c c c c e e e e e e c c c h h h h h h

**ATRLGDTLGDGEAVGVAHSRAASYHLAQAIAPQDGDEEDEHGAPCRVDAAAKALSRAAATKLFRESVRVA**

h h h h c c c c c c c c e e e e h h h h h h h h h h h h h h c c c c c c c c c c c c c c h h h h h h h h h h h h h h h h h h h h c c c c c

**TPYQLLRAALAGPAHSGILIDAANSVFILDELHAYDARRLGYILASARLWERLGGRITVLSATLPRALAD**

c h h h h h h h h h c c c c c c c h h h h h h c h h h h h h h h h h h h h h h h h h h h h h h h h h h h h h h c c c c e e e e e e c c c h h h h h h

**LFESTLTAPITFLDTPDLGLPARHLLHTRGHHLTDPATLEEIRLRLSRDESVLVIANNVSQAIALYEQLA**

h h h *c c c c c c c* e e e *c c c c c c c c* h h h h h h h *c c c c c c c c* h h h h h h h h h *c c c c c* e e e e e *c c* h h h h h h h h h h *c c*

**PDVCERFGQDAALLLHSRFRRMDRSRIEQKIADRFATVAPDAQNSRKPGLVVATQVVEVSLDVDFDVLFT**

*c c* h h h h h *c c* h h h h h h h h h h h h h h h h h h h h h h h h h h h h *c c c c c c c c c c c* e e e e e e e e e e *c c c c c* h h h h h *c*

**GAAPLEALLQRFGRTNRVGARPPADVIVHHPAWTTRRRQPGEYADGIYPREPVESAWHILTRNHGRVIDE**

*c c c* h h h h h h h h *c c c c c c c c c c c c c c* e e e *c c c c c c c c c c c c c c c c c c c c c c c c c c c c c* h h h h h h *c c c c c* h h h h

**ADATAWLDEVYATDWGRQWHREVLERRERFDRAFLQFRYPFEDRTDLADTFDELFDGSEAILAEDQDAYS**

h h h h h h h *c* e e e e *c c c c c* h h h h h h h h h h h h h h h h h h h *c c c c c c c c c c c* h h h h h h h h *c c* h h h h h h h h h h h h h

**AALAAPDGDHPGAGRLLAEEYLIPVPHWASPLSRYEKQLKVRVINGDYHPDHGLMAVRGLPQPAYRAGEV**

h h h h *c c c c c c c c c c* h h h h h h h h h h *c c c c c c c c* h h h h h h h h e e e e e *c c c c c c c c* h h h h h *c c c c c c c c c c c c* e e

**L**

e

c: random coil, e: extended strand, h: alpha helix, letters with red underline represent HD domain, letters with red dash-line represent Hel domain

**Figure 1. The architecture of the subtype I-B-*Svi* CRISPR-Cas system in strain *S. virginiae* IBL14 and the predicted secondary structure of *Svi*Cas3 in the strain IBL14. (A)** The architectural diagram of the subtype I-B-*Svi* CRISPR-Cas system loci. **(B)** The predicted secondary structure of *Svi*Cas3.

Furthermore, the biochemical properties of the seven *Svi*Cas proteins in the subtype I-B-*Svi cas* gene operon were investigated (Table S3) [4] (http://web.expasy.org/protparam). All the *Svi*Cas proteins are hydrophilic as expected, i.e., their grand average of hydropathicity (GRAVY) values are negative. In the six definite *Svi*Cas proteins (*Svi*Cas1, 2, 3, 4, 5 and 7), the numbers of cysteine residues, respectively, are 2, 1, 2, 5, 1 and 0, and the numbers of methionine residues are 6, 0, 4, 0, 2 and 2. *That is,* there are no disulfide bonds in *Svi*Cas prtoeins 2, 5 and 7 which are generally responsible for RNA processing [1,4,13,14]. Strikingly, among the seven *Svi*Cas proteins, the signature protein *Svi*Cas3 has the largest theoretical molecular weight (MW) of 84352.40 Da (771 aa), the lowest theoretical isoelectric point (pI) of 5.78, the lowest instability index (I-ind) of 35.49 and the highest GRAVY of -0.153.

The results of Nucleotide BLAST for the seven *cas* genes showed that *cas*1, *cas*3, *cas*4 and *cas*7 had the highest identity with the four genes (74.51%, 74.18%, 75.09% and 70.78%, in proper order): Sequence ID-KM527027.1 [*Salinispora pacifica* strain DSM 45549 Cas1 protein I-B (*cas*1) gene, complete cds], Sequence ID-CP054919.1 [*Kitasatospora sp*. NA04385 chromosome, complete genome], Sequence ID-CP061007.1 [*Saccharopolyspora spinosa* strain CCTCC M206084 chromosome, complete genome] and Sequence ID-CP016076.1 [*Actinoalloteichus fjordicus* strain ADI127-7 chromosome, complete genome], while the results of Nucleotide BLAST for the three genes *cas*2, *cas*5, and *gvgl*007770 were "No significant similarity found" (http://www.ncbi.nlm.nih.gov). Furthermore, we performed Protein BLAST on the seven *Svi*Cases and found that *Svi*Cas1, *Svi*Cas2, *Svi*Cas3, *Svi*Cas4, *Svi*Cas5, *Svi*Cas7 and Gvgl007770 shared the highest identity with the seven proteins (85.89%, 85%, 64.77%, 84.85%, 75.71%, 78.92% and 59.23%, respectively): Sequence ID-WP_123471944.1 [type I-B CRISPR-associated endonuclease Cas1, *Streptomyces sp*. CEV 2-1], Sequence ID-WP_065001230.1 [CRISPR-associated endonuclease Cas2, *Streptomyces sp*. H-KF8], Sequence ID-WP_121433360.1 [CRISPR-associated helicase Cas3', *Actinomadura pelletieri*], Sequence ID-WP_055591556.1 [CRISPR-associated protein Cas4, *Streptomyces sp*. H-KF8], Sequence ID-WP_158013542.1 [CRISPR-associated protein Cas5, partial, *Streptomyces thermoautotrophicus*],

Sequence ID-WP_045691965.1 [type I-B CRISPR-associated protein Cas7/Cst2/DevR, *Streptomyces rubellomurinus*] and Sequence ID-WP_147449255.1 [hypothetical protein, *Actinomadura pelletieri*] (http://www.ncbi.nlm.nih.gov) [32]. In the subtype I-B-*Svi* Cas system, the unknown protein Gvgl7770 is neither Cas6 nor Cas8. What is the function of Gvgl7770? The classic *Sp*Cas9 from *Streptococcus pyogene* has the highest identity (85.84%) to Cas9 (Sequence ID: WP_138125798.1) from *Streptococcus dysgalactiae* and the second highest (85.09%) to Cas9 (Sequence ID: WP_159583349.1) from *Streptococcus halichoeri*, indicating that *Svi*Cas3 (64.77%) is more distinctive than *Sp*Cas9 and probably performs some functions that neither *Sp*Cas9 nor other Cas3 proteins share. In summary, the subtype I-B-*Svi* Cas system has unique features.

Cas3 superfamily proteins typically contain an N-terminal histidine-aspartate (HD) domain (belonging to a superfamily of metal-dependent phosphohydrolases) and a C-terminal ATP-dependent superfamily 2 helicase (Hel) domain [13,33-35]. The secondary structure of *Svi*Cas3 was predicted (Figure 1B) by online software (GOR4) (https://npsa-prabi.ibcp.fr/cgi-bin/npsa_automat.pl?page=npsa_gor4.html). *Svi*Cas3 includes 49.42% (381 aa) alpha helix (h) (55.92% in *Sp*Cas9), 9.99% (77 aa) extended strand (e) (8.55% in *Sp*Cas9) and 40.60% (313 aa) random coil (c) (35.53% in *Sp*Cas9). Notably, it contains seven repeat motifs, consisting of four (AALL, AHQL, GLPA and LEEI) or three (GRL, RFG and VAT) amino acid residues. The motifs AALL, AHQL and LEEI are only present in the alpha helix, and the motif GLPA is present in the random coil. However, the functions of these repeat motifs are unclear.

Cas9 in type II systems contains two nuclease domains, i.e., an HNH-like nuclease domain in the middle of the protein that cleaves the crRNA complementary strand and a RuvC-like nuclease domain near the amino (N) terminus that cleaves the noncomplementary strand, which generates blunt-ended double-strand breaks (DSB), typically 3 nt in front of the 3' end of a proto-spacer (a sequence in the target genomic DNA that is consistent with a spacer) adjacent motif (PAM) [1,4,36]. Cas9 is also involved in the trimming of pre-crRNA to form crRNA via a Cas9-crRNA-tracrRNA ternary complex and in the silencing of invasive DNA [32,37]. Unlike Cas9 cutting both the target DNA strand (TS) and the nontarget DNA strand (NTS) using its two nuclease domains, Cas3 can perform a crRNA-guided, PAM-dependent cleavage sequentially on both the NTS and TS using its HD domain with the help of its Hel domain in the presence of ATP [13,23,32,38-40], implying that it probably produces a local single strand broken (SSB) DNA or double-strand broken (DSB) DNAs with sticky-ends. Analytical results show *Svi*Cas3 exhibits an HD-like nuclease (HD domain) at the N terminus (1-186 aa) and a DNA helicase (Hel domain) in the middle of the protein (232-591 aa) (http://www.ebi.ac.uk/interpro/sequencesearch/iprscan5-S20170430-040754-0890-51239778-oy), implying that the carboxyl termini of Cas3 and Cas9 are both nonconserved.

**Self-targeting genome editing**

Self-targeting genome editing refers to endogenous genome editing that is triggered by delivering an engineered gene editing plasmid into a cell and further implemented by the Cas system(s) and homology-directed repair (HDR) or the non-homologous end joining (NHEJ) systems of the cell itself, in which the spacer designed in the guide DNA (g-DNA) fragment fully matches a portion of the complete genomic sequences of the cell itself. Since 2015, we have repeatedly demonstrated that the endogenous subtype I-B-*Svi* Cas system can be programmed for self-targeting genome editing by transforming a gene editing plasmid into the cells of *S. virginiae* IBL14 (https://kns.cnki.net/kns8?dbcode=CDMD).

To develop self-targeting gene editing tools and investigate the mechanism of antibiotic metabolism in *S. virginiae* IBL14, we selected seven endogenous genes participating in penicillin metabolism as targets (*sviipe* with 1143 nt: encoding isopenicillin-N epimerase, EC 5.1.1.17; *svipam*1 with 2901 nt and *svipam*2 with 2049 nt: encoding penicillin amidase, EC 3.5.1.11; *svibla*1 with 2082 nt, *svibla*2 with 1542 nt , *svibla*3 with 1245 nt and *svibla*4 with 921 nt: encoding beta-lactamase, EC 3.5.2.6; corresponding accession numbers in the NCBI database: KY243071, KY243072, KY243073, KY243074, KY243075, KY243076 and KY243077) for self-targeting genome editing, and chose plasmid pKC1139 [41] as a vector to construct gene editing plasmids (Figure S1A and Table S1). Additionally, we arbitrarily designed the fragments of template DNA donor (t-DNA) (upstream homologous arm / UHA plus downstream homologous arm / DHA plus insertion sequence if present: 201+391+807=1399 nt, 945+1102=2047 nt, 414+616=1030 nt, 219+322=541 nt, 418+648=1066 nt, 414+424=838 nt and 727+832=1559 nt for the genes *sviipe*, *svipam*1, *svipam*2, *svibla*1, *svibla*2, *svibla*3 and *svibla*4, respectively), the insertion fragment (807 nt for chloramphenicol acetyltransferase biosynthesis / *sviipe*) and the deletion fragments (327 nt / *svipam*1, 447 nt / *svipam*2, 577 nt / *svibla*1, 402 nt *svibla*2, 318 nt / *svibla*3 and 513 nt / *svibla*4) (Figure 2C and Table S2).

Experimentally, we first extracted the *S. virginiae* IBL14 genome (~8.4 Mbp) and plasmid pKC1139 (6308 bp), and chemically synthesized g-DNA fragments (essential elements: promoter-repeat-spacer-repeat-terminator) (Figure S2A) (http://www.addgene.org/; http://www.ncbi.nlm.nih.gov). After preparing the t-DNA fragments (Figure S2B) of the seven genes of interest and constructing the seven gene editing plasmids (Figures S2C and S2D, Tables S1 and S2), we transformed these gene editing plasmids into the protoplasts of *S. virginiae* IBL14 following the procedures described in the section "Construction of gene-edited mutants of *S. virginiae* IBL14". For each gene editing, three randomly picked single colonies on MR2YE plates were verified as potential gene-edited mutants (*Strain name-edited gene abbreviation*) by DNA electrophoresis (Figure 2A, Table S2) following the procedures described in Materials and Methods [42]. Next, we constructed two double gene deletion mutants (*S. virginiae* IBL14-Δ*svipam*1Δ*sviipe*::*cat* and *S. virginiae* IBL14-Δ*svipam*1Δ*svibla*1) using the plasmid-removed mutant *S. virginiae* IBL14-Δ*svipam*1 as a host according to the same procedures as described above. The results demonstrated that the strategy of iterative gene editing in self-targeting genome editing of *S. virginiae* IBL14 is convenient and practicable (Figures 2B and 2C, Tables S1 and S2).

**A**

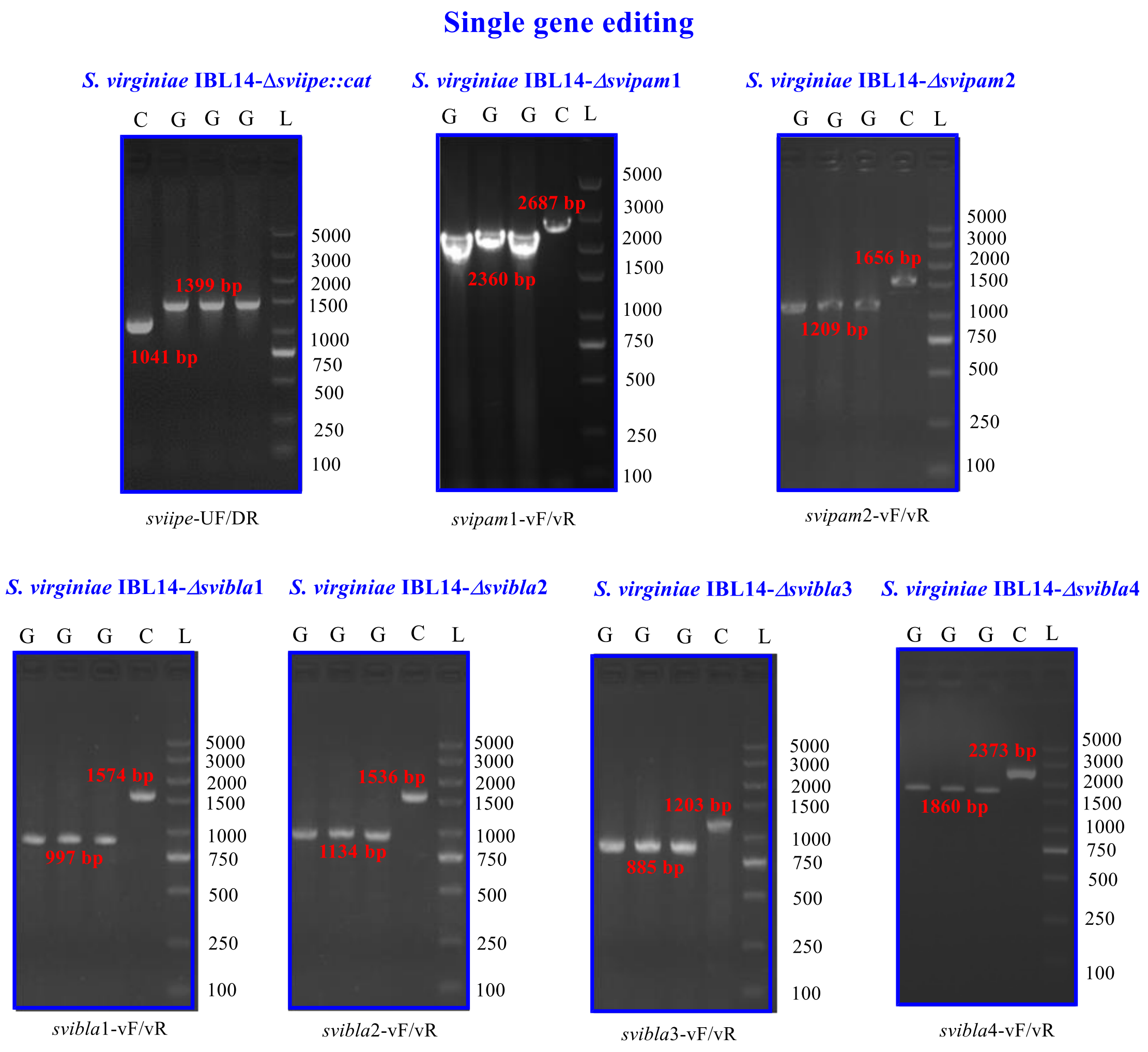

C: control- *S. virginiae* IBL14-pKC1139; G: gene-edited mutant; L: ladder

**B**

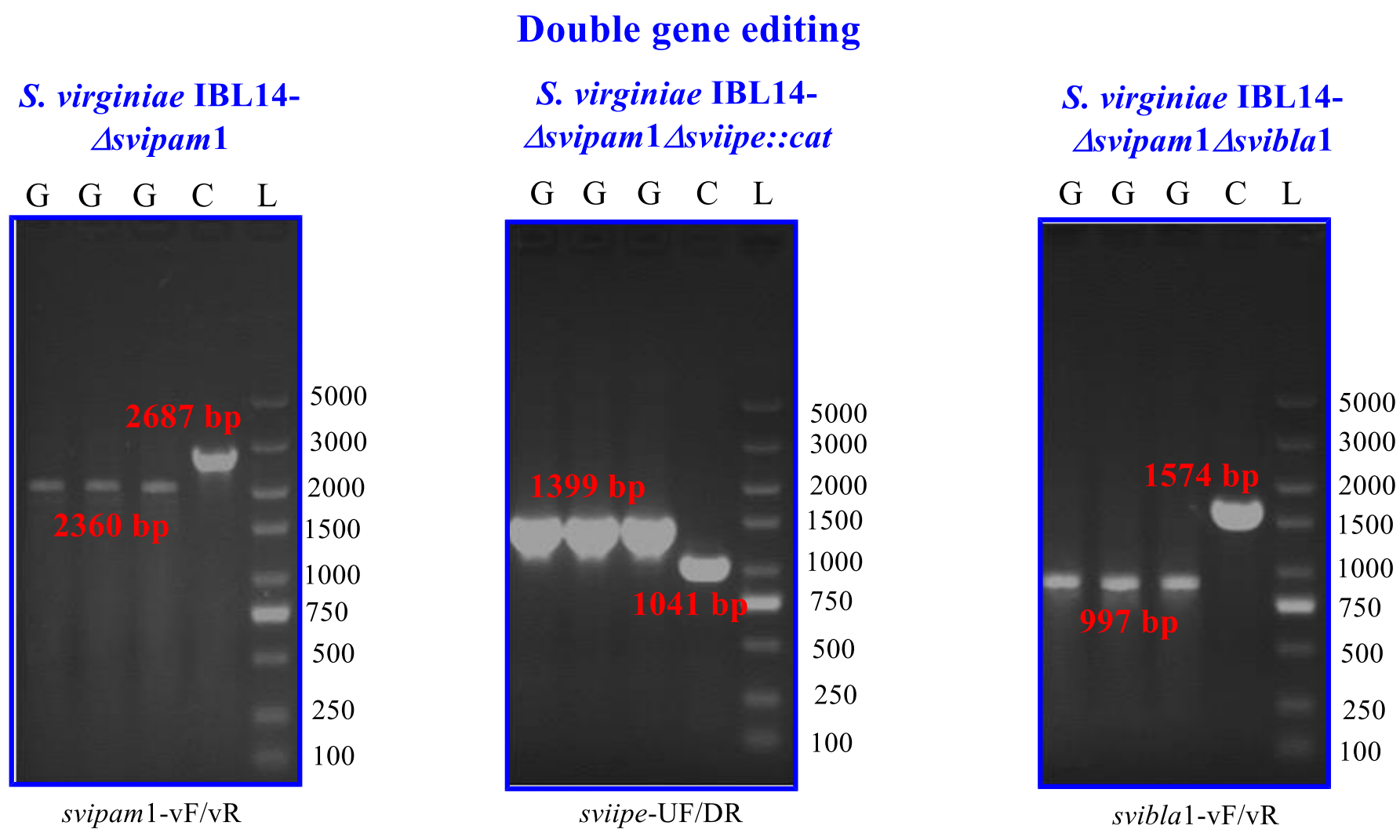

C: control-*S. virginiae* IBL14-pKC1139; G: gene-edited mutant; L: ladder

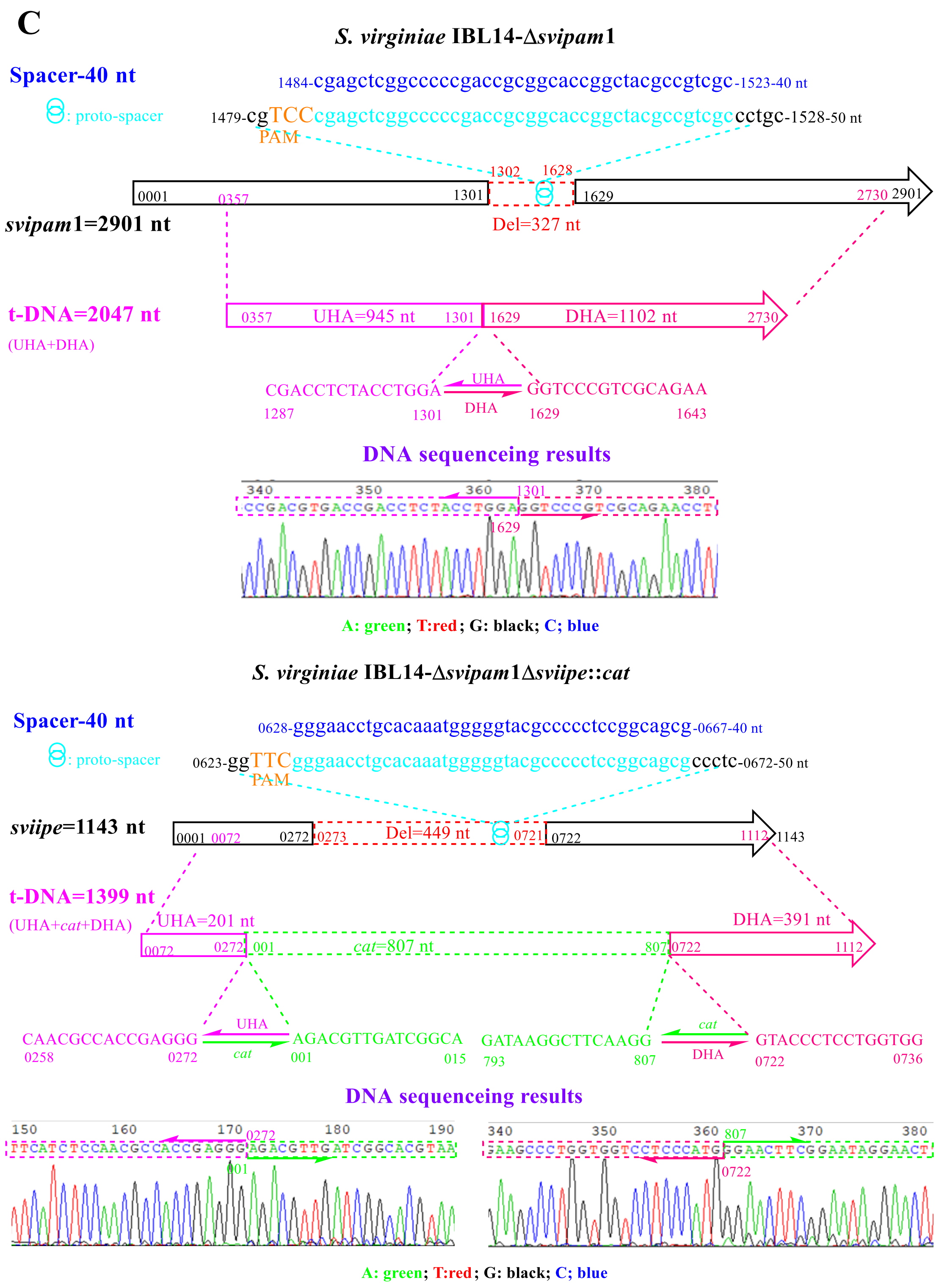
C
S. virginiae IBL14-Δsvipam1
Spacer-40 nt
1484-cgagctcggcccccgaccgcggcaccggctacgccgtcgc-1523-40 nt
: proto-spacer
1479-cgTCCcgagctcggcccccgaccgcggcaccggctacgccgtcgccctgc-1528-50 nt
PAM
1302
1628
0001
0357
1301
1629
2730
2901
svipam1=2901 nt
Del=327 nt
t-DNA=2047 nt
(UHA+DHA)
0357
UHA=945 nt
1301
1629
DHA=1102 nt
2730
CGACCTCTACCTGGA
UHA
DHA
GGTCCCGTCGCAGAA
1287
1301
1629
1643
DNA sequenceing results
340
350
360
1301
370
380
CCGACGTGACCGACCTCTACCTGGAGGTCCCGTCGCAGAACCT
1629
A: green; T:red; G: black; C; blue
S. virginiae IBL14-Δsvipam1Δsviipe::cat
Spacer-40 nt
0628-gggaacctgcacaaatgggggtacgccccctccggcagcg-0667-40 nt
: proto-spacer
0623-ggTTCgggaacctgcacaaatgggggtacgccccctccggcagcgccctc-0672-50 nt
PAM
sviipe=1143 nt
0001
0072
0272
0273
Del=449 nt
0721
0722
1112
1143
t-DNA=1399 nt
(UHA+cat+DHA)
UHA=201 nt
DHA=391 nt
0072
0272
001
cat=807 nt
807
0722
1112
CAACGCCACCGAGGG
UHA
cat
AGACGTTGATCGGCA
GATAAGGCTTCAAGG
cat
DHA
GTACCCTCCTGGTGG
0258
0272
001
015
793
807
0722
0736
DNA sequenceing results
150
160
170
0272
180
190
TTCATCTCCAACGCCACCGAGGGAGACGTTGATCGGCACGTAA
001
340
350
360
807
370
380
GAAGCCCTGGTGGTCCTCCCATGGGAACTTCGGAATAGGAACT
0722
A: green; T:red; G: black; C; blue

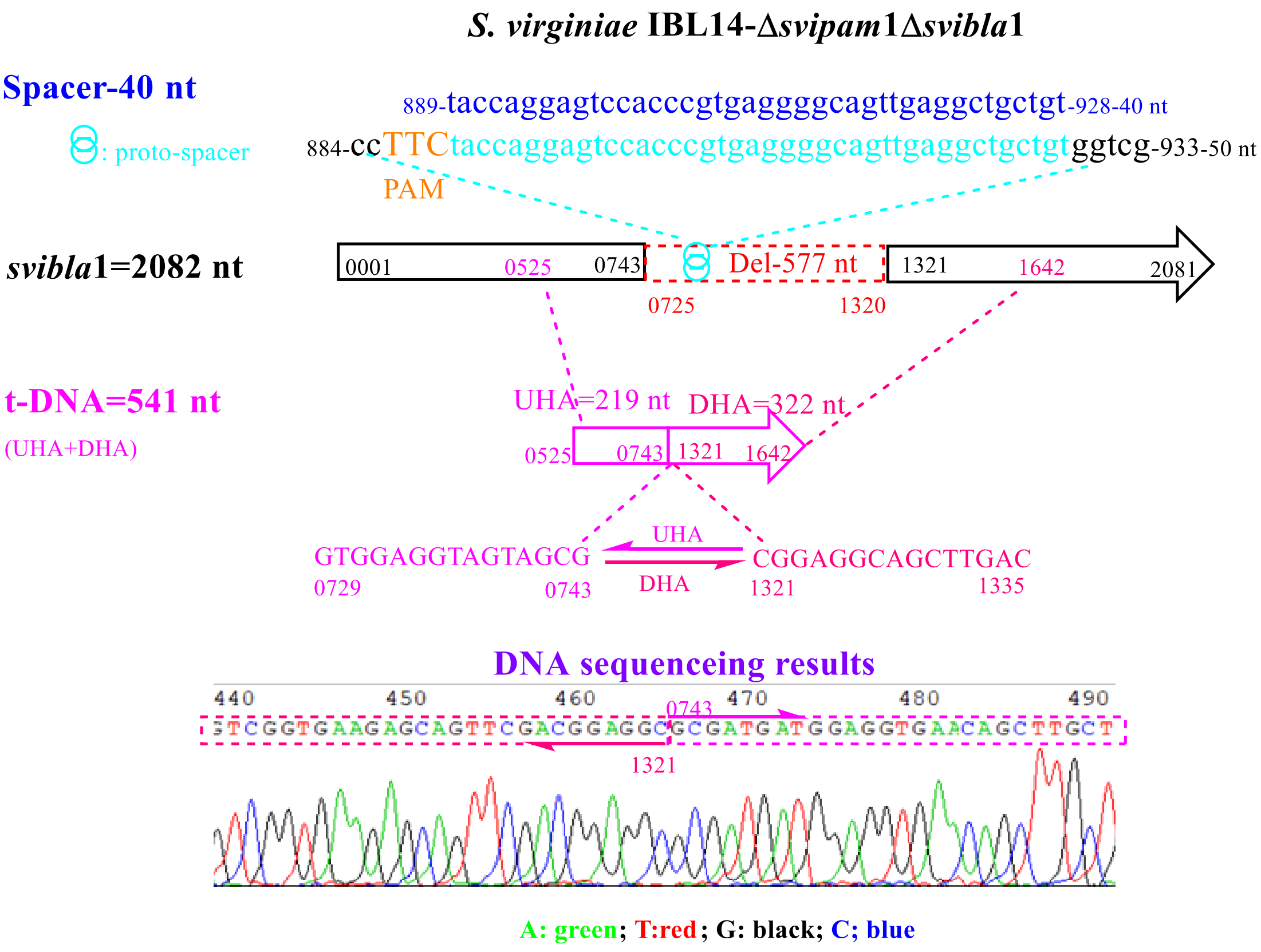

**Figure 2. RNA-guided self-targeting genome editing in *S. virginiae* IBL14. (A)** DNA gel electrophoresis of the PCR products of edited sequences in single gene-edited mutants. **(B)** DNA gel electrophoresis of the PCR products of edited sequences in double gene-edited mutants. **(C)** The selected proto-spacers, the engineered t-DNAs as well as the DNA sequencing results of the double gene-edited mutants (The junction sequences between UHA and DHA, UHA and *cat*, and *cat* and DHA are highlighted in Figure 2C).

In self-targeting genome editing, including the deletion and insertion of DNA sequences of the genes of interest, all the experimental results (Figure 2 and Figure S2, Table S2) were correct, as we anticipated, implying that it is not impossible for other class 1 (types I, III and IV) CRISPR-Cas systems to be utilized for self-targeting genome editing. In particular, all tested mutants were the correctly gene-edited mutants, motivating us to further design and construct gene editing tools based on the subtype I-B-*Svi* CRISPR-Cas system for non-self-targeting genome editing.

**Genome editing in *E. coli* JM109(DE3)**

To test whether the subtype I-B-*Svi* CRISPR-Cas system was effective in non-self-targeting genome editing, we selected the gene *lacZ* (3075 nt) encoding beta-D-galactosidase as a target in *E. coli* JM109 (DE3) genome for its advantages of its blue-white selection and a known CRISPR-Cas system, which is incapable of self-targeting genome editing (http://crispr.i2bc.paris-saclay.fr/) [8,43]. Following the procedures described in both Construction of gene editing plasmids and Construction of Cas expression plasmids, we tentatively designed and constructed a set of gene editing tools composed of a gene editing plasmid pKC1139-t/g-Δ*lacZ* and a Cas expression plasmid pCas-*cas*7-5-3-4-1-2 bearing all six definite *cas* genes of the subtype I-B-*Svi* Cas system (Figures S1 and S3, Tables S1 and S2). The arbitrarily designed fragment sizes of t-DNA (UHA plus DHA) and deletion in the gene editing of *lacZ* were 1144 nt (423+721 nt) and 420 nt, respectively, and the selected PAM was ttc (Figure 3C and Table S2). After transformation of the Cas expression plasmid pCas-*cas*7-5-3-4-1-2 and the gene editing plasmid pKC1139-t/g-Δ*lacZ* into *E. coli* JM109 (DE3) competent cells, three randomly selected white single colonies on LBPETs as the candidate gene-edited mutant *E. coli* JM109(DE3)-Δ*lacZ* (Figure 3A) were verified by DNA electrophoresis (Figure 3B) and DNA sequencing analysis (Figure 3C). The results in Figure 3 and Figure S3 suggested that the subtype I-B-*Svi* CRISPR-Cas system could also be developed as a gene editing tool for genome editing in other prokaryotic species.

**A**

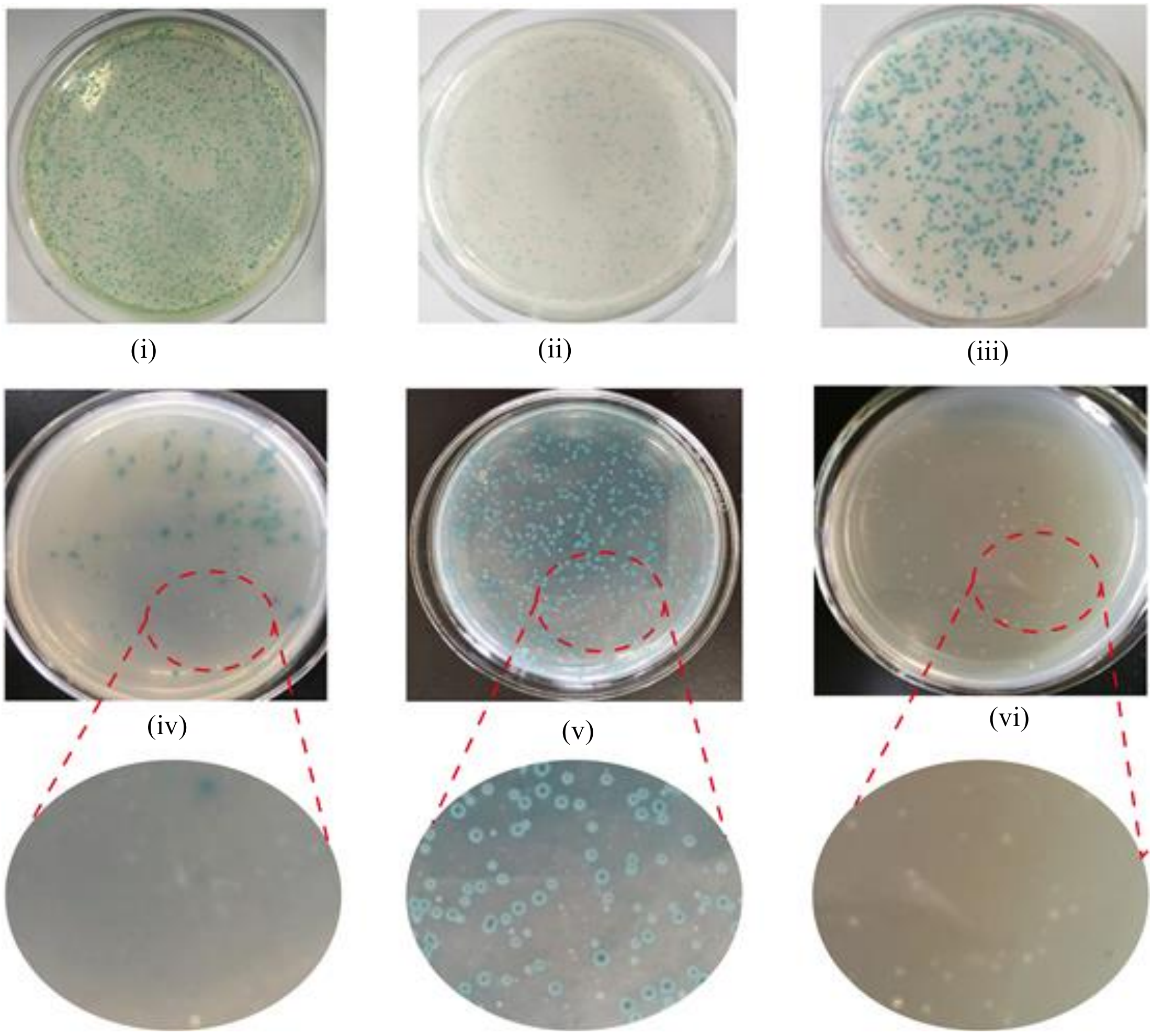

**i-iii:** *E. coli* JM109 (DE3) or its single plasmid transformants by pKC1139-t/g-Δ*lacZ*, pKC1139-t/g-Δ*lacZ*::*cat*, pCas-*cas*7-5-3-4-1-2, pCas-*cas*6-7-5-3, pCas-*cas*7-5-3 or pCas-*cas*3.
**iv-vi:** double plasmid transformants by pKC1139-t/g-Δ*lac*Z plus pCas-*cas*7-5-3-4-1-2, pCas-*cas*6-7-5-3, pCas-*cas*7-5-3 or pCas-*cas*3, respectively.

**B**

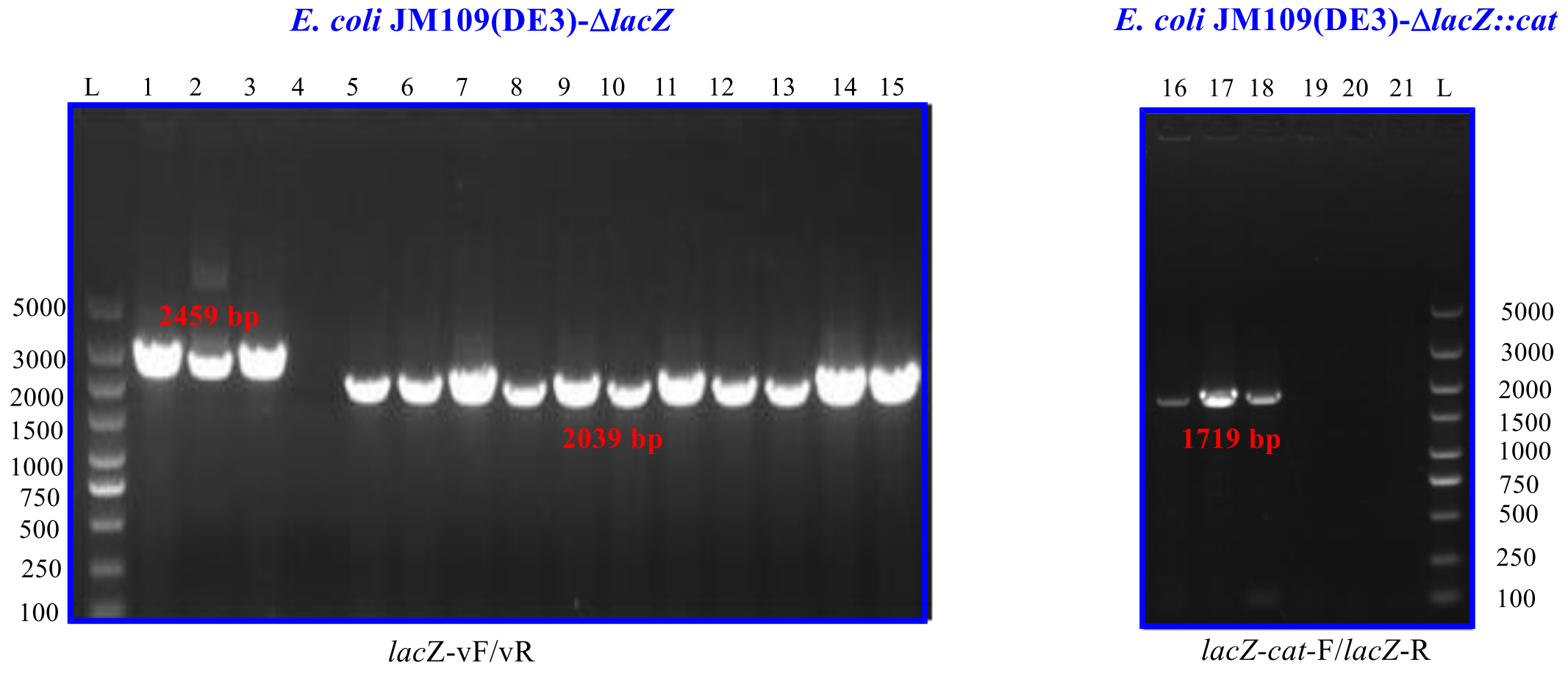

L: ladder
1, 21: *E. coli* JM109(DE3)
2: *E. coli* JM109(DE3) transformant by pKC1139-t/g-Δ*lacZ*
3, 20: *E. coli* JM109(DE3) transformant by pCas-*cas*3
4: dd$H_2O$
5-7: *E. coli* JM109(DE3)-Δ*lacZ* by pCas-*cas*3+pKC1139-t/g-Δ*lacZ*
8-10: *E. coli* JM109(DE3)-Δ*lacZ* by pCas-*cas*7-5-3+pKC1139-t/g-Δ*lacZ*
11-13: *E. coli* JM109(DE3)-Δ*lacZ* by pCas-*cas*7-5-3-4-1-2+pKC1139-t/g-Δ*lacZ*
14-15: *E. coli* JM109(DE3)-Δ*lacZ* by pCas-*cas*6-7-5-3+pKC1139-t/g-Δ*lacZ*
16-18: *E. coli* JM109(DE3)-Δ*lacZ*::*cat* by pCas-*cas*3+pKC1139-t/g-Δ*lacZ*::*cat*
19: *E. coli* JM109(DE3) transformant by pKC1139-t/g-Δ*lacZ*::*cat*

* 14-15: *E. coli* JM109(DE3)-Δ*lacZ* by pCas-*cas*6-7-5-3+pKC1139-t/g-Δ*lacZ*, where *cas*6 temporarily stands for *gvgl*007770, indicating that the product Gvgl007770 of gene *gvgl*007770 has no side effect in the genome editing of strain *E. coli* JM109(DE3).

C

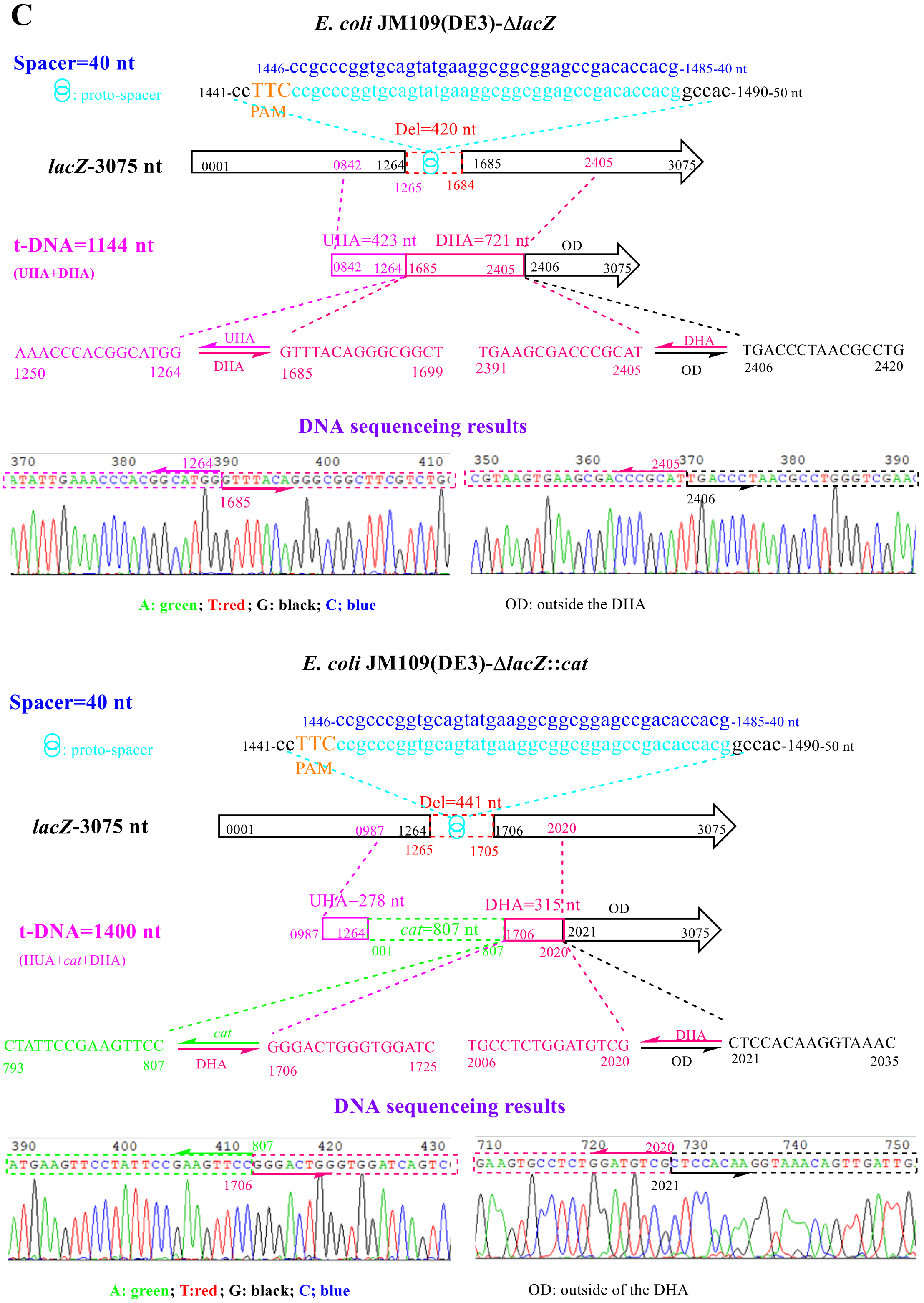

**Figure 3. RNA-guided genome editing in *E. coli* JM109(DE3). (A)** Typical features of both host *E. coli* JM109(DE3) and its transformants on LBPETs. **(B)** DNA gel electrophoresis of the PCR products of edited sequences in the genome editing of *E.*

*coli* JM109(DE3) (Lanes 16-21: PCR products occur only while the genome of the gene-edited mutant *E. coli* JM109(DE3)-Δ*lacZ*::*cat* is utilized as template because the primer *lac*Z-*cat*-F is a portion of the gene *cat* in t-Δ*lacZ*::*cat* and the primer *lacZ*-R is part of the gene *lacZ* but is not included in t-Δ*lacZ*::*cat*.). **(C)** The selected proto-spacers, the engineered t-DNAs as well as the DNA sequencing results of the PCR products of edited sequences from the two gene-edited mutants *E. coli* JM109 (DE3)-Δ*lacZ* and *E. coli* JM109 (DE3)-Δ*lacZ*::*cat* genomes, respectively (The junction sequences between UHA and DHA, *cat* and DHA, and DHA and the outside of DHA are highlighted in Figure 3C).

It has been reported [1,4,36,44] that proteins Cas1, Cas2 and Cas4 are, respectively, a metal-dependent deoxyribonuclease with a unique fold consisting of an N-terminal β strand domain and a C-terminal α-helical domain, a small endoribonuclease specific to U-rich regions and a RecB-like nuclease with a three-cysteine C-terminal cluster (the three-cysteine C-terminal cluster of *Svi*Cas4 with 165 aa is composed of Cys153, Cys156 and Cys162), and all of them are responsible for new spacer acquisition in the adaptation phase in CRISPR-Cas systems that direct microbial immunity to exogenous DNA. Accordingly, we designed and constructed the Cas expression plasmid: pCas-*cas*7-5-3, harbouring the three genes *cas*7, *cas*5 and *cas*3 (Figures S1B, S3B and S3C). Together with the plasmid pKC1139-t/g-Δ*lacZ*, the plasmid pCas-*cas*7-5-3 was delivered into *E. coli* JM109 (DE3) competent cells. White single colonies on LBPETs, potentially gene-edited mutant *E. coli* JM109(DE3)-Δ*lacZ*, were also obtained (Figure 3A) and further confirmed by DNA electrophoresis and sequencing analysis (Figures 3B and 3C). These results demonstrated that Cas1, Cas2 and Cas4 are not required in non-self-targeting genome editing based on the subtype I-B-*Svi* CRISPR-Cas system and suggested that the gene editing tools constructed based on *cas*7-5-3 can be simply and easily used for the genome editing of other prokaryotic microorganisms because of their monocistron and relatively low molecular weight advantages (a total of 143813.69 Da / 1323 aa for *Svi*Cas7, *Svi*Cas5 and *Svi*Cas3; 158441.41 Da / 1368 aa for the *Sp*Cas9).

Cas5, Cas6, Cas7 and Cas8 are subunits of the Cascade/Csy complex in the type I CRISPR-Cas system, which are responsible for recognizing the repeat-derived 5' handle of mature crRNA, yielding mature crRNA, forming the backbone of the Cascade complex, and recognizing the PAM sequence and recruiting Cas3 (or Cas2/3) [1,4,14,20,21,44,45], respectively. Therefore, we further simplified and constructed a Cas expression plasmid pCas-*cas*3 harbouring only the single gene *Svicas*3 due to the lack of definite Cas6 and Cas8 proteins in this subtype I-B-*Svi*Cas system, and a gene editing plasmid pKC1139-t/g-Δ*lacZ*::*cat* (the same as pKC1139-t/g-Δ*lacZ* except for t-Δ*lacZ*::*cat* instead of t-Δ*lacZ*) (Figure 3C, Figures S1 and S3, Tables S1 and S2). After transforming the plasmids pCas-*cas*3 plus pKC1139-t/g-Δ*lacZ* or pCas-*cas*3 plus pKC1139-t/g-Δ*lacZ*::*cat* (t-DNA=UHA+*cat*+DHA: 1400 bp=278 bp+807 bp+315 bp; deletion fragment=441 bp; PAM: ttc) into *E. coli* JM109 (DE3) competent cells, we obtained white gene-edited mutants of *E. coli* JM109(DE3)-Δ*lacZ* and *E. coli* JM109(DE3)-Δ*lacZ*::*cat*, respectively (Figure 3A), and verified them by DNA electrophoresis and DNA sequencing analysis. Luckily, the band size of DNA electrophoresis and the result of DNA sequencing were exactly as designed (Figures 3B and 3C, Table S2), demonstrating that in the non-self-targeting genome editing of *E. coli* JM109(DE3), *Svi*Cas3 is necessary while both *Svi*Cas7 and *Svi*Cas5 are auxiliary. This finding breaks the view that Cascade is necessary in genome editing directed by type I CRISPR-Cas systems [22-24] but is consistent with the view that a full R-loop formation (the R-loop–forming Cascade) enables Cas3 to bind to the NTS bulge and thus reduces off-target effects [20,23,39]. Notably, during CRISPR interference in type I-C CRISPR-Cas systems, Cas3-mediated target DNA recognition and cleavage are independent of the composition and architecture of the Cascade surveillance complex [40], which also indirectly supports our results.

The main disadvantage in the genome editing of *E. coli* JM109(DE3) was that many blue or blue-white colonies occurred in addition to white colonies on LBPETs, and many small white colonies switched to blue-white colonies and even to blue colonies (Figure 3A), which resulted in two bands in the DNA electrophoresis of colony PCR. The reason is that the cells of the correctly gene-edited mutant *E. coli* JM109 (DE3)-Δ*lacZ* initially hydrolyzsed the antibiotics apramycin and kanamycin, and then the host *E. coli* JM109 (DE3) cells grew, which was validated by our experiments that small white colonies early grown were correctly *lacZ*-edited mutants, but blue colonies later grown were not.

**Genome editing in *C. glutamicum***

To further verify the effectiveness of the single *Svi*Cas3 enzyem in genome editing of prokaryotic microorganisms, an important industrial (producing diverse amino acids) strain *Corynebacterium glutamicum* ATCC 13032 lacking the CRISPR system (because Cas9 is toxic to *Corynebacterium glutamicum* ATCC 13032 cells, the genome of the cells is difficult to be edited) (http://crispr.i2bc.paris-saclay.fr/) [46] was selected as the host of interest and the gene *ldh* with 945 nt (encoding lactate dehydrogenase, which catalyses the reversible conversion of lactic acid to pyruvate) was selected as the gene of interest. In the genome editing of the strain *C. glutamicum* ATCC 13032, we designed and constructed the Cas expression plasmid pEC-XK99E-*cas*3 (Figures S4B and S4C, Tables S1 and S2) and a gene editing plasmid pXMJ19-t/g-Δ*ldh*::*egfp* (*egfp* encoding an enhanced green fluorescent protein) (Figures S4A and S4C). The arbitrarily designed fragment sizes of t-DNA (UHA plus *egfp* plus DHA) and deletion in the gene editing of *ldh* were 2402 nt (776+726+900 nt) and 1998 nt, respectively, and the selected PAM was ttc (Figure 4C and Table S2). After delivering the two plasmids pXMJ19-t/g-Δ*ldh*::*egfp* and pEC-XK99E-*cas*3 into *C. glutamicum* ATCC 13032 competent cells, we obtained potential gene-edited mutants on LBHIS plates with 25 μg/ml chloramphenicol and 50 μg/ml kanamycin and tested them by basic PCR (primers: *ldh*-vF / vR and *ldh*-*egfp*-vF / *ldh*-vR, respectively) and consequent sequencing analysis (Figure 4) following the procedures described in the section "Construction of gene-edited mutants of *C. glutamicum* ATCC 13032". Unfortunately, transformants of both *C. glutamicum* ATCC 13032-pXMJ19-t/g-Δ*ldh*::*egfp* and *C. glutamicum* ATCC 13032-Δ*ldh*::*egfp* showed green fluorescence, suggesting that the arbitrarily designed 776 bp-HUA fragment in pXMJ19-t/g-Δ*ldh*::*egfp* may have a promoter (as there is no promoter designed for *egfp* gene transcription in the plasmid pXMJ19-t/g-Δ*ldh*::*egfp*.).

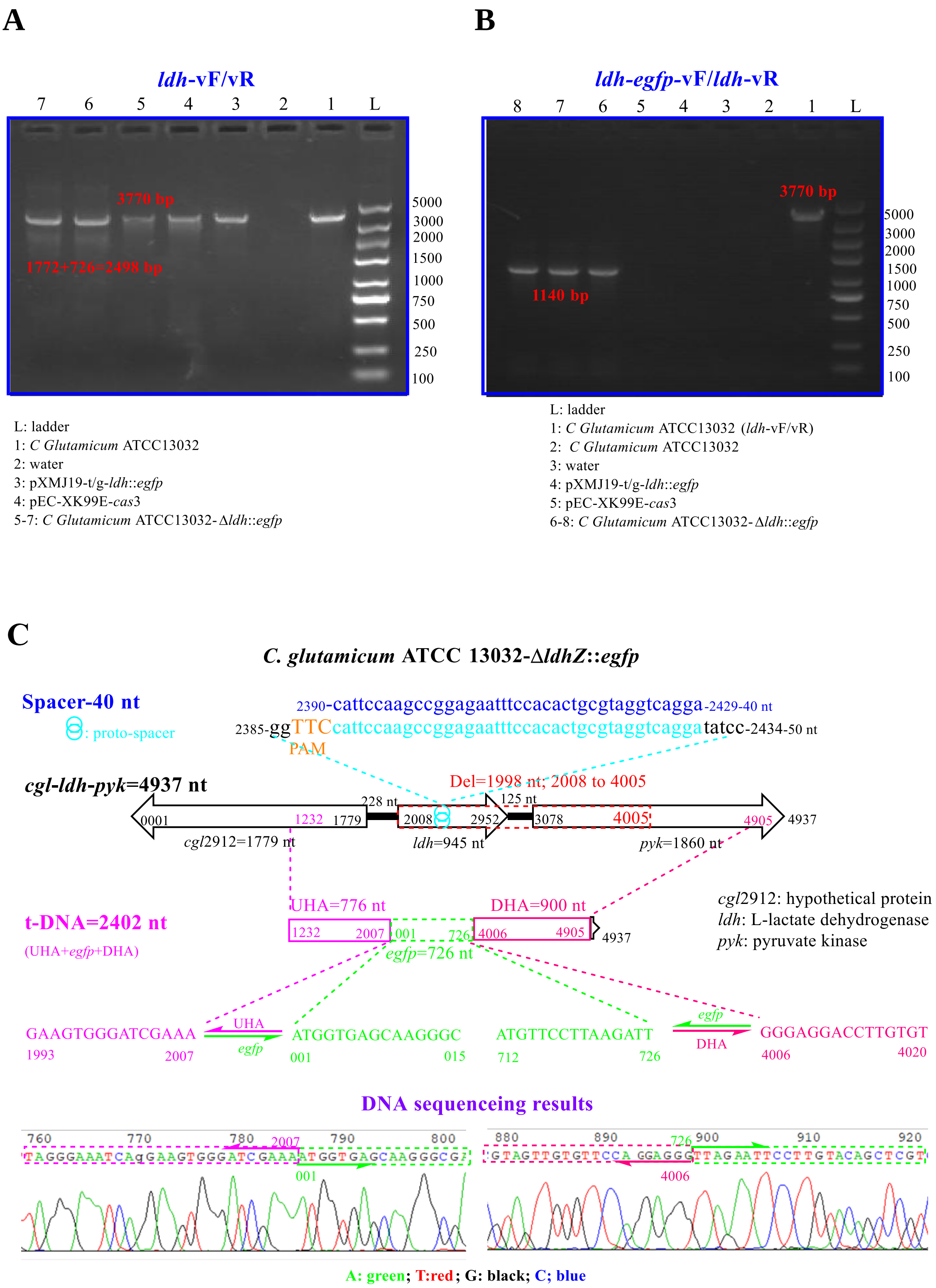

**Figure 4. RNA-guided genome editing in *C. glutamicum* ATCC 13032. (A and B)**

DNA gel electrophoresis of the PCR products of edited sequences in the genome editing of *C. glutamicum* ATCC 13032 (The primers in (A) and (B) are respectively *ldh*-vF / vR and *ldh-egfp*-vF / *ldh*-vR, where the primer *ldh-egfp*-vF is a portion of the gene *egfp* in the t-Δ*ldh*::*egfp* fragment and the primer *ldh*-vR is part of the gene *ldh* but is not contained in the t-Δ*ldh*::*egfp*. Therefore, the PCR products can occur only when the genome of the correctly gene-edited mutant *C. glutamicum* ATCC 13032-Δ*ldh*::*egfp* is used as template). **(C)** The selected proto-spacer, the engineered t-DNA as well as the DNA sequencing results of the PCR products of the edited sequences of the gene-edited mutant *C. glutamicum* ATCC 13032-Δ*ldh*::*egfp* genome (The junction sequences between UHA and *egfp* and between *egfp* and DHA are highlighted in Figure 4C).

Nevertheless, we obversed from Figure 4A that two of three lanes of the DNA gel electrophoresis of the PCR products of edited sequences in the gene-edited mutant *C. glutamicum* ATCC 13032-Δ*ldh*::*egfp* genome had an inconspicuous band of 2498 bp (gene-edited mutant *C. glutamicum* ATCC 13032-Δ*ldh*::*egfp*) in addition to an obvious band of 3770 bp (wild-type host *C. glutamicum* ATCC 13032) when the primer pair *ldh*-vF/vR was used (template: genomic DNA), implying that the gene editing tools developed based on the subtype I-B-*Svi* CRISPR-Cas system were effective although the efficiency was not very high in the genome editing of the strain *C. glutamicum* ATCC 13032. The conclusion was further verified by DNA gel electrophoresis of the PCR products of the edited sequences when the primer pair *ldh*-vF/vR (Figure 4A) was replaced with the primer pair *ldh-egfp*-vF/*ldh*-vR (Figure 4B) and subsequently subjected to DNA sequencing analysis (Figure 4C). In addition, dual plasmid transformation with pXMJ19-t/g-Δ*ldh::egfp* plus pEC-XK99E-*cas*9 was carried out for comparison in the genome editing of *C. glutamicum* ATCC 13032, and unfortunately, none of the colonies grew on LBHIS plates, suggesting that the biocompatibility between *Svi*Cas3 and the host cells of *C. glutamicum* ATCC 13032 is much better than that between *Sp*Cas9 and the host cells[46] and indicating that it is possible for the *Svi*Cas3 enzyme to be applied in the genome editing of other prokaryotic species where Cas9 is difficult to use.

In the genome editing of *E. coli* JM109(DE3) and *C. glutamicum* ATCC 13032, the crRNA fragment composed of R-S-R can search for both the PAM sequence and spacer's base-pairing sequence (equivalent to proto-spacer) in a complementary strand of the endogenous DNA, guide *Svi*Cas3 (*Svicas*3 without codon optimization) to perform a site-specific cleavage of the target gene and hence genome editing mediated by the HDR of the cells themselves, indicating that both the architecture of crRNA and the codon optimization of *Svicas*3 (GC%: *Svicas*3-69%, *E. coli* genome-50.8%; *C. glutamicum* genome-53.8%) may not be required in the genome editing of prokaryotes conducted by the single *Svi*Cas3.

## Discussion

It was reported that type I CRISPR-Cas systems could be repurposed for self-targeting genome editing [endogenous subtype I-A and subtype III-B CRISPR-Cas systems in monoploid thermophilic archaeon *Sulfolobus islandicus* with a normal growth temperature of 80-90°C [47]; endogenous subtype I-B CRISPR-Cas system in polyploid haloarchaeon *Haloarcula hispanica* [48], *Lactobacillus crispatus* VMC3 from human isolates cultured at anaerobic conditions[31] and *Clostridium pasteurianum* [49] with a normal growth temperature of 37°C] by delivering a genome-editing plasmid (carrying an artificial self-targeting CRISPR array and a donor DNA) into cells themselves and for a spectrum of long-range genomic deletions in human embryonic stem cells (hESCs) by delivering ribonucleoprotein (RNP) complexes (composed of *Tfu*Cascade and crRNA) and *Tfu*Cas3 (a type I-E CRISPR-Cas system in *Thermobifida fusca* / *Tfu*; growth normally at 55°C) into hESCs via electroporation [23]. Our experimental results (Figure 2) indicated that the subtype I-B-*Svi* CRISPR-Cas system could be utilized for self-targeting genome editing too (a culture temperature range from 25-42°C). In the self-targeting and non-self-targeting genome editing, we successfully used different fragment sizes of spacer (40 nt), t-DNA (541 to 2402 nt), insertion (726 to 807 nt) and deletion (318 to 1998 nt) as well as four PAMs with "tnc" motif ("n" indicating any of four bases a, c, g and t; "tac" used in *svu*016, "tcc" in *svipam*1, "tgc" in *sviomt*07 and "ttc" in the other genes) (https://kns.cnki.net/kns8?dbcode=CDMD). Among the four PAMs, "tac", "tcc" and "tgc" are three new PAMs compared with the PAMs (ttc, act, taa, tat, tag, and cac) of the subtype I-B CRISPR-Cas system of *H. volcanii* in interference against invading DNAs [16,32]. Also the crRNA structure composed of R-S-R in g-DNA is effective for genome editing conducted by the *Svi*Cas3 (Figures 2-4 and Figures S2-S4). As is well known, a tracrRNA is often required for immune and genome editing in type II CRISPR-Cas systems, but not in type I CRISPR-Cas systems [4,32]. This conclusion was validated by our experimental results again (Figurers 2-4 and Figure S6).

In summary, the *Svi*Cas3 enzyme has the following advantages over *Sp*Cas9: (i) a small molecular weight suitable for vector carrying (*Svicas*3/2316 bp, *Spcas*9/4107 bp) and protein expression (*Svi*Cas3/84352.40 Da, *Sp*Cas9/158441.41 Da); (ii) similar to DNA and RNA, an acidic isoelectric point (*Svi*Cas3/pI5.78, *Sp*Cas9/pI8.98) carrying negative charges at normal physiological pH condition; (iii) a bacterial protein of nonanimal origin with a low possibility of eliciting antibody of *Svi*Cas3; and (iv) in particular, no off-target effects or indel (insertion and deletion) formation detected in the analysis of potential off-target sites (Figure S5). Given that the gene editing tools developed based on the subtype I-B-*Svi* CRISPR-Cas3 system can be effectively used for template-based genome editing of prokaryotic microbial cells through HDR system of host cells, we anticipate that the gene editing tools developed based on this CRISPR-Cas system will be widely used in all fields related to biotechnology, including basic biology, bio-pharmaceutical and clinical medicine, agriculture, environmental modification, and so on.

## Matrials and methods

### Strains, plasmids and gene editing elements

*Streptomyces virginiae* IBL14 isolated from activated sludge from a steroidal drug factory was deposited in our laboratory and at the China Center for Type Culture collection, Wuhan, China (accession number: CCTCC M 206045) (*20*) and in this study is the root of the gene editing tools and the strain of interest for self-targeting genome editing. *Escherichia coli* JM109 (DE3) and *Corynebacterium glutamicum* ATCC 13032 were used as hosts of interest for non-self-targeting genome editing and *E. coli* DH5α was used as hosts for plasmid amplification and storage.

Plasmid pKC1139 [used in genome editing of *S. virginiae* IBL14 and *E. coli* JM109 (DE3)] harboring a DNA sequence *repts* (a replicon of temperature sensitivity, used for removing the plasmid) [41] and pXMJ19 (used in genome editing of *C. glutamicum* ATCC 13032) (BioVector NTCC Inc, Beijing, China) carrying a *cat* gene encoding chloramphenicol acetyltransferase for chloramphenicol hydrolysis were selected as vectors for construction of gene editing plasmids (named plasmid-t/g-*target gene abbreviation*), which carry an engineered g-DNA and an engineered t-DNA, respectively, for crRNA transcription and editing template.

Commercialized plasmids pCas (Addgene, Beijing Zhongyuan, Ltd, Beijing, China) bearing the gene *cas*9 derived from *Streptococcus pyogene* (http://www.addgene.org/) and pEC-XK99E (BioVector NTCC Inc, Beijing, China) carrying a *aph* gene encoding aminoglycoside phosphotransferase conferring resistance to kanamycin / neomycin were used for construction of Cas expression plasmids carrying a *cas*(es) encoding a Cas protein(s) in genome editing of *E. coli* JM109 (DE3) and *C. glutamicum* ATCC 13032, respectively. All strains and the plasmids used in this study are listed in Table S1.

In this study, primers were designed by Snapgene® 2.3.2 or Primer Premier 5.0, verified by Oligo 7.0 and chemically sysnthesized by General Biosystems, Inc.,

Chuzhou, Anhui province, China. All components of g-DNAs and t-DNAs were arbitrarily selected, including PAMs, spacers, repeats, promoters and terminators and DNA fragment sizes of deletion, replacement and insertion. All the details of the primers, g-DNAs and t-DNAs used in this study are summarized in Table S2.

**Solutions, media and chemicals**

Inorganic salts solution (ISS) [200 mg $FeCl_3 \cdot 6H_2O$, 40 mg $ZnCl_2$, 10 mg $MnCl_2 \cdot 4H_2O$, 10 mg $CuCl_2 \cdot 2H_2O$, 10 mg $Na_2B_4O_7 \cdot 10H_2O$ and 10 mg $(NH_4)_6Mo_7O_{24} \cdot 4H_2O$ in 1 liter $ddH_2O$, filter sterilization] and TES solution (TES) [57.3 g N-[tris(hydroxymethyl)methyl]-2-amino-ethane sulfonic acid in 1 liter $ddH_2O$, pH 8.0, filter sterilization] were first prepared as stock solution. A protoplast preparation buffer (PPB) [20 g sucrose, 25 mg $K_2SO_4$, 0.5 ml 1 M $MgCl_2 \cdot 6H_2O$, 0.2 ml ISS and 89 ml $ddH_2O$, 121 $^oC$, 20 min; then 10 ml TES, 20 μl 1 M $CaCl_2 \cdot 2H_2O$ and 1 ml 0.5% (W/V) $KH_2PO_4$ supplemented before use] was prepared for the protoplast preparation of *S. virginiae* IBL14. A chemical transformation solution for *S. virginiae* IBL14 (CTSS) was prepared following the procedures: (i) preparing 1 M tris(hydroxymethyl)aminomethane (Tris)-maleate solution (100 ml, pH 8.0, filter sterilization), (ii) preparing solution A (1 ml 2.5% $K_2SO_4$, 25 ml 10.3% sucrose, 0.2 ml ISS and 73.8 ml $ddH_2O$, mixed well, sterilized at 121 $^oC$ for 20 min), (iii) preparing solution B (18.6 ml solution A, 1 ml 1 M Tris-maleate solution and 0.4 ml 5 M $CaCl_2$, filter-sterilization) and (iv) preparing CTSS (5 ml PEG3350 sterilized at 121 $^oC$ for 20 min and 15 ml solution B, mixed well). All solutions prepared above were stored at 4 $^oC$ in a refrigerator.

Sporulation medium (SM) (1 liter: 10.0 g glycerol, 10.0 g starch, 1.0 g glutamine, 0.5 g $NaNO_3$, 0.25 g $K_2HPO_4 \cdot 3H_2O$, 0.25 g $MgSO_4 \cdot 7H_2O$, 0.25 g proline, 0.01 g $FeSO_4 \cdot 7H_2O$ and 15 g agar added in tap water, pH 7.0~7.4, 115$^o$C, 30 min; then 0.01 g filter-sterilized Vitamin B supplemented before pouring plates) and cell growth medium (CGM) [1 liter: 3.0 g $NH_4Cl$, 1.55 g $K_2HPO_4 \cdot 3H_2O$, 0.85 g $NaH_2PO_4 \cdot 2H_2O$, 0.2 g $MgSO_4 \cdot 7H_2O$, 0.1 ml 0.1 % (W/V) $CaCl_2 \cdot 2H_2O$, 0.1 ml 0.01 % (W/V) $FeSO_4 \cdot 7H_2O$ and 0.1 ml 0.001 % (W/V) $ZnSO_4$, dissolved well with about 800 ml tap water, pH 7.0, then 3.0 g glucose, 3.0 g yeast extract, 3.0 g maize slurry, 3.0 g β-cyclodextrin and about 200 ml tap water supplemented, 115$^o$C, 30 min; 20 g agar added for solid medium] were prepared respectively for the spore formation and vegetative cell growth of *S. virginiae* IBL14. Protoplast growth medium (PGM) (1 liter: CGM medium supplemented with 5 g glycine and 103 g sucrose) and modified R2YE medium (MR2YE) [1 liter: 103 g sucrose, 10.12 g $MgCl_2 \cdot 6H_2O$, 10 g glucose, 0.25 g $K_2SO_4$, 0.1 g Casamino acids, 5 g Yeast extract, 2 ml ISS, 100 ml TES and 22 g agar added in tap water, pH7.2, 115$^o$C, 30 min; melting the medium and then 10 ml 0.54% (W/V) $KH_2PO_4$, 8 ml 3.68% (W/V) $CaCl_2 \cdot 2H_2O$ and 15 ml 20% (W/V) L-proline supplemented before use] were prepared for the protoplast preparation and regeneration of *S. virginiae* IBL14.

Luria-Bertani medium (LB) (1 liter: 5 g yeast extract, 10 g peptone and 10 g NaCl added in tap water, pH 7.0~7.2, 121$^o$C, 20 min; 15 g agar added for solid medium) was prepared for the growth of *E. coli*. LB plates for *E. coli* transformation (LBPET) were prepared for the selection of transformants (before use) following the preparation procedures: (i) preparing a mixture with final concentrations of 20 mM $MgCl_2 \cdot 6H_2O$, 40 mg/l arabinose, 50 mg/l 5-bromo-4-chloro-3-indolyl-β-D-galactopyranoside (X-gal) and 50 mg/l isopropyl-β-D-thiogalactoside (IPTG) in an Eppendorf (EP) tube (the total volume of the mixture depends on the plate number, generally 50 μl for each plate), (ii) spreading the mixture evenly to LB plates with final concentrations of 100 mg/l kanamycin (for pCas) and 100 mg/l apramycin (for pKC1139) and (iii) subsequently placing the plates into an incubator to stand at 30°C for 2 h (keep out of light).

LB-glucose medium (LBG) (1 liter: 5 g yeast extract, 10 g trypton, 10 g NaCl and 5 g glucose added in tap water, pH 7.0~7.2, 115°C, 30 min; 15 g agar added for solid medium) and LB-Glycine-Tween medium (LBGT) (1 liter: 5 g yeast extract, 10 g peptone, 10 g NaCl, 30 g glycine and 1 g Tween 80 added in tap water, 115°C, 30 min) were used as seed medium and growth medium in preparation of *C. glutamicum* ATCC 13032 competent cell, respectively. LB-brain heart infusion-sorbitol medium (LBHIS) (1 liter: 2.5 g yeast extract, 5 g trypton, 5 g NaCl, 18.5 g brain heart infusion powder and 91 g sorbitol added in tap water, 121°C, 20 min; 15 g agar added for solid medium) was prepared for plasmid transformation in *C. glutamicum* ATCC 13032.

All reagents used in this study were of biochemical grade, analytical grade or higher, and most of them were purchased from Sangon Biotech Co., Ltd., Shanghai, China.

## Bioinformatics analysis of the subtype I-B-*Svi* CRISPR-Cas system

The whole genome sequencing of *S. virginiae* IBL14 was carried out using Illumina HiSeq 2000 sequencing platform [42] by BGI Tech Solutions Co., Ltd., Shenzhen, China. From the analytical results of CRISPRfinder program online by entering the whole genome sequence of *S. virginiae* IBL14 (http://crispr.i2bc.paris-saclay.fr/), we found that the *S. virginiae* IBL14 genome encompasses 18 CRISPR candidates, among which CRISPR1, CRISPR2, and CRISPR3 are three confirmed CRISPRs (Table 1).

After a total of 7862 genes in the *S. virginiae* IBL14 genome were predicted by using Glimmer 3.02 and the function annotation was accomplished by blasting the protein sequences of the genes with the following databases: KEGG 59 (Kyoto Encyclopedia of Genes and Genomes), COG 20090331 (Cluster of Orthologous Groups of proteins), SwissProt 201206, NR 20121005 and GO 1.419 (Gene Ontology), we found that in the strain IBL14 genome there is an operon composed of seven *cas* genes, i.e., *gvgl*007770 (KY243078), *cas*7(KY176013), *cas*5 (KY176012), *cas*3 (KY176010), *cas*4 (KY176011), *cas*1 (KY176008) and *cas*2 (KY176009) in transcription direction, among which six are definite *cas* genes (http://www.ncbi.nlm.nih.gov). Further, based on the gene composition and architecture of the six *cas* genes [4,32], the CRISPR-Cas system in the *S. virginiae* IBL14 genome was temporarily named subtype I-B-*Svi* CRISPR-Cas system (Figure 1A).

**Construction of gene editing plasmids**

In self-targeting genome editing, the four steps to construct gene editing plasmids (pKC1139-t/g-*target gene abbreviation*) are as follows (Figures S1A and S2, Tables S1 and S2). (i) Preparation of an upstream homologous arm (UHA) and a downstream homologous arm (DHA) [adding an inserted DNA sequence (IDS) as a part of t-DNA if required, e.g., *cat* in t-Δ*sviipe*::*cat*]: a basic PCR procedure [pre-denaturation at 94°C for 10 min, 32 cycles of denaturation (at 94°C for 30 s), annealing (at Tm-5°C for 30 s), extension (at 72°C for 60±30 s) and a final extension at 72°C for 10 min] was employed (TransTaq™ DNA Polymerase High Fidelity, TransGen Biotech, Beijing, China) respectively to amplify a full size of target gene or a fragment of target gene (primer: *gene abbreviation*-vF / vR, template: host genomic DNA extracted by an Ezup Column Bacteria Genomic DNA Purification Kit, Sangon Biotech Co., Ltd., Shanghai, China), UHA and DHA of the gene of interest [primers: *gene abbreviation*-UF / UR (UF harboring an *Hind*III restriction site) and *gene abbreviation*-DF / DR (DR harboring a *Xba*I restriction site), template: the purified PCR-amplified gene of interest] as well as IDS if required. (ii) Synthesis of t-DNA by an overlap PCR: a conventional overlap PCR method was used for the synthesis and amplification of a designed t-DNA (containing complementary sequences with pKC1139, restriction site, UHA plus DHA as well as IDS if needed) under the same operational conditions as described in the basic PCR above. (iii) Chemical synthesis of g-DNA: a g-DNA fragment composed successively of a complementary sequence with pKC1139, restriction site (e.g., *BamH*I at 5'-end), promoter, repeat, spacer, repeat, terminator, restriction site (e.g., *EcoR*I at 3'-end) and the other complementary sequence with the pKC1139 was directly chemically synthesized by General Biosystems, Inc., Chuzhou, Anhui province, China. (iv) Construction of gene editing plasmid: after restriction digestion (*Hind*III / *Xbal*I for t-DNA, *BamH*I / *EcoR*I for g-DNA) and purification of pKC1139, t-DNA and g-DNA, the purified linear pKC1139 was ligated one by one with the purified t-DNA and the purified g-DNA using T4 ligase with a suitable ratio of t-DNA / g-DNA to pKC1139 (often 3:1, mol / mol) under the conditions of 22 $^{o}$C for 120 min and further 65 $^{o}$C for 10 min. The resultant product, i.e., gene editing plasmid pKC1139-t/g-*target gene abbreviation*, was transformed, tested and preserved in *E. coli* DH5α (extracted when required by an AxyPrep Easy-96 Plasmid Kit, Axygen Biosciences, Hangzhou, China). All of the DNA fragments and plasmids generated in this study were tested, purified and collected by a DNA electrophoresis in 1.5±0.5 % agarose gels and an AxyPrep Gel DNA Isolation Kit (Corning Incorporated, Shanghai, China).

The construction procedures of pKC1139-t/g-Δ*lacZ* and pKC1139-t/g-Δ*lacZ*::*cat* in genome editing of *E. coli* JM109 (DE3) and pXMJ19-t/g-Δ*ldh*::*egfp* in genome editing of *C. glutamicum* ATCC 13032 are the same as those described above excepting plasmid, promoter, spacer, terminator, template and primer (Figures S3 and S4, Tables S1 and S2).

**Construction of Cas expression plasmids**

Four Cas expression plasmids (pCas-*cas gene abbreviation*) in genome editing of *E. coli* JM109(DE3), i.e., pCas-*cas*7-5-3-4-1-2, pCas-*cas*6-7-5-3, pCas-*cas*7-5-3 and pCas-*cas*3 respectively carrying six (*cas*7, *cas*5, *cas*3, *cas*4, *cas*1 and *cas*2), four (*cas*6, *cas*7, *cas*5 and *cas*3), three (*cas*7, *cas*5 and *cas*3) and one (*cas*3) genes, were constructed and verified by PCR, DNA electrophoresis and sequencing analysis (Figures S1B and S3, Tables S1 and S2). Briefly, four linear DNA fragments (*cas*7-5-3-4-1-2, *cas*6-7-5-3, *cas*7-5-3 and *cas*3) and the linear fragment of plasmid pCas absent of gene *cas*9 were firstly prepared (TransStart FastPfu DNA Polymerase, TransGen Biotech, Beijing, China) by the basic PCR (template: the genomic DNA of *S. virginiae* IBL14 and pCas, primers: *cas*7-5-3-4-1-2-F / *cas*7-5-3-4-1-2-R, *cas*6-7-5-3-F / *cas*6-7-5-3-R, *cas*7-5-3-F / *cas*7-5-3-R, *cas*3-F / *cas*3-R, pCas-*cas*7-5-3-4-1-2-F / pCas-*cas*7-5-3-4-1-2-R, pCas-*cas*6-7-5-3-F / pCas-*cas*6-7-5-3-R, pCas-*cas*7-5-3-F / pCas-*cas*7-5-3-R and pCas-*cas*3-F / pCas-*cas*3-R), and then purified and collected by the same methods as described above, respectively. Further, the purified linear fragment of the pCas without *cas*9 was ligated respectively with the purified fragments *cas*7-5-3-4-1-2, *cas*6-7-5-3, *cas*7-5-3 and *cas*3 using an Exnase (ClonExpress® Entry One Step Cloning Kit, Sangon Biotech Co., Ltd., Shanghai, China) under the conditions of 37 °C for 30 min and then on ice for 5 min. The ligated products, i.e., Cas expression plasmids: pCas-*cas*7-5-3-4-1-2, pCas-*cas*6-7-5-3, pCas-*cas*7-5-3 and pCas-*cas3*, were finally transformed, tested and preserved in *E. coli* DH5α (Figures S3B and S3C, Tables S1 and S2) following the procedures described in Construction of gene editing plasmids.

In the construction of Cas expression plasmid pEC-XK99E-*cas*3 for genome editing of *C. glutamicum* ATCC 13032, all the construction procedures excepting plasmid, promoter, terminator and corresponding primers were the same as described in the construction steps of pCas-*cas3* (Figures S4B and S4C, Tables S1 and S2). To be more precise, the plasmid, promoter and terminator in Cas expression plasmid pEC-XK99E-*cas*3 are pEC-XK99E, trc promoter and rrnB T1 terminator, respectively. In addition, the construction procedure of plasmid pEC-XK99E-*cas*3-t-Δ*ldh::egfp* (single-plasmid gene-editing tool ) was the same as that of pEC-XK99E-*cas*3, except that the t-Δ*ldh::egfp* fragment was inserted in it (Figure S6) (Note: In the prokaryotic genome editing reported in this paper, the *Svicas*3 was not codon optimizied; that is, it is a raw *cas*3 gene from the strain *S. virginiae* IBL14 genome)

**Construction of gene-edited mutants of *S. virginiae* IBL14**

To promote transforming efficiency of gene editing plasmids in self-targeting genome editing of *S. virginiae* IBL14, the protoplast of *S. virginiae* IBL14 was prepared. In the protoplast preparation, a single colony of *S. virginiae* IBL14 from a CGM plate was inoculated into a flask containing 30 ml of CGM and cultivated in a rotory shaker at 200 rpm at 30°C for 48 h. A volume of 300 μl of the culture was transferred into a flask (with a stainless steel spring or 8-10 glass beads in it) containing 30 ml PGM and cultivated at 200 rpm at 30°C for 48 h again. The mycelia in the culture were collected by a centrifugation at 4000 rpm for 5 min. After three cycles of re-suspension (with 20 ml 10.3% (W/V) sucrose, 10 ml PPB and 10 ml PPB again, respectively), placement and centrifugation, 10 ml re-suspended-mycelia mixture was obtained. About 100 μl of 200 mg/ml lysozyme solution (filter-sterilization) was added into the 10 ml re-suspended-mycelia mixture, which was placed into a water bath and incubated at 30°C for about 30 min, gently shaken up once per 5 min (until over 90% protoplasts). After a treatment with a sterilized spore filter, the mixture was centrifuged at 4000 rpm for 10 min. The pellet fraction (protoplasts) was washed by re-suspension (5 ml PPB) and centrifugation (at 4000 rpm for 5 min). The washed protoplasts were re-suspended with 2.5 ml PPB and stored in a refrigerator at -20°C.

A chemical transformation method was employed to obtain gene-edited mutants of *S. virginiae* IBL14. Firstly, 15 μl pKC1139-t/g-*target gene abbreviation* and 235 μl CTSS solution were added into an EP tube with 50 μl of the washed protoplasts. Then the mixture gently mixed well was spread evenly to three MR2YE plates (100 μl each) and incubated at 30°C in an incubator. When colonies just began to arise on the plates (about 24 h), 1 ml apramycin (a final concentration of 50 mg/l) was supplemented and spread evenly to the plates. After standing at 30°C for 1 h, the plates were inverted and incubated at 30°C for further 2-4 days. Three single colonies as potential gene-edited mutants (*S. virginiae* IBL14-*edited gene abbreviation*) on the MR2YE plates were randomly picked and the genomes of randomly picked three single colonies were respectively extracted and verified through DNA gel electrophoresis analysis (the original mutants) and later DNA sequencing analysis to the PCR products of edited sequences in the self-targeting genome editing (template: extracted mutant genome; primers: *gene abbreviation*-vF / vR or -UF / DR).

Three single colonies (the ninth generation mutants) of correctly gene-edited mutant *S. virginiae* IBL14-Δ*svipam*1 were respectively spread to three MR2YE plates containing a final concentration of 100 mg/l apramycin and then incubated at 42°C for 4-6 days to remove the gene editing plasmid pKC1139-t/g-Δ*svipam*1 utilizing its temperature-sensitive replication origin [41].

**Construction of gene-edited mutants of *E. coli* JM109(DE3)**

To improving transforming efficiency of Cas expression plasmid and gene editing plasmid in genome editing of *E. coli* JM109 (DE3), *E. coli* JM109 (DE3) competent cells were prepared following the preparation procedures below. Briefly, a single colony of *E. coli* JM109 (DE3) from an LB plate was inoculated into a flask containing 30 ml LB and incubated in a rotory shaker at 200 rpm for 12-14 h at 37°C. A volume of 600 μl of the seed was transferred to a flask containing 30 ml of fresh LB and cultivated at 200 rpm for ~2 h at 37°C (a value of $OD_{600}$ 0.5 or so). The culture was totally transferred to a pre-cooled EP tube (on ice for 15 min) and then centrifuged at 4000 rpm for 10 min at 4°C. The collected pellet was re-suspended with 15 ml pre-cooled 0.1 M $CaCl_2$, placed on ice for 10 min and centrifuged again. After three cycles of re-suspension (with 15, 10 and 1 ml 0.1 M $CaCl_2$, orderly), placement and centrifugation, 1 ml of the competent cell mixture of *E. coli* JM109 (DE3) was obtained and stored in a refrigerator at -20°C [Sometimes, *E. coli* JM109 (DE3) competent cells directly prepared by using SSCS solution; a High Efficiency *E. coli* Competent Cell Preparation Kit, Shanghai Generay Biotech Co., Ltd., Shanghai, China].

A chemical transformation method was used for the construction of gene-edited mutants of *E. coli* JM109 (DE3). Briefly, 10 μl Cas expression plasmid (pCas-*cas*7-5-3-4-1-2, pCas-*cas*6-7-5-3, pCas-*cas*7-5-3, or pCas-*cas*3) and 10 μl gene editing plasmid (pKC1139-t/g-Δ*lacZ* or pKC1139-t/g-Δ*lacZ*::*cat*) were added into an EP tube containing the competent cell mixture of 100 μl *E. coli* JM109 (DE3) and gently mixed well. The EP tube with 120 μl of the competent cell+plasmid mixture was placed on ice for 30 min, then into a water bath at 42°C for 90 s and on ice for 15 min again. A volume of 180 μl LB was added into the EP tube and incubated at 200 rpm for 1.5 h at 30°C. All the resulting mixture was spread evenly to three LBPETs (100 μl for each). The plates were placed to an incubator to incubate at 30°C for 2-3 days. Three white single colonies as potential gene-edited mutants were randomly selected and the genomes of randomly selected three white colonies were extracted and verified through DNA gel electrophoresis analysis and DNA sequencing analysis to the PCR products of edited sequences (primers: *lac*Z-vF / vR or *lac*Z-*cat*-F / *lac*Z-R) using the same methods as described above.

**Construction of gene-edited mutants of *C. glutamicum* ATCC 13032**

In the preparation of *C. glutamicum* ATCC 13032 competent cells, a single colony of *C. glutamicum* ATCC 13032 from an LBG plate was inoculated into a flask containing 30 ml LBG and incubated in a rotory shaker at 220 rpm at 30°C for 12-14 h. A volume of 0.5 ml the overnight culture was transferred to a flask with 50 ml fresh LBGT and cultivated at the same conditions for about 3-5 h ($OD_{600}$: 0.6-0.9). The culture was washed by centrifuging at 4000 rpm at 4 °C for 10 min. The washed pellet was resuspended with 30 ml pre-cooled 10% (V/V) glycerinum and centrifuged, repeatedly for three times. Finally the resultant pellet (competent cells) of *C. glutamicum* ATCC 13032 was resuspended with 0.2 ml pre-cooled 10% (V/V) glycerinum and stored in a refrigerator at -80°C. All operations were carried out on ice or under the temperature lower than 4°C.

In the gene editing of double-plasmid transformation (pXMJ19-t/g-Δ*ldh*::*egfp* plus pEC-XK99E-*cas*3 or pXMJ19-t-Δ*ldh*::*egfp* plus pEC-XK99E-*cas*3), the transformation process of gene editing plasmid pXMJ19-t/g-Δ*ldh*::*egfp* (or pXMJ19-t-Δ*ldh*::*egfp* ) was first carried out by adding 5 µl pXMJ19-t/g-Δ*ldh*::*egfp* (or pXMJ19-t-Δ*ldh*::*egfp* ) and 80 µl of *C. glutamicum* ATCC 13032 competent cells into a pre-cooled electroporation cuvette. After placed on ice for 5-10 min. the cell mixture was electro-transformed at 1.8 kV for 5 ms. The transformed cells were transferred to a pre-cooled eppendorf tube with 915 µl LBHIS and incubated orderly in a water bath at 46°C for 6 min and in a rotory shaker at 100 rpm at 30°C for 1 h. The transformed cell mixture was spread evenly to three LBHIS plates (100 µl for each) with 25 µg/ml chloramphenicol (Cm) and the plates were placed to an incubator to incubate at 30°C for 36 h. The pXMJ19-t/g-Δ*ldh*::*egfp* (or pXMJ19-t-Δ*ldh*::*egfp* ) plasmids from single colonies as potential recombinant *C. glutamicum* ATCC 13032-pXMJ19-t/g-Δ*ldh*::*egfp* (or *C. glutamicum* ATCC 13032-pXMJ19-t-Δ*ldh*::*egfp* ) were extracted and verified through the basic PCR (primers: *ldh*-UF / DR) and restriction digestion (*Hind*III plus *Xba*I or *BamH*I plus *EcoR*I) as described above. The correct recombinant of *C. glutamicum* ATCC 13032-pXMJ19-t/g-Δ*ldh*::*egfp* (or *C. glutamicum* ATCC 13032-pXMJ19-t-Δ*ldh*::*egfp*) was stored in a refrigerator at 4°C.

In the transformation of plasmid pEC-XK99E-*cas*3, the competent cell preparation of *C. glutamicum* ATCC 13032-pXMJ19-t/g-Δ*ldh*::*egfp* (or *C. glutamicum* ATCC 13032-pXMJ19-t-Δ*ldh*::*egfp*) followed the same procedures as described in the preparation of *C. glutamicum* ATCC 13032 competent cells above. Also the transformation of plasmid pEC-XK99E-*cas*3 followed the same procedures as described in the transformation process of gene editing plasmid pXMJ19-t/g-Δ*ldh*::*egfp* (or pXMJ19-t-Δ*ldh*::*egfp*), excepting LBHIS plates with 25 μg/ml chloramphenicol and 50 μg/ml kanamycin. The genomes of randomly selected three single colonies as potential gene-edited mutant *C. glutamicum* ATCC 13032-Δ*ldh*::*egfp* were extracted and verified through DNA gel electrophoresis and DNA sequencing analysis to the PCR products of edited sequences (primers: *ldh*-vF / *ldh*-vR or *ldh-egfp*-vF / *ldh*-vR or *ldh-egfp*-$vF_a$ / *ldh*-$vR_a$).

In the gene editing of single-plasmid transformation (pEC-XK99E-*cas*3-t-Δ*ldh::egfp*), the transformation process of plasmid pEC-XK99E-*cas*3-t-Δ*ldh::egfp* was the same as that of pEC-XK99E-*cas*3. However, the *C. glutamicum* ATCC 13032-pXMJ19-t/g-Δ*ldh*::*egfp* competent cells were replaced by the *C. glutamicum* ATCC 13032 competent cells.

**Off-target analysis**

The primers (some of forward primers in a 5'-xN-3'+5'-PAM-3'+-5'-seed-3'motif; xN: supplementary bases just used for primer design, x: the number of base, N: any of four bases a, c, g and t) for off-target analysis were designed and chemically syntheized firstly (Table S2). Then the off-target changes of *Svi*Cas3 in non-self-target genome editing was investigated by DNA gel electrophoresis of the PCR products with potential off-target sites (templates: extracted genomes of strains; primers: *strain name abbreviation*-otSN-F / -R) and further DNA sequencing (the same methods as described above).

### DNA electrophoresis

DNA electrophoresis in this study was performed in 1.5±0.5 % agarose gels at 110 V for 30 min [42].

### DNA sequencing

DNA sequencing was carried out by Sangon Biotech Co., Ltd., Shanghai, China.

### Data reporting

No statistical methods were used to predetermine sample size. The experiments were not randomized and the investigators were not blinded to allocation during experiments and outcome assessment.

## Acknowledgments

We are grateful to Zhi-Nan Xu, Wei Liu and Chang Dong, College of Chemical and Biological Engineering, Zhejiang University, Hangzhou, China, for generous gift of some plasmids and strains.

## Additional information

**Competing interests**

A patent application (CN107557373A / WO2019056848A1 / EP3556860A1 / US11286506 B2) has been filed for the content disclosed in this study.

**Funding**

The work was not funded by any agency.

**Author contributions**

Wang-Yu Tong contributed to the research design, the result interpretation and the paper writing. De-Xiang Yong, Xin Xu, Cai-Hua Qiu, Yan Zhang, Xing-Wang Yang, Ting-Ting Xia, Qing-Yang Liu, Su-Li Cao, Yan Sun and Xue Li carried out the experiments and participated in the result interpretation.

## Additional files

**Supplementary Infromation for Prokaryotic genome editing based on the subtype I-B-*Svi* CRISPR-Cas system**

Table of contents

4. DNA-guided gene editing in *C. glutamicum*

5. Signature Cases in different types of CRISPR-Cas systems

Figures S1 to S6

Tables S1 to S5

Supplementary references

**Data availability**

Almost all relevant data for this study are included in the article and the supplementary Information. The operon gene sequences of the Subtype I-B-*Svi* Cas system and the target gene sequences in self-targeting genome editing of *S. virginiae* IBL14 have been deposited in NCBI database (https://www.ncbi.nlm.nih.gov/genome/). The complete genome reference sequences of the strains (*Escherichia coli* str. K-12 substr. MG1655 / NC_002695.2; *Corynebacterium glutamicum* ATCC 13032 / NC_003450.3) in the off-target analyses are accessible to the NCBI database.

## References

1 Mohanraju, P. *et al.* Diverse evolutionary roots and mechanistic variations of the CRISPR-Cas systems. *Science (New York, N.Y.)* **353**, aad5147, doi:10.1126/science.aad5147 (2016).

2 Gootenberg, J. S. *et al.* Nucleic acid detection with CRISPR-Cas13a/C2c2. *Science (New York, N.Y.)*, doi:10.1126/science.aam9321 (2017).

3 Khanzadi, M. N. & Khan, A. A. CRISPR/Cas9: Nature's gift to prokaryotes and an auspicious tool in genome editing. *Journal of basic microbiology* **60**, 91-102, doi:10.1002/jobm.201900420 (2020).

4 Makarova, K. S. *et al.* Evolution and classification of the CRISPR-Cas systems. *Nat Rev Microbiol* **9**, 467-477, doi:10.1038/nrmicro2577 (2011).

5 Terao, M. *et al.* Utilization of the CRISPR/Cas9 system for the efficient production of mutant mice using crRNA/tracrRNA with Cas9 nickase and FokI-dCas9. *Experimental Animals* **65**, 275-283, doi:10.1538/expanim.15-0116 (2016).

6 Choi, K. R. & Lee, S. Y. CRISPR technologies for bacterial systems: Current achievements and future directions. *Biotechnol Adv* **34**, 1180-1209, doi:10.1016/j.biotechadv.2016.08.002 (2016).

7 Fagerlund, R. D., Staals, R. H. & Fineran, P. C. The Cpf1 CRISPR-Cas protein expands genome-editing tools. *Genome Biol* **16**, 251, doi:10.1186/s13059-015-0824-9 (2015).

8 Jervis, A. J. *et al.* A plasmid toolset for CRISPR-mediated genome editing and CRISPRi gene regulation in Escherichia coli. *Microb Biotechnol*, doi:10.1111/1751-7915.13780 (2021).

9 Xu, X. *et al.* Engineered miniature CRISPR-Cas system for mammalian genome regulation and editing. *Molecular cell* **81**, 4333-4345.e4334, doi:10.1016/j.molcel.2021.08.008 (2021).

10 Pausch, P. *et al.* CRISPR-CasΦ from huge phages is a hypercompact genome editor. *Science (New York, N.Y.)* **369**, 333-337, doi:10.1126/science.abb1400 (2020).

11 Abudayyeh, O. O. *et al.* C2c2 is a single-component programmable RNA-guided RNA-targeting CRISPR effector. *Science (New York, N.Y.)* **353**, aaf5573, doi:10.1126/science.aaf5573 (2016).

12 Wang, Q. *et al.* The CRISPR-Cas13a Gene-Editing System Induces Collateral Cleavage of RNA in Glioma Cells. *Adv Sci (Weinh)* **6**, 1901299, doi:10.1002/advs.201901299 (2019).

13 Majumdar, S., Ligon, M., Skinner, W. C., Terns, R. M. & Terns, M. P. Target DNA recognition and cleavage by a reconstituted Type I-G CRISPR-Cas immune effector complex. *Extremophiles* **21**, 95-107, doi:10.1007/s00792-016-0871-5 (2016).

14 Koonin, E. V., Makarova, K. S. & Zhang, F. Diversity, classification and evolution of CRISPR-Cas systems. *Current opinion in microbiology* **37**, 67-78, doi:10.1016/j.mib.2017.05.008 (2017).

15 Elmore, J., Deighan, T., Westpheling, J., Terns, R. M. & Terns, M. P. DNA targeting by the type I-G and type I-A CRISPR–Cas systems of *Pyrococcus furiosus*. *Nucleic acids research* **43**, 10353-10363, doi:10.1093/nar/gkv1140 (2015).

16 Maier, L. K. *et al.* Essential requirements for the detection and degradation of invaders by the *Haloferax volcanii* CRISPR/Cas system I-B. *RNA Biol* **10**, 865-874, doi:10.4161/rna.24282 (2013).

17 Laronde-Leblanc, N. A. Defense systems up: structure of subtype I-C/Dvulg CRISPR/Cas. *Structure* **20**, 1450-1452, doi:10.1016/j.str.2012.08.015 (2012).

18 Hrle, A. *et al.* Structural analyses of the CRISPR protein Csc2 reveal the RNA-binding interface of the type I-D Cas7 family. *RNA Biol* **11**, 1072-1082, doi:10.4161/rna.29893 (2014).

19 Goren, M. G. *et al.* Repeat Size Determination by Two Molecular Rulers in the Type I-E CRISPR Array. *Cell Rep* **16**, 2811-2818, doi:10.1016/j.celrep.2016.08.043 (2016).

20 Xiao, Y. *et al.* Structure Basis for Directional R-loop Formation and Substrate Handover Mechanisms in Type I CRISPR-Cas System. *Cell* **170**, 48-60 e11, doi:10.1016/j.cell.2017.06.012 (2017).

21 Wilkinson, M. E. *et al.* Structural plasticity and in vivo activity of Cas1 from the type I-F CRISPR-Cas system. *Biochem J* **473**, 1063-1072, doi:10.1042/bcj20160078 (2016).

22 Csörgő, B. *et al.* A compact Cascade-Cas3 system for targeted genome engineering. *Nature methods,* doi:10.1038/s41592-020-00980-w (2020).

23 Dolan, A. E. *et al.* Introducing a Spectrum of Long-Range Genomic Deletions in Human Embryonic Stem Cells Using Type I CRISPR-Cas. *Molecular cell* **74**, 936-950 e935, doi:10.1016/j.molcel.2019.03.014 (2019).

24 Morisaka, H. *et al.* CRISPR-Cas3 induces broad and unidirectional genome editing in human cells. *Nature communications* **10**, 1-13, doi:10.1038/s41467-019-13226-x (2019).

25 Osakabe, K. *et al.* Genome editing in plants using CRISPR type I-D nuclease. *Communications biology* **3**, 648, doi:10.1038/s42003-020-01366-6 (2020).

26 Cameron, P. *et al.* Harnessing type I CRISPR-Cas systems for genome engineering in human cells. *Nature biotechnology* **37**, 1471-1477, doi:10.1038/s41587-019-0310-0 (2019).

27 Hu, C. *et al.* Craspase is a CRISPR RNA-guided, RNA-activated protease. *Science (New York, N.Y.)*, eadd5064, doi:10.1126/science.add5064 (2022).

28 Hu, C. *et al.* Allosteric control of type I-A CRISPR-Cas3 complexes and establishment as effective nucleic acid detection and human genome editing tools. *Molecular cell* **82**, 2754-2768.e2755, doi:10.1016/j.molcel.2022.06.007 (2022).

29 Wang, F.-Q., Zhang, C.-G., Li, B., Wei, D.-Z. & Tong, W.-Y. New microbiological transformations of steroids by *Streptomyces virginiae* IBL-14. *Environmental science & technology* **43**, 5967-5974 (2009).

30 Stern, A., Keren, L., Wurtzel, O., Amitai, G. & Sorek, R. Self-targeting by CRISPR: gene regulation or autoimmunity? *Trends Genet* **26**, 335-340, doi:10.1016/j.tig.2010.05.008 (2010).

31 Hidalgo-Cantabrana, C., Goh, Y. J., Pan, M., Sanozky-Dawes, R. & Barrangou, R. Genome editing using the endogenous type I CRISPR-Cas system in Lactobacillus crispatus. *Proceedings of the National Academy of Sciences of the United States of America* **116**, 15774-15783, doi:10.1073/pnas.1905421116 (2019).

32 Gasiunas, G., Sinkunas, T. & Siksnys, V. Molecular mechanisms of CRISPR-mediated microbial immunity. *Cell Mol Life Sci* **71**, 449-465, doi:10.1007/s00018-013-1438-6 (2014).

33 Sinkunas, T. *et al.* Cas3 is a single-stranded DNA nuclease and ATP-dependent helicase in the CRISPR/Cas immune system. *EMBO J* **30**, 1335-1342, doi:10.1038/emboj.2011.41 (2011).

34 He, L., St John James, M., Radovcic, M., Ivancic-Bace, I. & Bolt, E. L. Cas3 Protein-A Review of a Multi-Tasking Machine. *Genes* **11**, 208, doi:10.3390/genes11020208 (2020).

35 Sinkunas, T., Gasiunas, G. & Siksnys, V. Cas3 nuclease-helicase activity assays. *Methods in molecular biology (Clifton, N.J.)* **1311**, 277-291, doi:10.1007/978-1-4939-2687-9_18 (2015).

36 Chylinski, K., Makarova, K. S., Charpentier, E. & Koonin, E. V. SURVEY AND SUMMARY Classification and evolution of type II CRISPR-Cas systems. *Nucleic acids research* **42**, 6091-6105 (2014).

37 Rani, R. *et al.* CRISPR/Cas9: a promising way to exploit genetic variation in plants. *Biotechnol Lett* **38**, 1991-2006, doi:10.1007/s10529-016-2195-z (2016).

38 Mulepati, S. & Bailey, S. Structural and biochemical analysis of nuclease domain of clustered regularly interspaced short palindromic repeat (CRISPR)- associated protein 3 (Cas3). *The Journal of biological chemistry* **286**, 31896-31903, doi:10.1074/jbc.M111.270017 (2011).

39 Xiao, Y., Luo, M., Dolan, A. E., Liao, M. & Ke, A. Structure basis for RNA-guided DNA degradation by Cascade and Cas3. *Science (New York, N.Y.)* **361**, 1-15, doi:10.1126/science.aat0839 (2018).

40 Nimkar, S. & Anand, B. Cas3/I-C mediated target DNA recognition and cleavage during CRISPR interference are independent of the composition and architecture of Cascade surveillance complex. *Nucleic acids research* **48**, 2486-2501, doi:10.1093/nar/gkz1218 (2020).

41 Huang, H., Zheng, G., Jiang, W., Hu, H. & Lu, Y. One-step high-efficiency CRISPR/Cas9-mediated genome editing in Streptomyces. *Acta Biochim Biophys Sin (Shanghai)* **47**, 231-243, doi:10.1093/abbs/gmv007 (2015).

42 Li, Z.-Z. *et al.* Identification and functional analysis of cytochrome P450 complement in *Streptomyces virginiae* IBL14. *BMC Genomics* **14**, 130 (2013).

43 Pyne, M. E., Moo-Young, M., Chung, D. A. & Chou, C. P. Coupling the CRISPR/Cas9 System with Lambda Red Recombineering Enables Simplified Chromosomal Gene Replacement in Escherichia coli. *Applied and environmental microbiology* **81**, 5103-5114, doi:10.1128/aem.01248-15 (2015).

44 Xue, C. & Sashital, D. G. Mechanisms of Type I-E and I-F CRISPR-Cas Systems in Enterobacteriaceae. *EcoSal Plus* **8**, doi:10.1128/ecosalplus.ESP-0008-2018 (2019).

45 Hochstrasser, M. L., Taylor, D. W., Kornfeld, J. E., Nogales, E. & Doudna, J. A. DNA Targeting by a Minimal CRISPR RNA-Guided Cascade. *Molecular cell* **63**, 840-851, doi:10.1016/j.molcel.2016.07.027 (2016).

46 Liu, J. *et al.* Development of a CRISPR/Cas9 genome editing toolbox for Corynebacterium glutamicum. *Microbial cell factories* **16**, 205, doi:10.1186/s12934-017-0815-5 (2017).

47 Li, Y. *et al.* Harnessing Type I and Type III CRISPR-Cas systems for genome editing. *Nucleic acids research* **44**, e34-e34, doi:10.1093/nar/gkv1044 (2016).

48 Cheng, F. *et al.* Harnessing the native type I-B CRISPR-Cas for genome editing in a polyploid archaeon. *Journal of genetics and genomics = Yi chuan xue bao* **44**, 541-548, doi:10.1016/j.jgg.2017.09.010 (2017).

49 Pyne, M. E., Bruder, M. R., Moo-Young, M., Chung, D. A. & Chou, C. P. Harnessing heterologous and endogenous CRISPR-Cas machineries for efficient markerless genome editing in Clostridium. *Scientific reports* **6**, 25666, doi:10.1038/srep25666 (2016).

# Table of contents

# Supplementary Infromation for

# Prokaryotic genome editing based on the subtype I-B-*Svi* CRISPR-Cas system

**Authors:**

Wang-Yu Tong*, De-Xiang Yong, Xin Xu, Cai-Hua Qiu, Yan Zhang, Xing-Wang Yang, Ting-Ting Xia, Qing-Yang Liu, Su-Li Cao, Yan Sun and Xue Li

**Affiliations:**

*Integrated Biotechnology Laboratory, School of Life Sciences, Anhui University, 111 Jiulong Road, Hefei 230601, China*

** Corresponding author: tongwy@ahu.edu.cn*

*Tel.: +86-551-63861282*

*Fax: +86-551-63861282*

# Table of contents

# Supplementary disscusion

## 1. Molecular principles of genome editing

Lethal double-strand break DNA (DSB-DNA) in the host genome can often be repaired by homology-directed repair (HDR) mechanisms (with lower frequencies and fewer errors) of cells themselves when homologous DNA sequences are present, which is the foundation of template-based genome editing by introducing an engineered template DNA (t-DNA) fragment with two sequences at both ends respectively complementary to the terminal sequences around the two DSB products. Non-homologous end joining (NHEJ) (with higher frequencies and more errors) often occurs in the absence of homologous DNA sequences and hence leads to frame shift mutations through nucleotide insertions and / or deletions, especially in genomic DSB products with blunt-end [1,2].

In genome editing directed by CRISPR-Cas systems, the molecular events after a set of gene editing tools enter host cells mainly include (i) duplication, transcription and translation of the *cas* gene, (ii) transcription and processing of guide-DNA / g-DNA to form crRNA, (iii) site-specific cleavage of genomic DNA by the Cas enzyme and (iv) homologous recombination between target DNA and t-DNA by HDR. Obviously, events (i) and (iv) are totally performed by the molecular components of host cells and plasmids, and only the events (ii) and (iii) should be accomplished by Cas(es) [1,2]. In fact, event (ii) is easily achieved through a vector with engineered g-DNA after crRNA optimization. Accordingly, event (iii) in genome editing becomes the key step, requiring a restriction endonuclease, such as *Sp*Cas9 or *Svi*Cas3, to perform RNA-guided site-specific genomic DNA cleavage.

## 2. Genome editing efficiency

According to the molecular principles and operating processes of genome editing directed by the CRISPR-Cas system discussed above [1,2], we know that genome editing efficiency (GEE) with high fidelity directed by the CRISPR-Cas system consists of the transformation or transfection efficiency (TE) of gene editing tools and the homologous recombination efficiency (HRE) of t-DNA. The former (easy to optimize) mainly depends on the characteristics of plasmids (e.g., genetic markers) and physiological status of cells (e.g., competent state of recipient cells), while the latter (difficult to optimize) depends on the characteristics of the HDR mechanisms of the host cells themselves, the features of DNA products (e.g., sticky end-DSB, blunt end-DSB or local SSB) of digestion by the Cas effector complex, and the biocompatibility between the Cas effector complex and host cells

To test the GEE value based on the subtype I-B-*Svi* CRISPR-Cas system, a lot of transformation experiments were respectively carried out following the procedures described in Materials and Methods. As seen from Table S4, GEE values have changed greatly in the genome editing of the three prokaryotes [*S. virginiae* IBL14, *E. coli* JM109 (DE3) and *C. glutamicum* ATCC 13032] based on the *Svi*Cas system, which were respectively: 1.0%-*S. virginiae* IBL14, $9.0\times10^{-5}$ for *E. coli* JM109 (DE3) and $8.6\times10^{-7}$ for *C. glutamicum* ATCC 13032, up to five orders of magnitude. Further analysis shows that the great change of GEEs (from $8.6\times10^{-7}$ to 1.0%) is significantly dependent on the change of TEs [ranging from $1.8\times10^{-6}$ to 1.1%, up to four orders of magnitude; respectively are: 1.1%-*S. virginiae* IBL14, $3.28\times10^{-4}$ for *E. coli* JM109 (DE3) and $1.8\times10^{-6}$ for *C. glutamicum* ATCC 13032], but not on the change of HREs [ranging from 27.3% to 92%, on the same order of magnitude; respectively are: 92%-*S. virginiae* IBL14, 27.3% for *E. coli* JM109 (DE3) and 48.0% for *C. glutamicum* ATCC 13032]. In the genome editing of the three prokaryotes based on the *Svi*Cas system, all of the HRE values were higher than 27.3% [Note: in the genome editing of *E. coli* JM109 (DE3), the plasmid pCas harbours genes *exo*, *bet* and *gam* that encode λ Red recombinases to enhance HRE], indicating that in template-based prokaryotic microbial genome editing, the critical factor affecting GEE is not HRE but TE. In particular, we found that in different batches of genome-editing experiments with the same strain, the effect of experimental strategies (e.g., genetic marker) and physiological status of cells (e.g., competent cells) on GEE was very significant, suggesting that in template-based microbial genome editing, HDR-system engineering in cells of different species is not a priority as long as exogenous endonucleases (e.g., Cas3, Cas9, Zinc Finger Nucleases, etc.) and their HDR mechanisms (including allelic crossing-over mechanisms) are working properly. In short, to improve GEE, the optimization of experimental strategies and operational conditions is indispensable.

## 3. Off-target analysis

Many strategies for minimizing off-target effects in CRISPR/Cas9-mediated genome engineering have been reported, including optimization of the crRNA sequence, concentration control of the Cas9-sgRNA complex, mutation of *cas*9 and so on, but potential off-target cleavage could still occur, especially at the genomic DNA sites consistent with a seed sequence of 6-11 nt (corresponding to the terminal sequence of a spacer directly adjacent to PAM) (This may be because the specific recognition of target DNA is exquisitely delegated to crRNA in which the length of a spacer is limited, normally less than 30 nt.) [3-6]. In particular, off-target mutations can cause genomic instability and functional failure of other normal genes, which is a major concern in biomedical and clinical applications [4]. To investigate the off-target effects of genome editing directed *Svi*Cas3, we stochastically chose four potential off-target sites for each gene-edited mutant (7 potential off-target sites in *E. coli* JM109(DE3)-Δ*lacZ* based on the motif: PAM+seed=3+8=TTC+CCGCCCGG; 18 potential off-target sites in *C. glutamicum* ATCC 13032-Δ*ldh*::*egfp* based on the motif: PAM+seed=3+7= TTC+CATTCCA) designed and chemically synthesized 8-pair-primers (Table S2) for off-target-analysis based on the principle that the targeting specificity of Cas9 in genome editing could be tightly controlled by suitable selection of both spacer and PAM[4,7]. Further the off-target-analytical experiments were implemented according to the procedures described in Off-target analysis.

Fortunately, no one off-target changes (coverage rates: 4/7=57% in *E. coli* JM109(DE3)-Δ*lacZ* and 4/18=22% in *C. glutamicum* ATCC 13032-Δ*ldh*::*egfp*, respectively) or indel-formation were found in the results of DNA gel electrophoresis or further DNA sequencing of the PCR products of potential off-target sites in the non-self-targeting genome editing conducted by *Svi*Cas3 (Extended Data Fig. 12). This may be because: (i) Cas9 lacks an R-loop locking to verify a complete R-loop formation to trigger cleavage, while Cas3 requires a locked full R-loop to trigger cleavage [8,9]; and (ii) even in the presence of t-DNA, Cas9 still produces DSBs with blunt-end (an HNH-like nuclease domain to cleave TS and a RuvC-like nuclease domain to cleave NTS) [1,10,11], while the *Svi*Cas3 enzyme probably generates SSBs (HD domain cleave NTS and TS in sequence) [2,12,13]. In addition, these facts that (i) Cascade is not needed in the RNA-guided genome editing conducted by *Svi*Cas3 alone and (ii) the subtype I-B-*Svi* CRISPR-Cas system lacks Cas8 (the PAM recognition component) [8,14-16] indirectly support the results that "no off-target effects or indel-formation were detected in target sequencing" and the view that for the most part, Cascade-Cas3 interaction features a conformation-capture rather than an induced-fit mechanism and thus reduces off-target [8,9,13,14,17]. Besides, in the self-targeting genome editing of strain *S. virginiae* IBL14 (hundreds of trials), the PCR and DNA sequencing results of the gene-edited mutants of 22 genes of interest [7 genes participating in penicillin metabolism [18,19], 9 cytochrome P450 superfamily genes [20,21], 6 O-methyltransferase genes [22]] accord with the pre-design at all, indicating again that the genome editing conducted by the *Svi*Cas3 enzyme has a high degree of targeting and specificity. It is worth noting that in addition to the off-target effects of CRISPR–Cas editing per se (the specific genome editing protocol used), off-target effects might be cell-type specific, depending on the integration of DSBs in the genome by the HDR pathway of a particular cell type [7].

## 4. DNA-guided gene editing in *C. glutamicum*

In the study of gene editing of *Saccharomyces cerevisiae* LYC4 based on the subtype I-B-*Svi* CRISPR-Cas system, we unexpectedly found that g-DNA, i.e., crRNA, was unnecessary (see the sister article: Template-based eukaryotic genome editing directed by the *Svi*Cas3). To verify the effectiveness of the DNA-guided genome editing conducted by *Svi*Cas3, we first designed and constructed the gene editing plasmid-pXMJ19-t-Δ*ldh*::*egfp* (without g-DNA compared to the plasmid-pXMJ19-t/g-Δ*ldh*::*egfp*). In contrast to the designed fragment sizes of t-DNA (UHA+*egfp*+DHA=776+726+900=2402 nt) and deletion (1998 nt) in plasmid pXMJ19-t/g-Δ*ldh*::*egfp*, the designed fragment sizes of t-DNA and deletion in plasmid pXMJ19-t-Δ*ldh*::*egfp* were 2001 nt (375+726+900 nt) and 1623 nt, respectively (Figures S6A and S6C). After successively transforming the two plasmids pEC-XK99E-*cas*3 and pXMJ19-t-Δ*ldh*::*egfp* into *C. glutamicum* ATCC 13032 competent cells, we obtained potential gene-edited mutants on LBHIS plates containing 25 μg/ml chloramphenicol and 50 μg/ml kanamycin (Table S1). The results of DNA gel electrophoresis and DNA sequencing of the target sequence (Figure S6B) were consistent with the designed DNA fragment size and sequence (Figure S6C and Table S2), demonstrating that the potential gene-edited mutants were the correctly gene-edited mutants *C. glutamicum* ATCC 13032-Δ*ldh*::*egfp*).

Furthermore, we designed and constructed an all-in-one gene editing tool- pEC-XK99E-*cas*3-t-Δ*ldh*::*egfp* (pEC-XK99E-*cas*3 instead of pPXMJ19 as a starting plasmid) (Figure S6A). After transforming the all-in-one plasmid pEC-XK99E-*cas*3-t-Δ*ldh*::*egfp* into *C. glutamicum* ATCC 13032 competent cells, we obtained potential Δ*ldh*::*egfp*-edited colonies of *C. glutamicum* ATCC 13032 on LBHIS plates and verified them through DNA gel electrophoresis and DNA sequencing analysis of the PCR products of edited sequences. In the gene-edited mutants (*C. glutamicum* ATCC 13032-Δ*ldh*::*egfp*, Table S1) obtained by single-plasmid transformation (pEC-XK99E-*cas*3-t-Δ*ldh*::*egfp*) and double-plasmid transformation (pEC-XK99E-*cas*3 plus pXMJ19-t-Δ*ldh*::*egfp*), the results of DNA gel electrophoresis analysis (Figure S6B) and DNA sequencing of the target sequence (Figure S6C) were consistent with the designed fragment size and the designed sequence (Figure S6C and Table S2), indicating that the all-in-one gene editing tool pEC-XK99E-*cas*3-t-Δ*ldh*::*egfp* is practical in the DNA-guided, template-based genome editing of *C. glutamicum* ATCC 13032, as is the dual-plasmid gene editing tool pEC-XK99E-*cas*3 plus pXMJ19-t-Δ*ldh*::*egfp*.

In brief, the success of the DNA-guided *Svi*Cas3-based genome editing in *C. glutamicum* provides further support for the initial finding that *Svi*Cas3 can direct DNA-guided genome editing in *S. cerevisiae*. The results also support the idea that if there is no biocompatibility problem between an enzyme and host cells, the enzyme effective for a substrate in prokaryotes should also be effective for the substrate in eukaryotes, and vice versa.

## 5. Signature Cas proteins in different types of CRISPR-Cas systems

The main biochemical characteristics of the typical signature proteins (all for DNA excision except Cas13a for RNA excision) in six types of CRISPR-Cas systems excepting type IV are summarized in Table S5 [23-25] (http://web.expasy.org/protparam; http://www.addgene.org/; http://www.ncbi.nlm.nih.gov;). We can see that the contents of cysteine in all the proteins are lower than 1% (0.1% to 0.8%; 2 to 8 cysteines) and that the contents of negatively charged aspartic acid (5.5% to 7.5%) and glutamic acid (6.4% to 10.4%) in all the proteins vary by less than 40%. However, the contents of positively charged arginine (3.9% to 9.2%) and lysine (1.3% to 15.3%) in all the proteins vary greatly (more than 235%). To be more precise, the content of arginine in *Svi*Cas3 (9.2%) is significantly higher than those in *Sp*Cas9 (5.6%) from *Streptococcus pyogenes*, *Cj*Cas9 (4.3%) from *Campylobacter jejuni*, *Ll*Cas10 (5.3%) from *Lactococcus lactis* subsp. Lactis, *As*Cas12a (4.4%) from *Acidaminococcus sp*. BV3L6 and *Ls*Cas13a (3.9%) from *Leptotrichia shahii* and the lysine content in *Svi*Cas3 (1.3%) is much lower than those in *Sp*Cas9 (11.0%), *Cj*Cas9 (14.7%), *Ll*Cas10 (7.5%), *As*Cas12a (9.0%) and *Ls*Cas13a (15.3%. Additianally, we can see that the asparaginate content in *Svi*Cas3 (1.2%) is much lower than those in *Sp*Cas9 (5.1%), *Cj*Cas9 (6.8%), *Ll*Cas10 (6.7%), *As*Cas12a (5.7%) and *Ls*Cas13a (10.7%). In addition, among the six proteins, *Svi*Cas3 has the highest contents of Ala (15.4%), Arg (9.2), Gly (7.7%), His (3.8), Leu (12.3%), Pro (5.7%), Trp (1.7%) and Val (7.1%) and the lowest contents of Asn (1.2), Glu (6.4%), Ile (3.2%), Lys (1.3%), Met (0.5%), Phe (3.4%), Ser (4.0%) and Tyr (2.2%), suggesting that *Svi*Cas3 probably has more distinct functions from the other Cas signature proteins.

In particular, compared to *Sp*Cas9 (1368 aa, type II Cas system) (http://www.addgene.org/) and *As*Cas12a (1307 aa, type V Cas system) [24] that have been used for genome editing, we can see that the significant differences between *Svi*Cas3 and them are their MWs (*Svi*Cas3: 84352.40 Da, *Sp*Cas9: 158441.41 Da and *As*Cas12a: 151206.55 Da) and pIs (*Svi*Cas3: 5.78, *Sp*Cas9: 8.98 and *As*Cas12a: 8.01). The MW of *Svi*Cas3 is slightly over half of the MWs of *Sp*Cas9 (53.2%) and *As*Cas12a (55.8%), even lower than the MW (73.4%) of the smallest *Cj*Cas9 (*Cj*Cas9: 984 aa, 114896.12 Da, pI 9.37) [23], which suggests that *Svi*Cas3 alone may be more easily loaded into vectors than *Sp*Cas9, *Cj*Cas9 and *As*Cas12a and is therefore suitable for genome editing in eukaryotic cells.

## Figures S1 to S6

**Figure S1. Construction process of gene editing tools based on the subtype I-B-*Svi* CRISPR-CAS system.**

**Figure S2. Construction of gene editing plasmids in the self-targeting genome editing of *S. virginiae* IBL14.**

**Figure S3. Construction of gene editing tools in the genome editing of *E. coli* JM109 (DE3).**

**Figure S4. Construction of gene editing tools in the genome editing of *C. glutamicum* ATCC 13032.**

**Figure S5. Off-target analyses in the non-self-targeting genome editing of prokaryotic microorganisms conducted by the *Svi*Cas3.**

**Figure S6. DNA-guided genome editing of *C. glutamicum* ATCC 13032.**

A

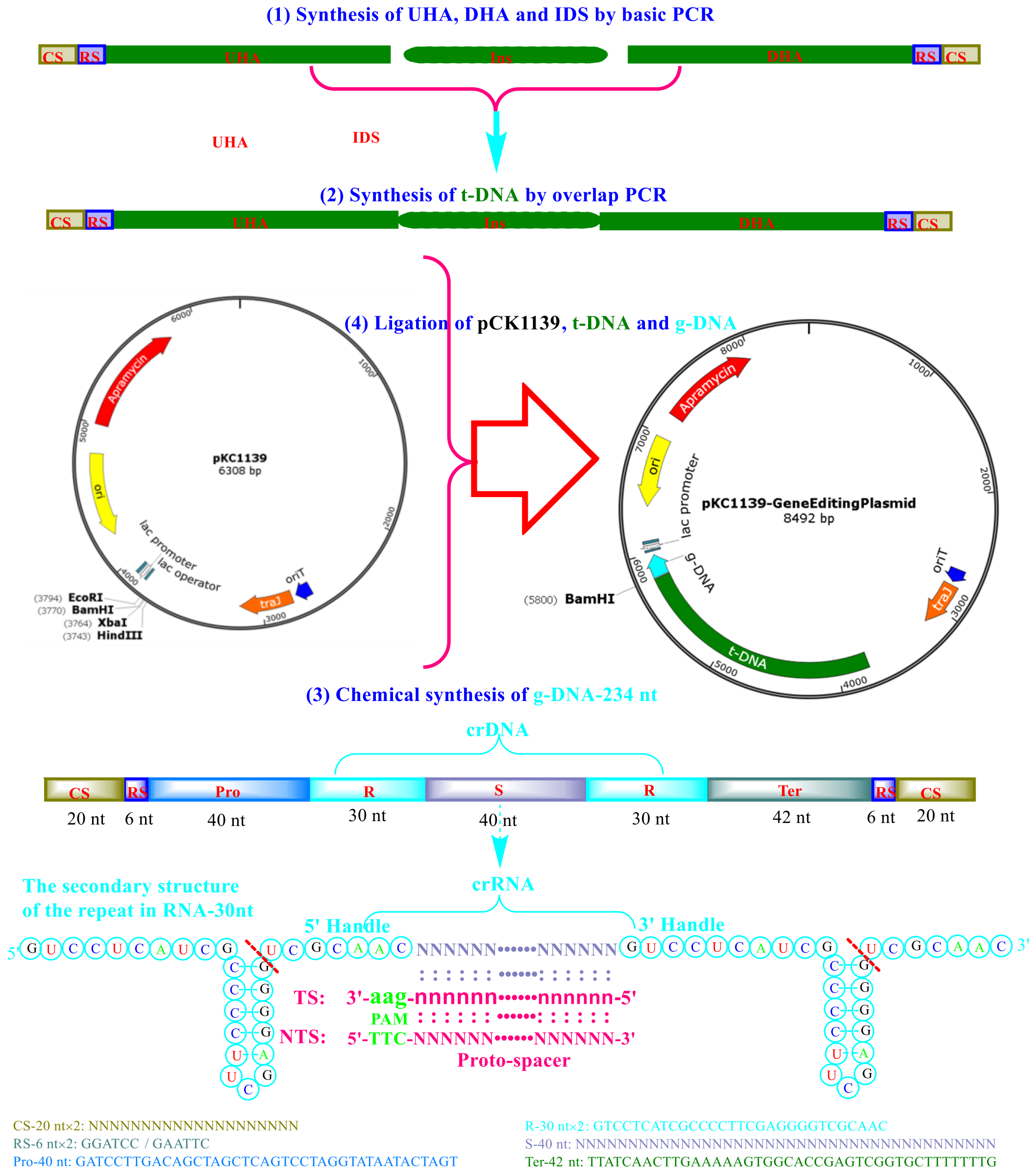

CS: complementary sequence
DHA: downstream homologous arm
Ins: Inserted DNA sequence
NTS: non-target strand DNA
Pro: promoter-pJ23119
R: repeat-the CRISPR3
RS: restriction site
S: spacer
TS: target strand DNA
Ter: terminator-*S. pyogenes*
UHA: upstream homologous arm

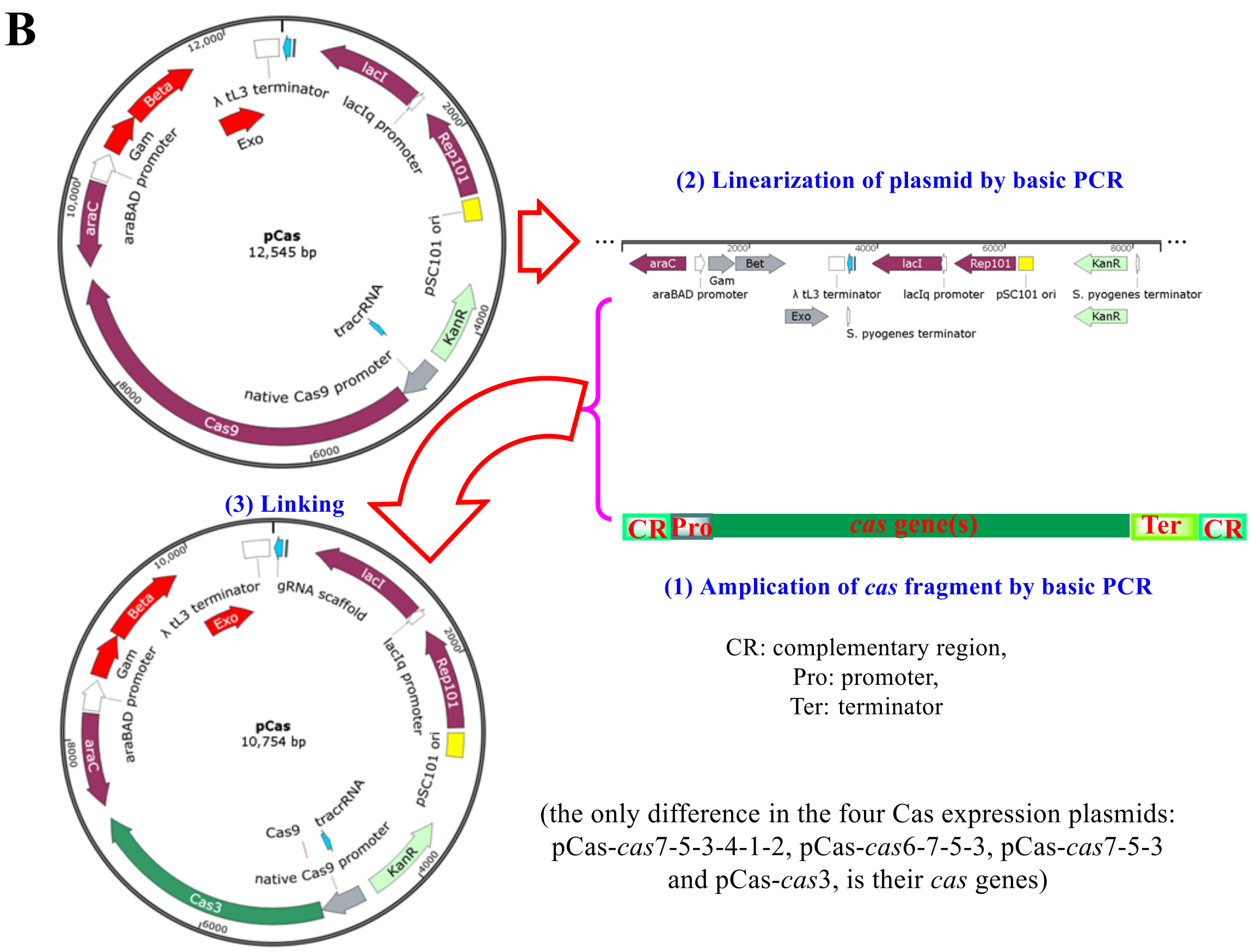

**Figure S1. Construction process of gene editing tools based on the subtype I-B-*Svi* CRISPR-CAS system. (A)** A schematic diagram of gene editing plasmid construction process. **(B)** A schematic diagram of Cas expression plasmid construction process.

A

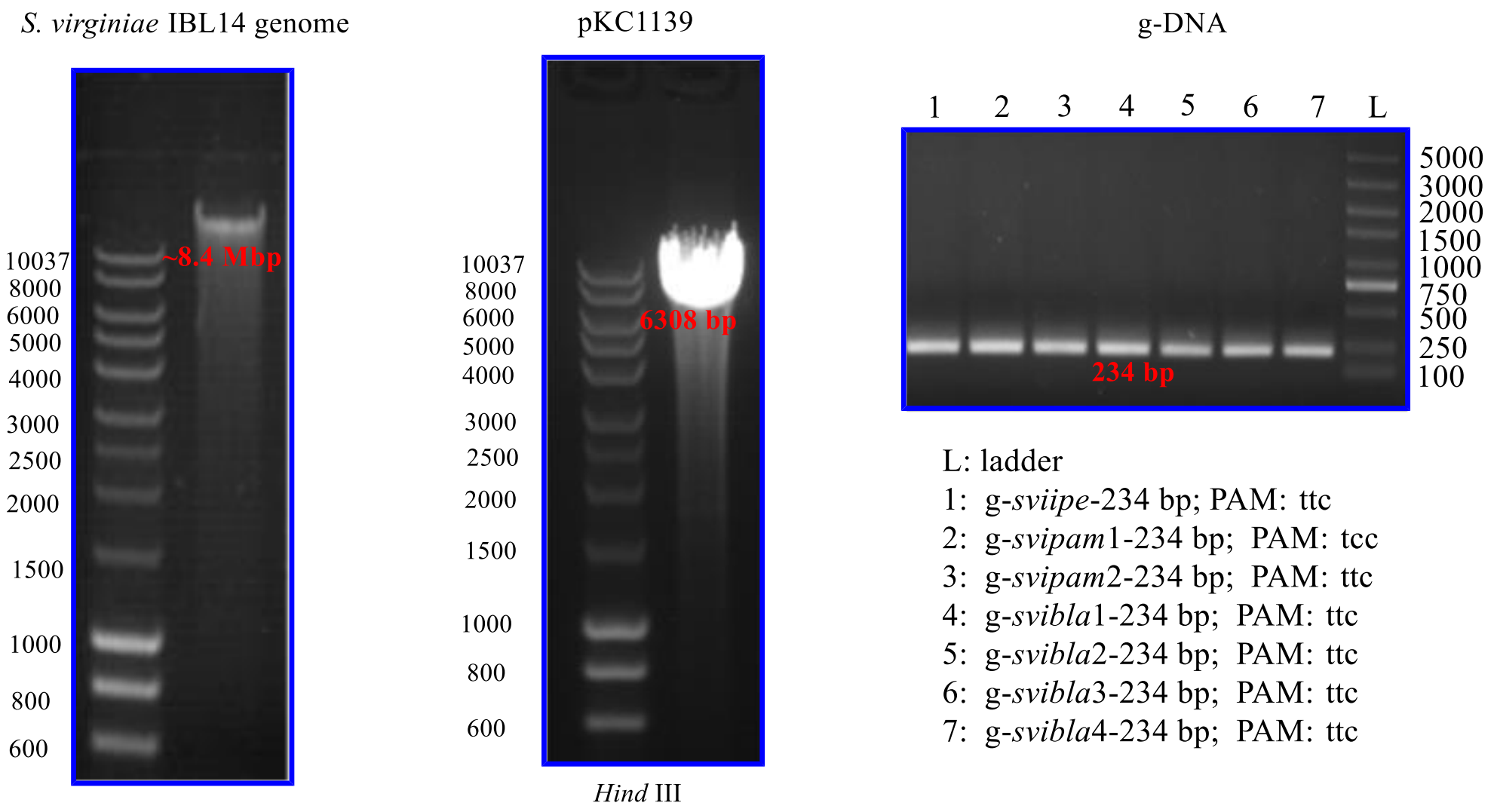

S. virginiae IBL14 genome
10037
8000
6000
5000
4000
3000
2500
2000
1500
1000
800
600
~8.4 Mbp
pKC1139
10037
8000
6000
5000
4000
3000
2500
2000
1500
1000
800
600
6308 bp
Hind III
g-DNA
1 2 3 4 5 6 7 L
5000
3000
2000
1500
1000
750
500
250
100
234 bp
L: ladder
1: g-sviipe-234 bp; PAM: ttc
2: g-svipam1-234 bp; PAM: tcc
3: g-svipam2-234 bp; PAM: ttc
4: g-svibla1-234 bp; PAM: ttc
5: g-svibla2-234 bp; PAM: ttc
6: g-svibla3-234 bp; PAM: ttc
7: g-svibla4-234 bp; PAM: ttc

B

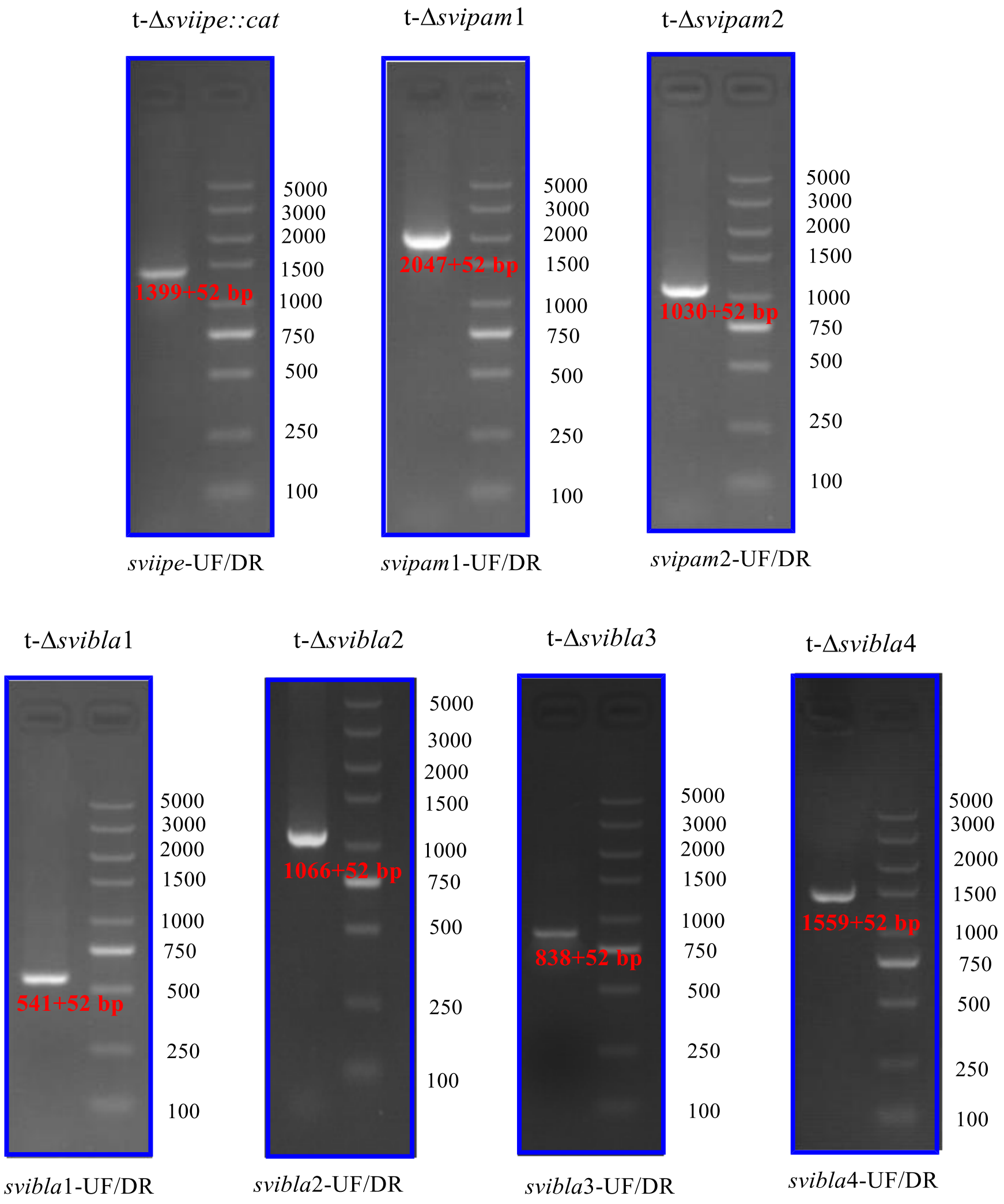

**C**

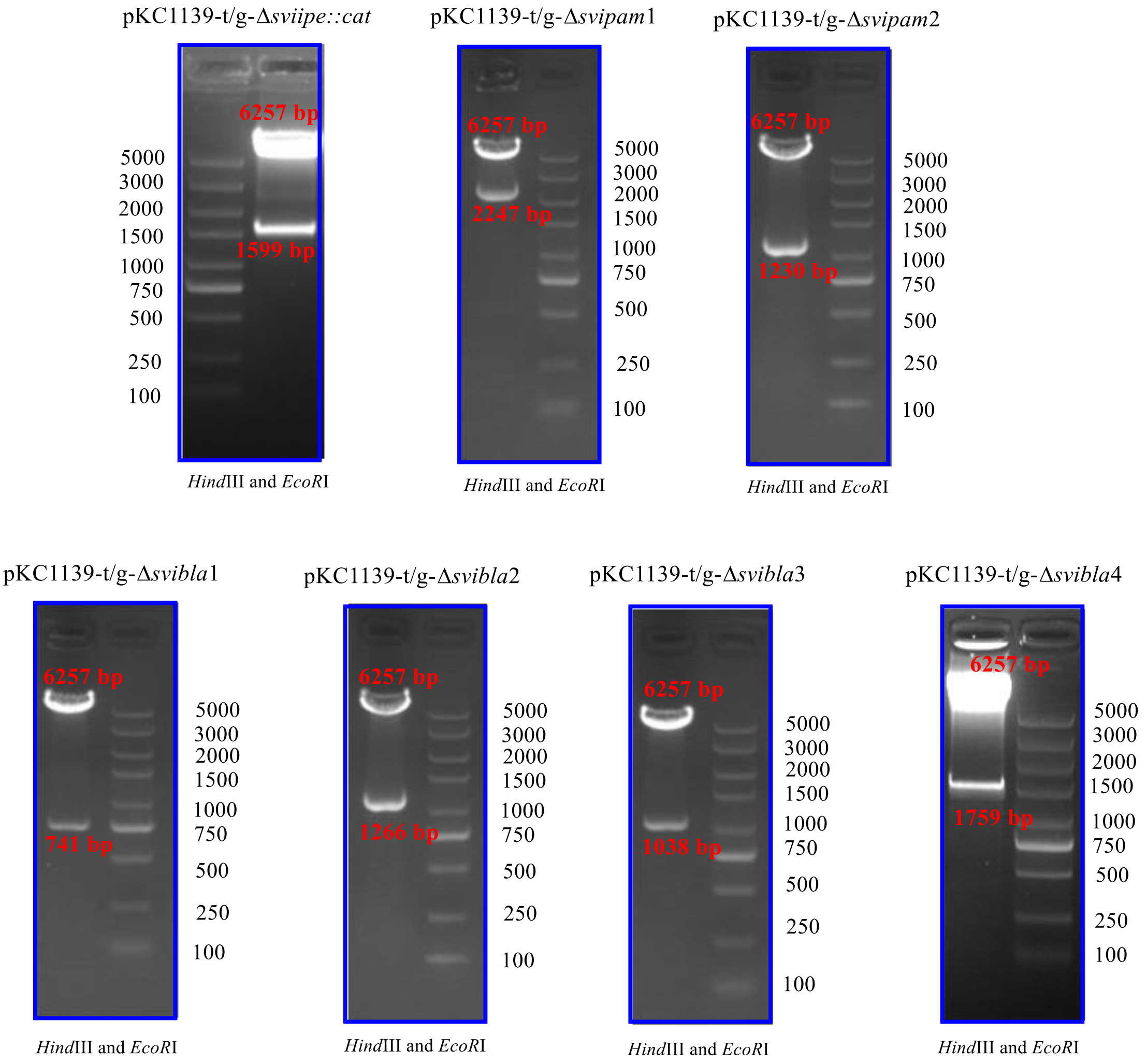

**D**

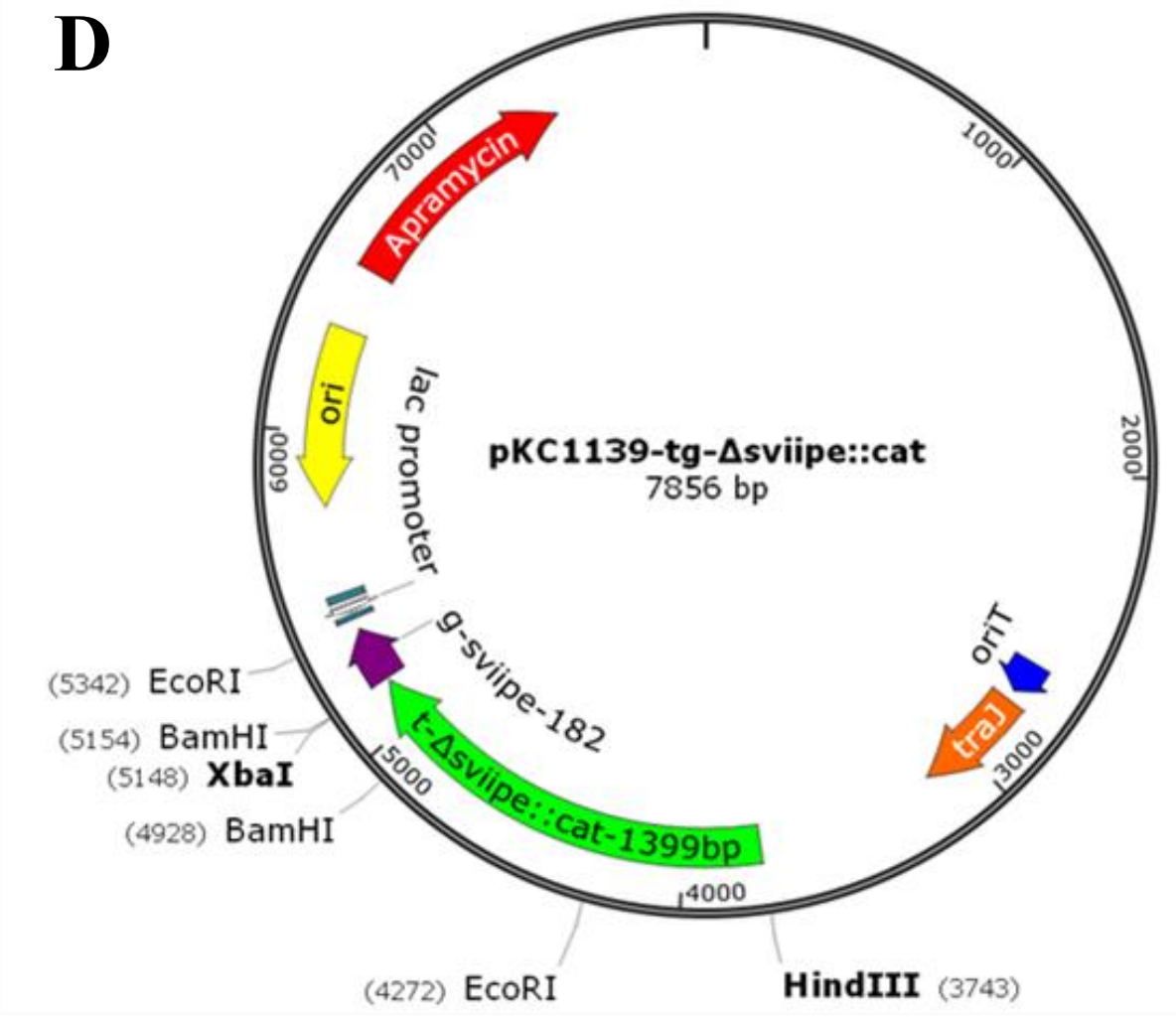

**Figure S2. Construction of gene editing plasmids in the self-targeting genome editing of *S. virginiae* IBL14. (A)** The extracted *S. virginiae* IBL14 genome, linearized pKC1139 by *Hind*III and chemically synthesized g-DNA fragments. **(B)** The t-DNA fragments amplified by overlap PCR. **(C)** The results of the constructed gene editing plasmids digested by *Hind*III and *Eco*RI. **(D)** The map of gene editing plasmid pKC1139-t/g-Δ*sviipe*::*cat* (In the self-targeting senome editing of *S. virginiae* IBL14, other six gene editing plasmids except the corresponding g-DNAs and t-DNAs were the same as this map).

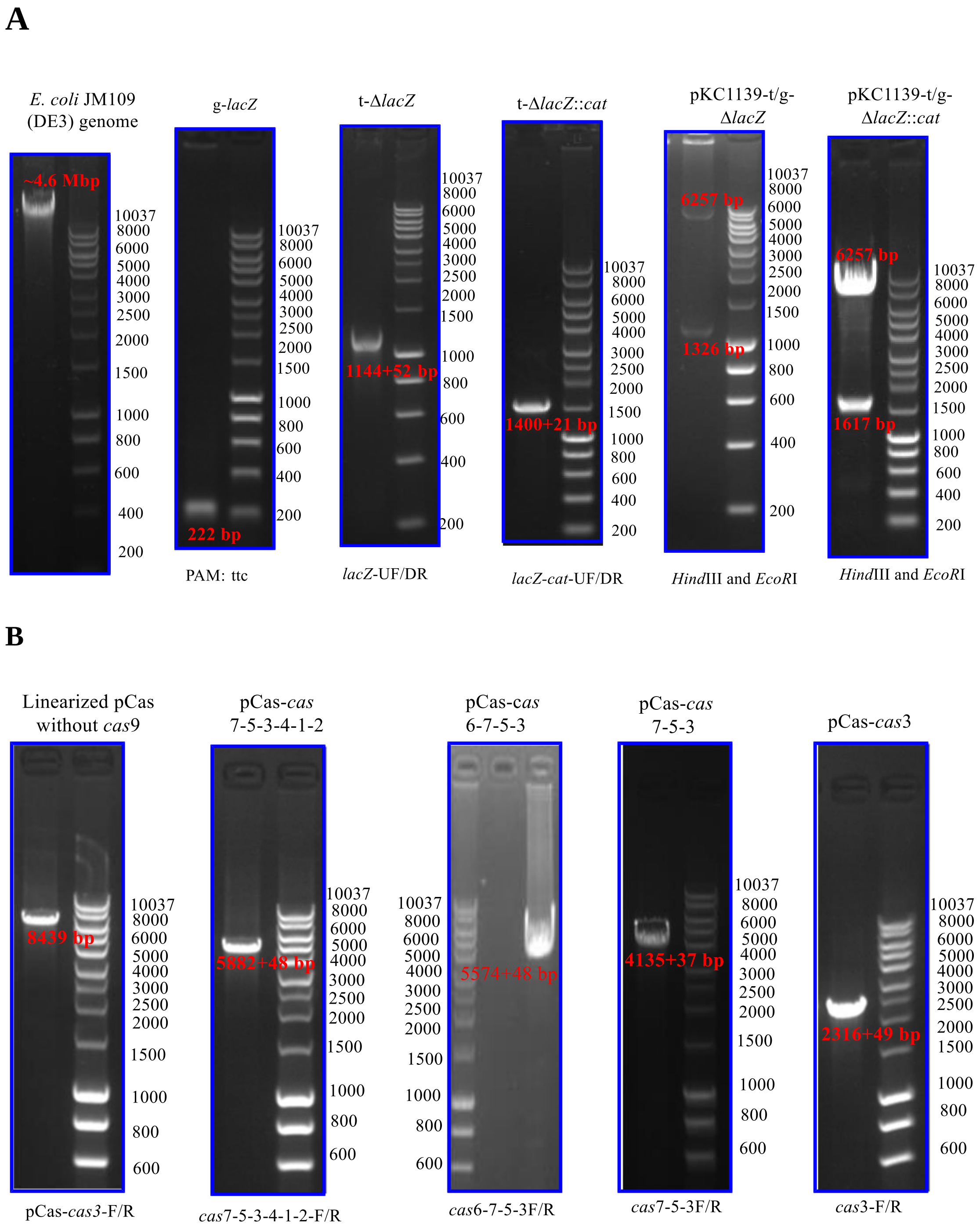

C

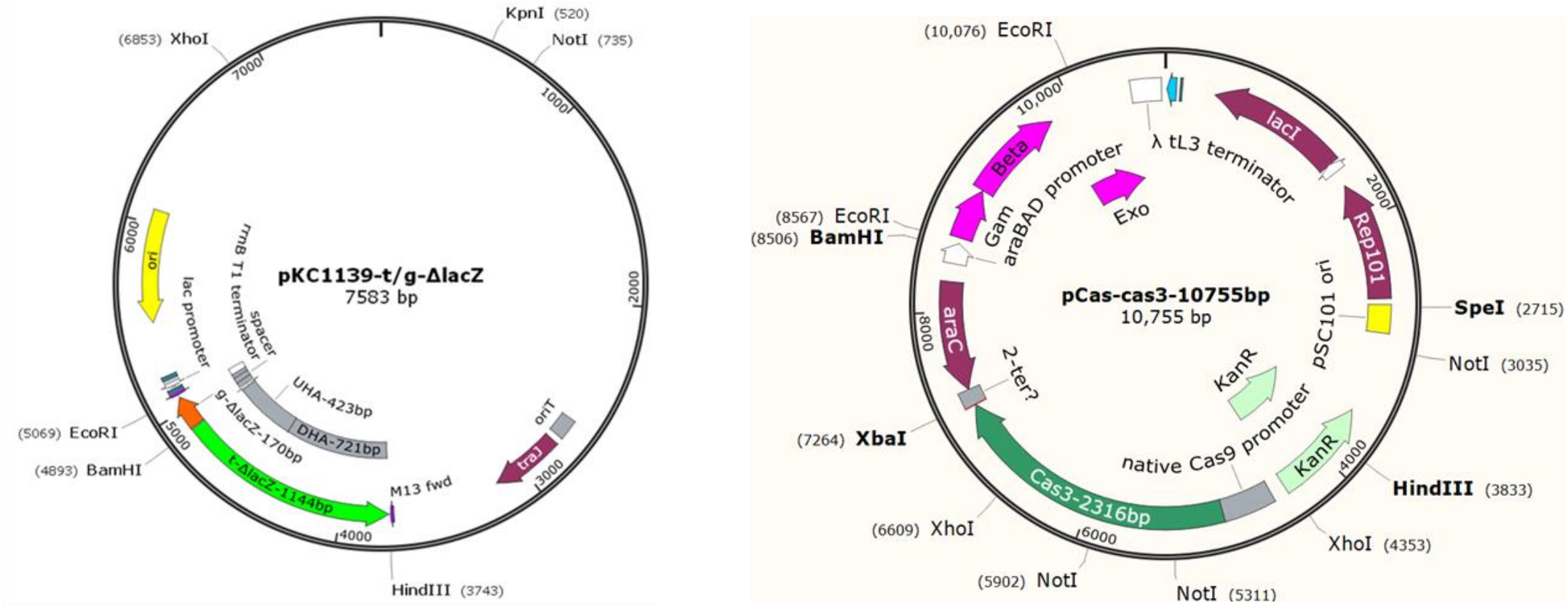

**Figure S3. Construction of gene editing tools in the genome editing of *E. coli* JM109 (DE3). (A)** The extracted *E. coli* JM109 (DE3) genome, chemically synthesized g-*lacZ*, overlap PCR amplified t-Δ*lacZ* and t-Δ*lacZ*::*cat* and the results of gene editing plasmids pKC1139-t/g-Δ*lacZ* and pKC1139-t/g-Δ*lacZ*::*cat* digested by *Hind*III and *EcoR*I, respectively. **(B)** The linearized pCas without *cas*9 by PCR and the fragments of genes *cas*7-5-3-4-1-2, *cas*6-7-5-3, *cas*7-5-3 and *cas*3 amplified by PCR from pCas-*cas*7-5-3-4-1-2, pCas-*cas*6-7-5-3, pCas-*cas*7-5-3 and pCas-*cas*3, respectively. **(C)** The maps of gene editing plasmid pKC1139-*lac*Z and Cas expression plasmid pCas-*cas*3 (In the genome editing of *E. coli* JM109 (DE3), other gene editing tools are the same as these two maps except for the corresponding t/g-DNAs and the *cas* genes).

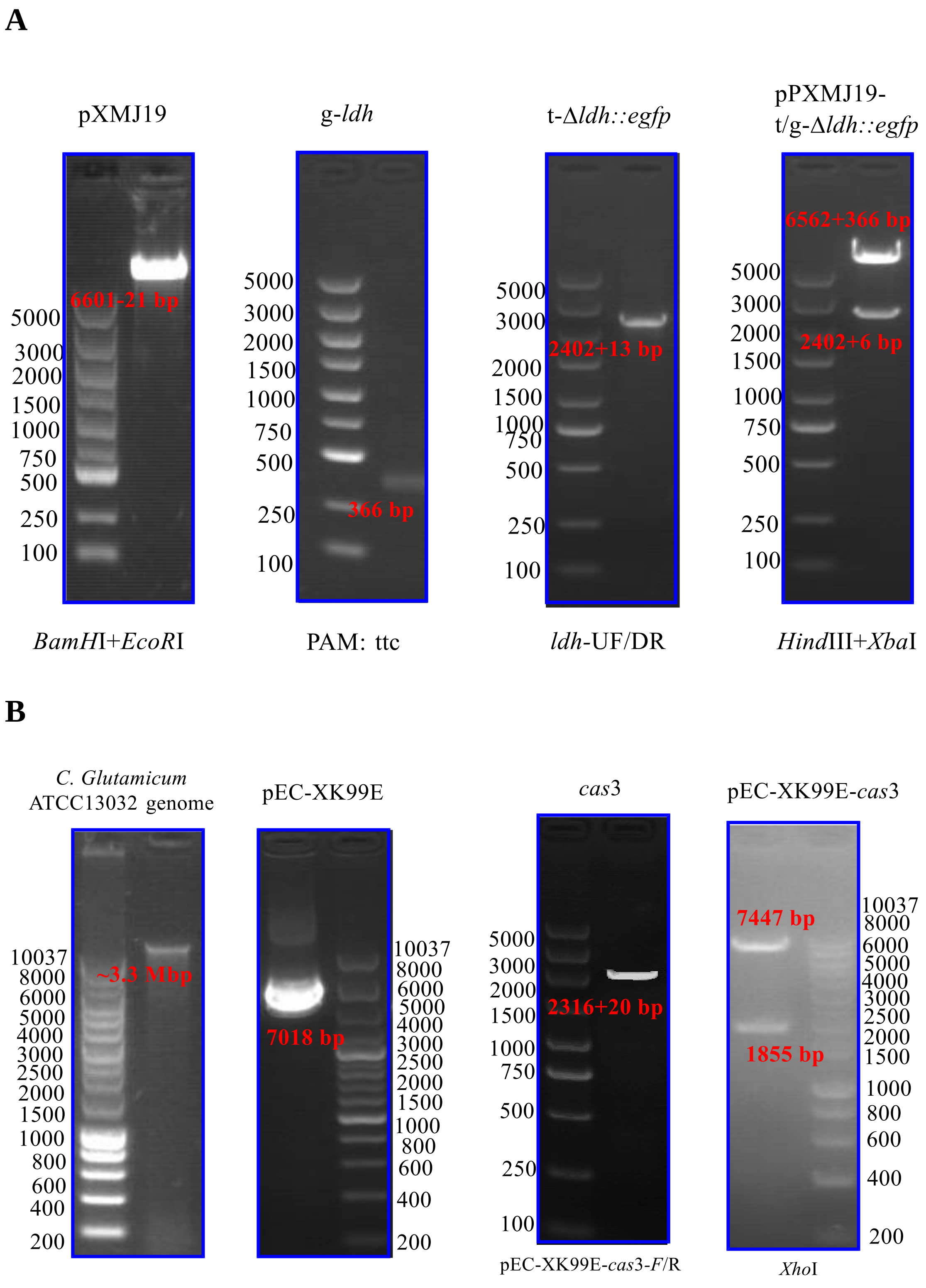

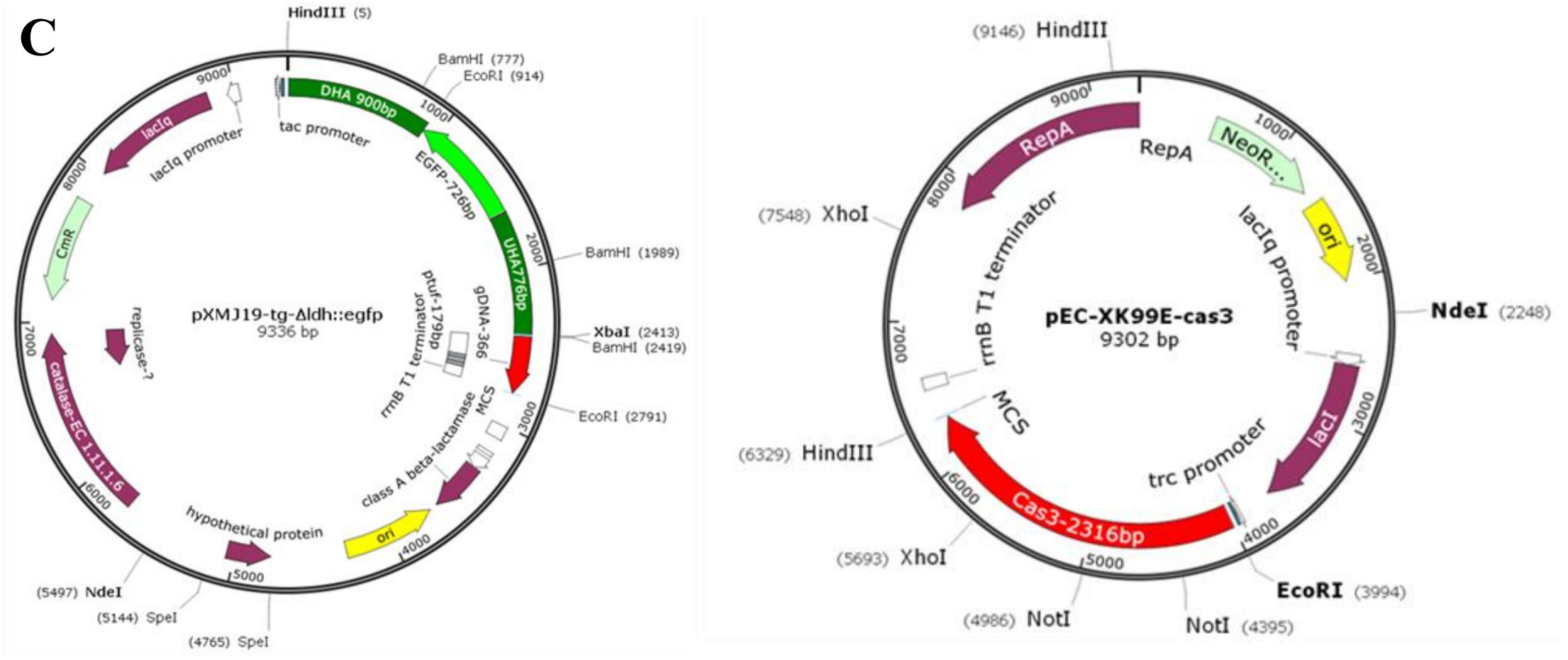

**Figure S4. Construction of gene editing tools in the genome editing of *C. glutamicum* ATCC 13032. (A)** The plasmid pXMJ19 digested by *BamH*I and *EcoR*I, overlap PCR amplified t-Δ*ldh*::*egfp*, chemically synthesized g-*ldh* and the result of the gene editing plasmid pXMJ19-t/g-Δ*ldh*::*egfp* digested by *Hind*III and *Xba*I. **(B)** The extracted *C. glutamicum* ATCC 13032 genome, extracted plasmid pEC-XK99E, PCR amplified *cas*3 from pEC-XK99E-*cas*3 and the result of pEC-XK99E-*cas*3 digested by *Xho*I (two *Xho*I restriction sites in this plasmid), respectively. **(C)** The maps of gene editing plasmid pXMJ19-t/g-Δ*ldh*::*egfp* and Cas expression plasmid pEC-XK99E-*cas*3.

**A**

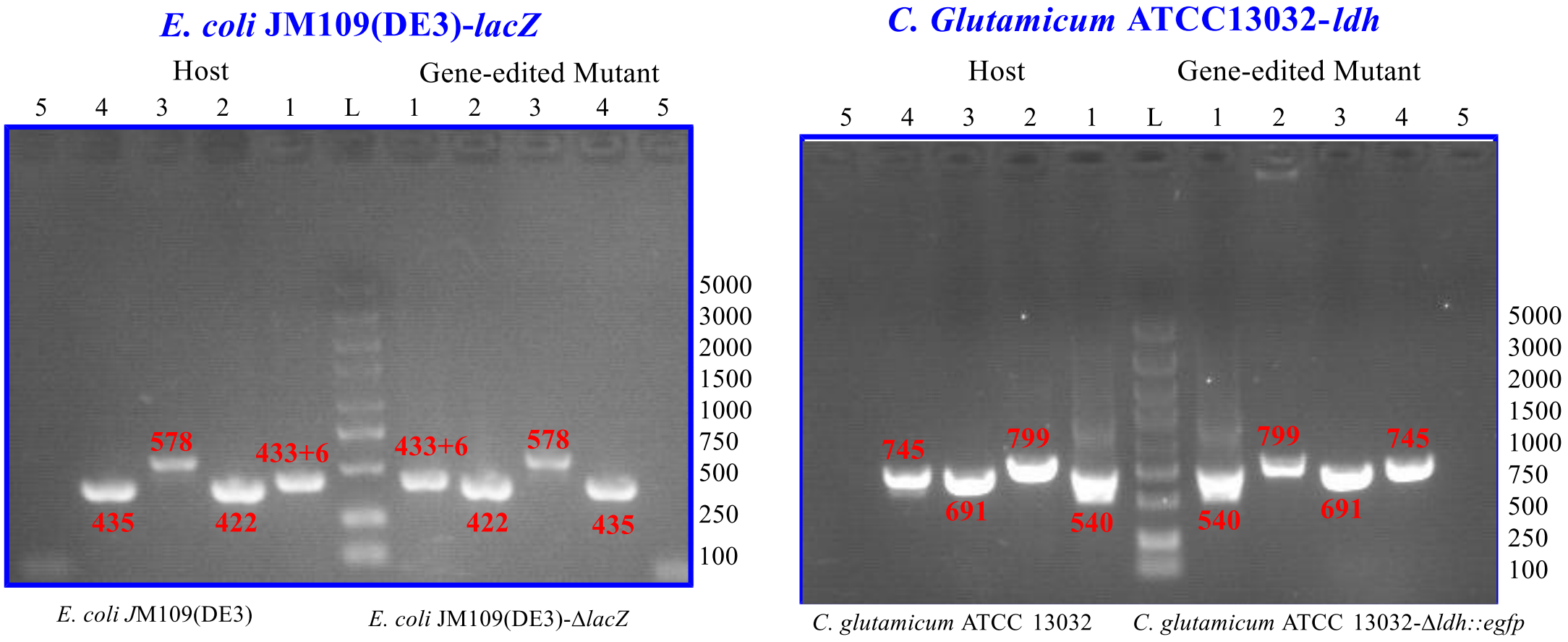

L: ladder: 5000 bp; 1: *Ec-lacZ*-ot01-F/R
2: *Ec-lacZ*-ot02-F/R; 3: *Ec-lacZ*-ot03-F/R
4: *Ec-lacZ*-ot04-F/R; 5: $ddH_2O$

L: ladder: 5000 bp; 1: *Cg-ldh*-ot01-F/R
2: *Cg-ldh*-ot02-F/R; 3: *Cg-ldh*-ot03-F/R
4: *Cg-ldh*-ot04-F/R; 5: $ddH_2O$

**B**

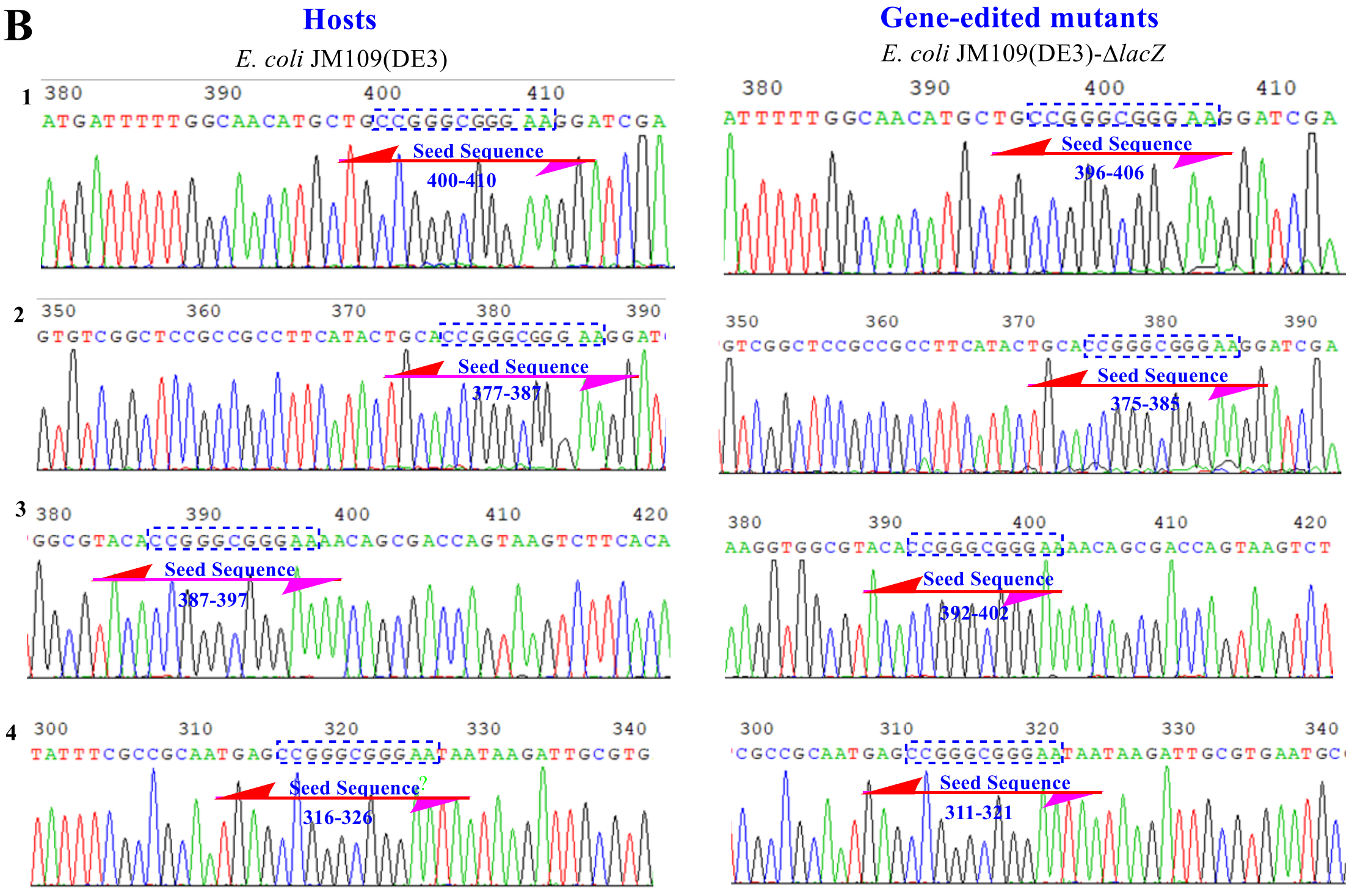

A: green; T:red; G: black; C; blue

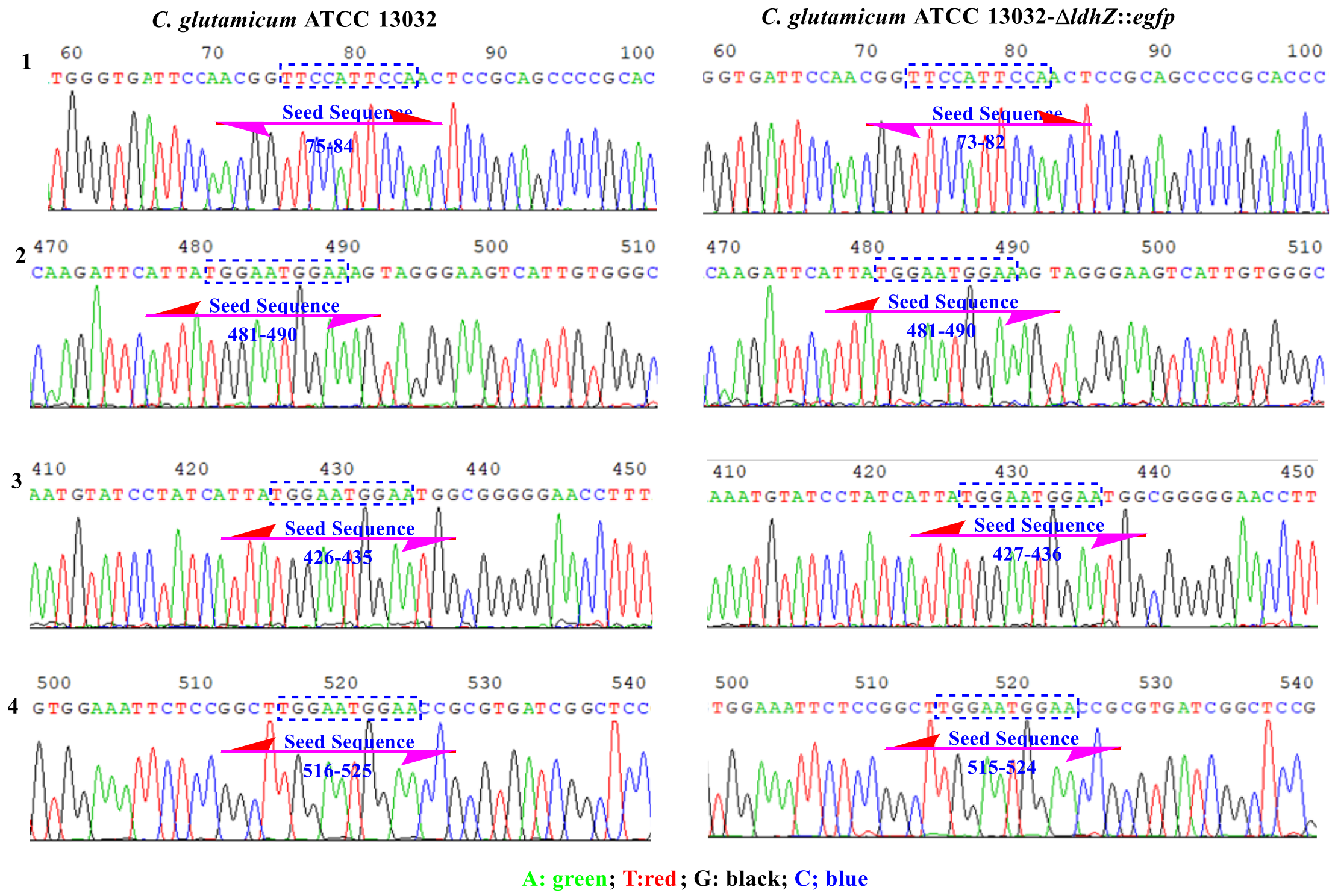

**Figure S5. Off-target analyses in the non-self-targeting genome editing of prokaryotic microorganisms conducted by the *Svi*Cas3. (A)** DNA gel electrophoresis of the PCR products of potential off-target sites in the gene-edited mutants. **(B)** The DNA sequencing results of the PCR products of potential off-target sites in the gene-edited mutants.

A

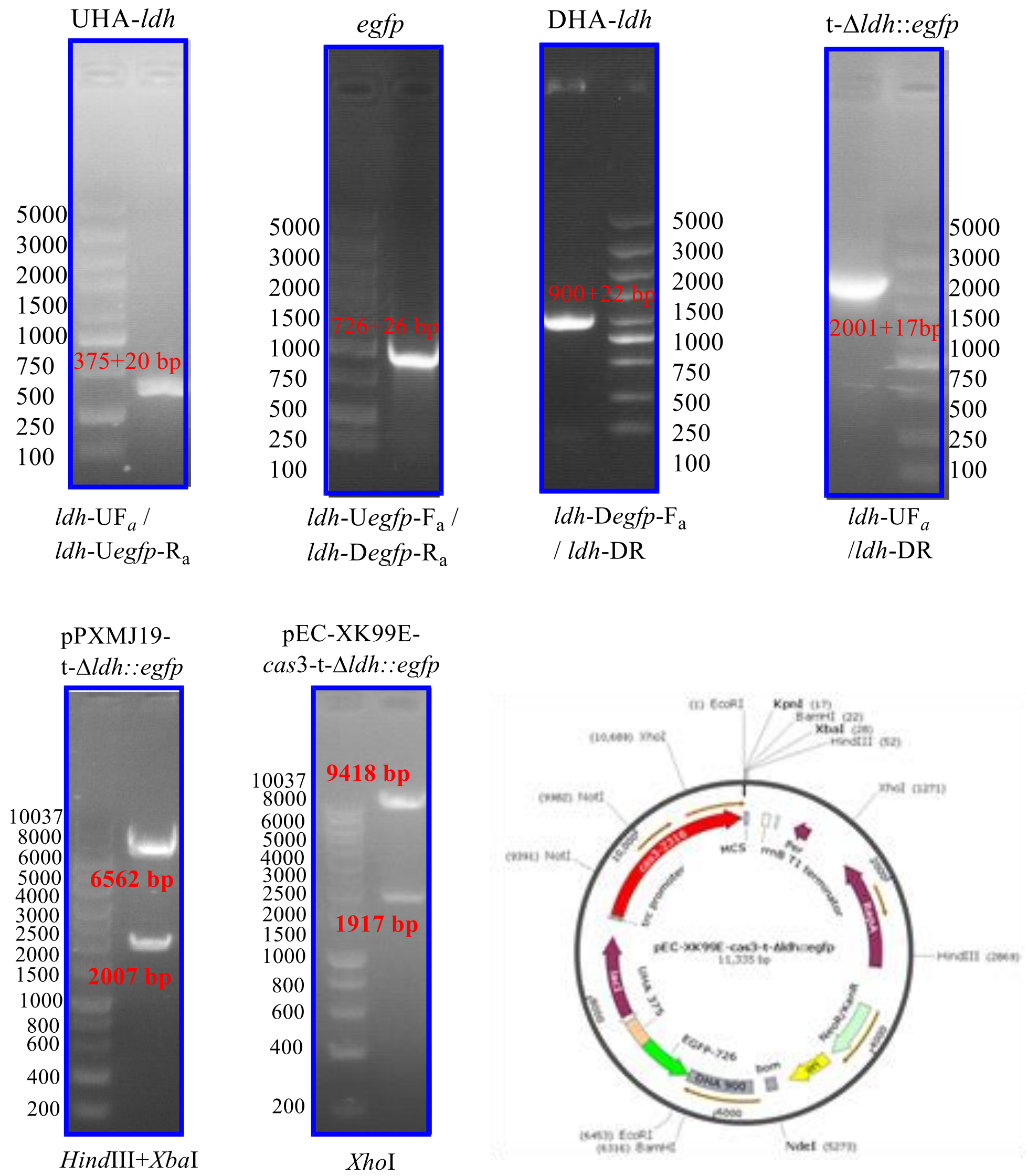

B

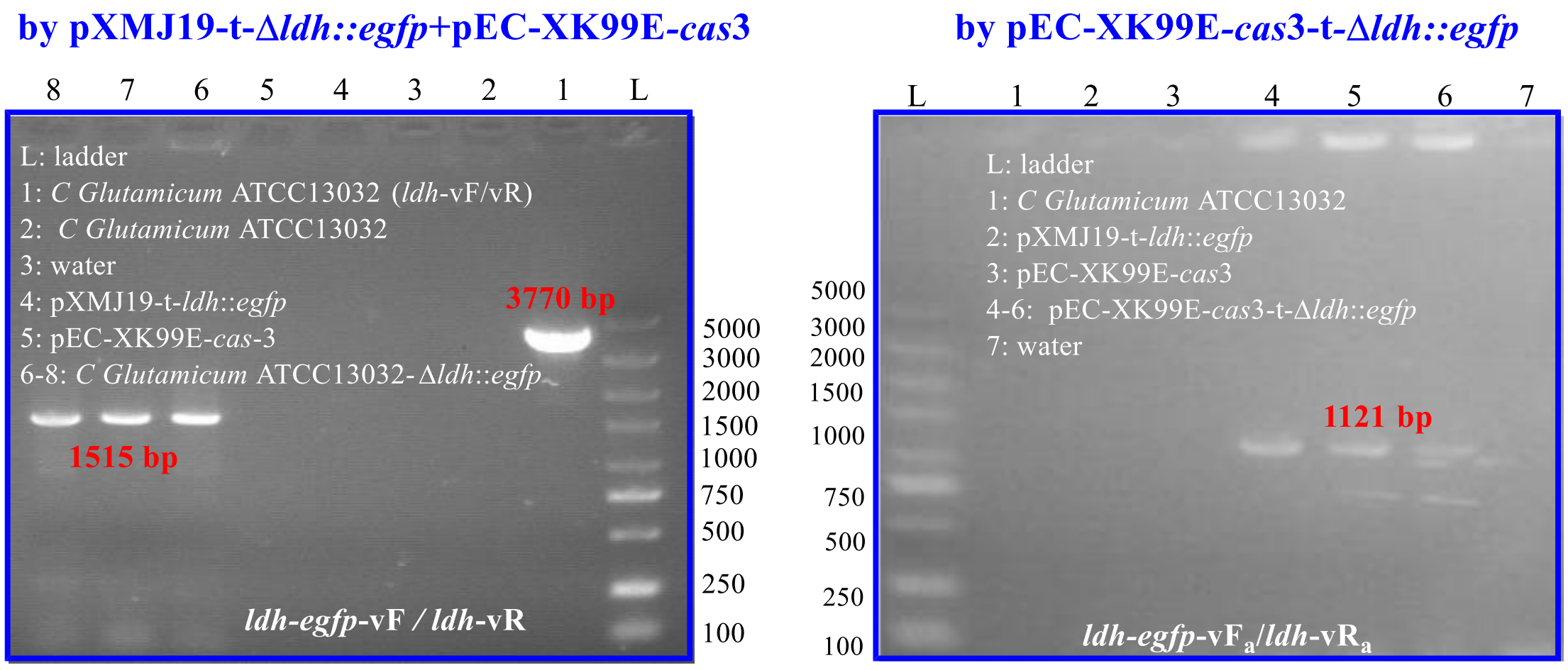
by pXMJ19-t-Δ*ldh::egfp*+pEC-XK99E-*cas*3
8 7 6 5 4 3 2 1 L
L: ladder
1: *C Glutamicum* ATCC13032 (*ldh*-vF/vR)
2: *C Glutamicum* ATCC13032
3: water
4: pXMJ19-t-*ldh::egfp*
5: pEC-XK99E-*cas*-3
6-8: *C Glutamicum* ATCC13032-Δ*ldh::egfp*
3770 bp
1515 bp
5000
3000
2000
1500
1000
750
500
250
100
*ldh-egfp*-vF / *ldh*-vR
by pEC-XK99E-*cas*3-t-Δ*ldh::egfp*
L 1 2 3 4 5 6 7
L: ladder
1: *C Glutamicum* ATCC13032
2: pXMJ19-t-*ldh::egfp*
3: pEC-XK99E-*cas*3
4-6: pEC-XK99E-*cas*3-t-Δ*ldh::egfp*
7: water
1121 bp
*ldh-egfp*-vF$_a$/*ldh*-vR$_a$

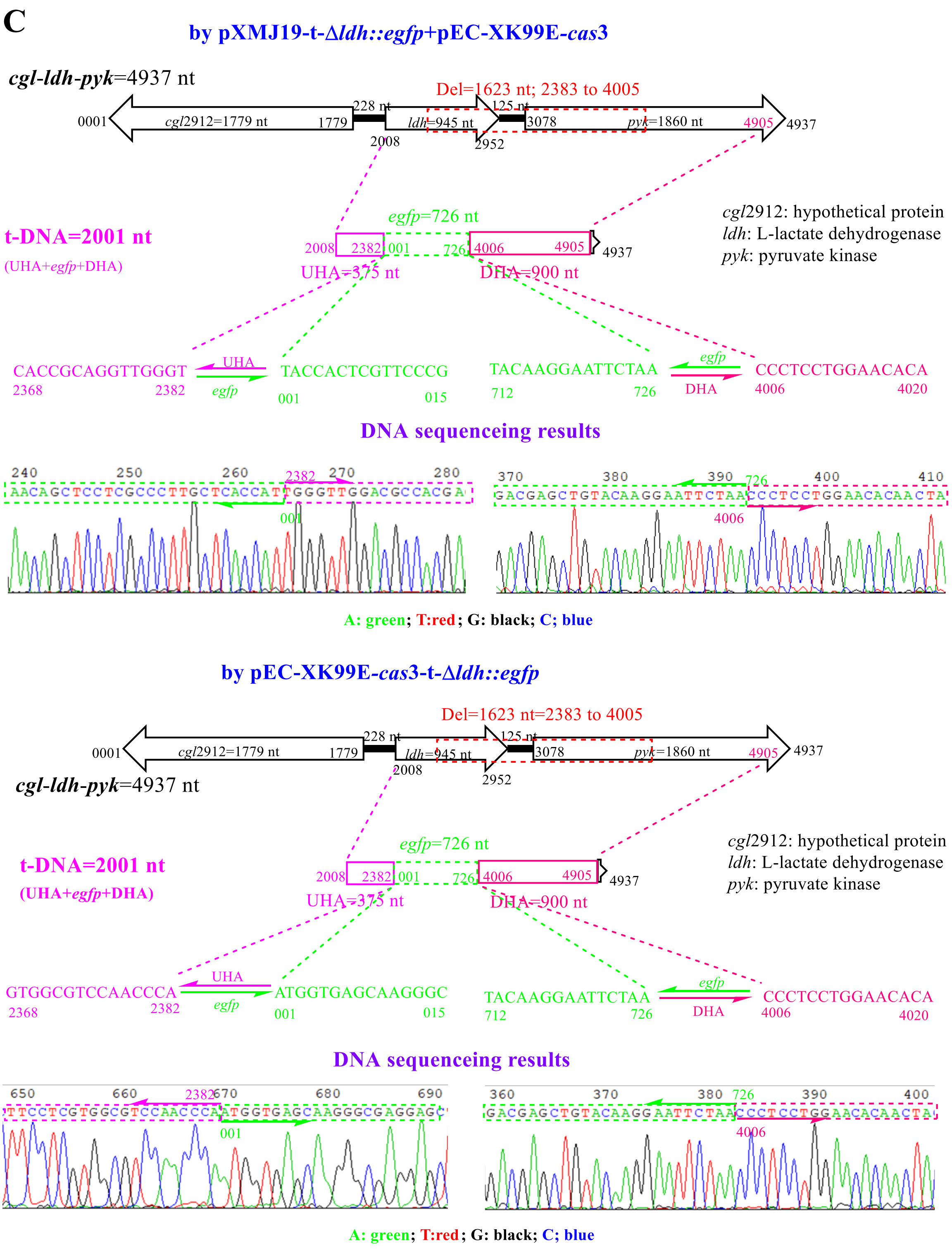

**Figure S6. DNA-guided genome editing in *C. glutamicum* ATCC 13032. (A)** The construction of DNA-guided genome editing tools (The PCR amplified UHA-*ldh*, *egfp*, DHA-*ldh* and the overlap PCR amplified t-Δ*ldh*::*egfp*, the results of plasmids pXMJ19-t-Δ*ldh*::*egfp* and pEC-XK99E-*cas*3-t-Δ*ldh*::*egfp* respectively digested by *Hind*III+*Xba*I and *Xho*I, as well as the map of plasmid pEC-XK99E-*cas*3-t-Δ*ldh*::*egfp*). **(B)** The results of DNA gel electrophoresis of the PCR product of target sequences in the DNA-guided genome editing of *S C. glutamicum* ATCC 13032 by pEC-XK99E-*cas*3 plus pXMJ19-t-Δ*ldh*::*egfp* and pEC-XK99E-*cas*3-t-Δ*ldh*::*egfp*, respectively. **(C)** The engineered t-DNAs as well as the results of DNA sequencing of the PCR product of target sequences in the DNA-guided genome editing of *S C. glutamicum* ATCC 13032 by pEC-XK99E-*cas*3-t-Δ*ldh*::*egfp* (PCR products only occur in gene-edited mutants *C. glutamicum* ATCC 13032-Δ*ldh*::*egfp* because the primers *ldh-egfp*-vF and *ldh-egfp*-$vF_a$ are a portion of gene *egfp* in t-Δ*ldh*::*egfp* and the primers *ldh*-vR and *ldh*-$vR_a$ are a portion of gene *ldh* that is not contained in t-Δ*ldh*::*egfp*. ).

## Tables S1 to S5

**Table S1.** Strains and plasmids used in this study

**Table S2.** Primers, g-DNAs and t-DNAs used in this study

**Table S3.** Main biochemical properties of *Svi*Cas proteins in the subtype I-B-*Svi* Cas system

**Table S4.** Genome editing efficiency of microbial and mammalian cells based on the *Svi*Cas system

**Table S5.** Main biochemical characteristics of the typical signature proteins in various type CRISPR-Cas systems

**Table S1.** Strains and plasmids used in this study (http://www.addgene.org/; http://www.ncbi.nlm.nih.gov)

| Item | Description | Application | Source |
|---|---|---|---|
| **Strain** | | | |
| *C. glutamicum* ATCC 13032 | wild-type | *C. glutamicum* ATCC 13032 | Our lab |
| *C. glutamicum* ATCC 13032-pXMJ19 | pXMJ19 | *C. glutamicum* ATCC 13032 | This study |
| *C. glutamicum* ATCC 13032-pEC-XK99E | pEC-XK99E | *C. glutamicum* ATCC 13032 | This study |
| *C. glutamicum* ATCC 13032-pXMJ19-t/g-Δ*ldh::egfp* | pXMJ19-t/g-Δ*ldh::egfp* | *C. glutamicum* ATCC 13032 | This study |
| *C. glutamicum* ATCC 13032-pEC-XK99E-*cas*3 | pEC-XK99E-*cas*3 | *C. glutamicum* ATCC 13032 | This study |
| *C. glutamicum* ATCC 13032-Δ*ldh::egfp* | Δ*ldh::egfp* | *C. glutamicum* ATCC 13032 | This study |
| *E. coli* DH5α | $F^-$，φ80d*lac*ZΔM15，Δ(*lac*ZYA-*arg*F)U169，*deo*R，*rec*A1，*end*A1，*hsd*R17(rk$^-$，mk$^+$)，*pho*A，*sup*E44，λ-，*thi*-1，*gyr*A96，*rel*A1 | | Our lab |
| *E. coli* DH5α-pCas | pCas | *E. coli* JM109 (DE3) | This study |
| *E. coli* DH5α-pCas-*cas*7-5-3-4-1-2 | pCas-*cas*7-5-3-4-1-2 | *E. coli* JM109 (DE3) | This study |
| *E. coli* DH5α-pCas-*cas*6-7-5-3 | pCas-*cas*6-7-5-3 | *E. coli* JM109 (DE3) | This study |
| *E. coli* DH5α-pCas-*cas*7-5-3 | pCas-*cas*7-5-3 | *E. coli* JM109 (DE3) | This study |
| *E. coli* DH5α-pCas-*cas*3 | pCas-*cas*3 | *E. coli* JM109 (DE3) | This study |
| *E. coli* DH5α-pEC-XK99E | pEC-XK99E | *C. glutamicum* ATCC 13032 | This study |
| *E. coli* DH5α-pEC-XK99E-*cas*3 | pEC-XK99E-*cas*3 | *C. glutamicum* ATCC 13032 | This study |
| *E. coli* DH5α-pEC-XK99E-*cas*3-t-Δ*ldh::egfp* | pEC-XK99E-*cas*3-t-Δ*ldh::egfp* | *C. glutamicum* ATCC 13032 | This study |
| *E. coli* DH5α-pKC1139 | pKC1139 | *S. virginiae* IBL14 | This study |
| *E. coli* DH5α-pKC1139-Δ*sviipe::cat* | pKC1139-t/g-Δ*sviipe::cat* | *S. virginiae* IBL14 | This study |
| *E. coli* DH5α-pKC1139-Δ*svipam*1 | pKC1139-t/g-Δ*svipam*1 | *S. virginiae* IBL14 | This study |
| *E. coli* DH5α-pKC1139-Δ*svipam*2 | pKC1139-t/g-Δ*svipam*2 | *S. virginiae* IBL14 | This study |
| *E. coli* DH5α-pKC1139-Δ*svibla*1 | pKC1139-t/g-Δ*svibla*1 | *S. virginiae* IBL14 | This study |
| *E. coli* DH5α-pKC1139-Δ*svibla*2 | pKC1139-t/g-Δ*svibla*2 | *S. virginiae* IBL14 | This study |
| *E. coli* DH5α-pKC1139-Δ*svibla*3 | pKC1139-t/g-Δ*svibla*3 | *S. virginiae* IBL14 | This study |
| *E. coli* DH5α-pKC1139-Δ*svibla*4 | pKC1139-t/g-Δ*svibla*4 | *S. virginiae* IBL14 | This study |
| *E. coli* DH5α-pKC1139-t/g-Δ*lacZ* | pKC1139-t/g-*lacZ* | *E. coli* JM109 (DE3) | This study |
| *E. coli* DH5α-pKC1139-t/g-Δ*lacZ*::*cat* | pKC1139-t/g-Δ*lacZ*::*cat* | *E. coli* JM109 (DE3) | This study |
| *E. coli* DH5α-pXMJ19 | pXMJ19 | *C. glutamicum* ATCC 13032 | This study |
| *E. coli* DH5α-pXMJ19-t/g-Δ*ldh*::*egfp* | pXMJ19-t/g-Δ*ldh*::*egfp* | *C. glutamicum* ATCC 13032 | This study |
| *E. coli* DH5α-pXMJ19-t-Δ*ldh*::*egfp* | pXMJ19-t-Δ*ldh*::*egfp* | *C. glutamicum* ATCC 13032 | This study |
| *E. coli* JM109 (DE3) | *end*A1, *rec*A1, *gyr*A96, *thi*-1, *hsd*R17 ($rk^-$, $mk^+$), *rel*A1, *sup*E44, Δ(*lac*-*pro*AB)/[F', *tra*D36, *pro*AB, *lac*I$^q$ZΔM15], λ–, lDE3 | *E. coli* JM109 (DE3) | Our lab |
| *E. coli* JM109 (DE3)-Δ*lacZ* | Δ*lacZ* | *E. coli* JM109 (DE3) | This study |
| *E. coli* JM109 (DE3)-Δ*lacZ*::*cat* | Δ*lacZ*::*cat* | *E. coli* JM109 (DE3) | This study |
| *E. coli* JM109 (DE3)-pKC1139 | pKC1139 | *E. coli* JM109 (DE3) | This study |
| *E. coli* JM109 (DE3)-pCas | pCas | *E. coli* JM109 (DE3) | This study |
| *E. coli* JM109 (DE3)-pCas-*cas*7-5-3-4-1-2 | pCas-*cas*7-5-3-4-1-2 | *E. coli* JM109 (DE3) | This study |
| *E. coli* JM109 (DE3)-pCas-*cas*6-7-5-3 | pCas-*cas*6-7-5-3 | *E. coli* JM109 (DE3) | This study |
| *E. coli* JM109 (DE3)-pCas-*cas*7-5-3 | pCas-*cas*7-5-3 | *E. coli* JM109 (DE3) | This study |
| *E. coli* JM109 (DE3)-pCas-*cas*3 | pCas-*cas*3 | *E. coli* JM109 (DE3) | This study |
| *S. virginiae* IBL14 | wild-type | *S. virginiae* IBL14 | Our lab |
| *S. virginiae* IBL14-pKC1139 | pKC1139 | *S. virginiae* IBL14 | This study |
| *S. virginiae* IBL14-Δ*sviipe*::*cat* | Δ*sviipe*::*cat* | *S. virginiae* IBL14 | This study |
| *S. virginiae* IBL14-Δ*svipam*1 | Δ*svipam*1 | *S. virginiae* IBL14 | This study |
| *S. virginiae* IBL14-Δ*svipam*2 | Δ*svipam*2 | *S. virginiae* IBL14 | This study |
| *S. virginiae* IBL14-Δ*svibla*1 | Δ*svibla*1 | *S. virginiae* IBL14 | This study |
| *S. virginiae* IBL14-Δ*svibla*2 | Δ*svibla*2 | *S. virginiae* IBL14 | This study |
| *S. virginiae* IBL14-Δ*svibla*3 | Δ*svibla*3 | *S. virginiae* IBL14 | This study |
| *S. virginiae* IBL14-Δ*svibla*4 | Δ*svibla*4 | *S. virginiae* IBL14 | This study |
| *S. virginiae* IBL14-Δ*svipam*1Δ*sviipe*::*cat* | Δ*svibla*1Δ*sviipe*::*cat* | *S. virginiae* IBL14 | This study |
| *S. virginiae* IBL14-Δ*svipam*1Δ*svibla*1 | Δ*svibla*1Δ*svipam*1 | *S. virginiae* IBL14 | This study |
| **Plasmid** | | | |
| pCas(12545 bp) | *cas*9, lambda-Red recombinase expression plasmid, $Kan^R$, ParaB promoter | *E. coli* JM109 (DE3) | Addgene |
| pCas-*cas*7-5-3-4-1-2 | *cas*7-5-3-4-1-2 | *E. coli* JM109 (DE3) | This study |
| pCas-*cas*6-7-5-3 | *Cas*6-7-5-3 | *E. coli* JM109 (DE3) | This study |
| pCas-*cas*7-5-3 | *cas*7-5-3 | *E. coli* JM109 (DE3) | This study |
| pCas-*cas*3 | *cas*3 | *E. coli* JM109 (DE3) | This study |
| pEC-XK99E (7018bp) | Shuttle vector (*C. glutamicum* / *E. coli*), ori pUC, KanR, lacIq, *rep*A | *C. glutamicum* ATCC 13032 | Biovector Inc |
| pEC-XK99E-*cas*3 | *cas*3 | *C. glutamicum* ATCC 13032 | This study |
| pEC-XK99E-*cas*3-t-Δ*ldh::egfp* | *cas*3-t-Δ*ldh::egfp* | *C. glutamicum* ATCC 13032 | This study |
| pKC1139 (6308 bp) | Expression vector in *Streptomyces*, $Apramycin^R$, oriT, lac promoter | *S. virginiae* IBL14 / *E. coli* JM109 (DE3) | Our lab |
| pKC1139-t/g-Δ*sviipe::cat* | t/g-Δ*sviipe::cat* | *S. virginiae* IBL14 | This study |
| pKC1139-t/g-Δ*svipam*1 | t/g-Δ*svipam*1 | *S. virginiae* IBL14 | This study |
| pKC1139-t/g-Δ*svipam*2 | t/g-Δ*svipam*2 | *S. virginiae* IBL14 | This study |
| pKC1139-t/g-Δ*svibla*1 | t/g-Δ*svibla*1 | *S. virginiae* IBL14 | This study |
| pKC1139-t/g-Δ*svibla*2 | t/g-Δ*svibla*2 | *S. virginiae* IBL14 | This study |
| pKC1139-t/g-Δ*svibla*3 | t/g-Δ*svibla*3 | *S. virginiae* IBL14 | This study |
| pKC1139-t/g-Δ*svibla*4 | t/g-Δ*svibla*4 | *S. virginiae* IBL14 | This study |
| pKC1139-t/g-Δ*lacZ* | t/g-Δ*lacZ* | *E. coli* JM109 (DE3) | This study |
| pKC1139-t/g-Δ*lacZ::cat* | t/g-Δ*lacZ::cat* | *E. coli* JM109 (DE3) | This study |
| pXMJ19 (6601 bp) | Shuttle vector (*C. glutamicum* / *E. coli*), CmR, ori pUC, lacIq, *cat* encoding Catalase | *C. glutamicum* ATCC 13032 | Biovector Inc |
| pXMJ19-t/g-Δ*ldh::egfp* | t/g-Δ*ldh::egfp* | *C. glutamicum* ATCC 13032 | This study |
| pXMJ19-t-Δ*ldh::egfp* | t-Δ*ldh::egfp* | *C. glutamicum* ATCC 13032 | This study |

**Table S2.** Primers, g-DNAs and t-DNAs used in this study*

| Items | Sequence (5´ to 3´) | Note |
|---|---|---|
| **Primers** | | |
| ***S. virginiae* IBL14** | | |
| *sviipe*-vF | ccaggccaatcccgcccaagcggcccttt | Δ*sviipe*::*cat* |
| *sviipe*-vR | cccgcaagggctggatcatggacacctat | Δ*sviipe*::*cat* |
| *sviipe*-UF | gtaaaacgacggccagtgccAAGCTTgcacctcaaccacggctcctacgg | UHA |
| *sviipe*-U*cat*-R | tcttacgtgccgatcaacgtctccctcggtggcgttggagatgaa | UHA-*cat* |
| *sviipe*-U*cat*-F | ttcatctccaacgccaccgagggagacgttgatcggcacgtaaga | UHA-*cat* |
| *sviipe*-D*cat*-R | agccctggtggtcctcccatgggaacttcggaataggaacttcattta | DHA-*cat* |
| *sviipe*-D*cat*-F | taaatgaagttcctattccgaagttcccatgggaggaccaccagggct | DHA-*cat* |
| *sviipe*-DR | tcgcgcgcggccgcggatccTCTAGAaggcgttcgtactcctcgggc | DHA |
| *svipam*1-vF | tcgttcccgcagaccaccggctccctcaa | Δ*svipam*1 |
| *svipam*1-vR | ttggcccacagggtcgtctggtcggtgtagtt | Δ*svipam*1 |
| *svipam*1-UF | gtaaaacgacggccagtgccAAGCTTcgagaccgacgccttcct | UHA |
| *svipam*1-UR | gatgaggttctgcgacgggacctccaggtagaggtcggtcacg | UHA |
| *svipam*1-DF | cgtgaccgacctctacctggaggtcccgtcgcagaacctcatc | DHA |
| *svipam*1-DR | tcgcgcgcggccgcggatccTCTAGAgttgaggttgacgaccatccg | DHA |
| *svipam*2-vF | gtgaccaccgaggcctacga | Δ*svipam*2 |
| *svipam*2-vR | gaggaggggcgtggtgagga | Δ*svipam*2 |
| *svipam*2-UF | gtaaaacgacggccagtgccAAGCTTaccaccgaggcctacgaggtct | UHA |
| *svipam*2-UR | tctccaggatctcgacctcgacgaagaggatgtgggtgccgat | UHA |
| *svipam*2-DF | atcggcacccacatcctcttcgtcgaggtcgagatcctggaga | DHA |
| *svipam*2-DR | tcgcgcgcggccgcggatccTCTAGAtgctgtcggcctccatgtgc | DHA |
| *svibla*1-vF | acggcgacgagggactgtg | Δ*svibla*1 |
| *svibla*1-vR | ccgcccgtggtgcggaccgtcgagaaggggga | Δ*svibla*1 |
| *svibla*1-UF | gtaaaacgacggccagtgccAAGCTTctcccccgaacacccctactac | UHA |
| *svibla*1-UR | aagagcagttcgacggaggcgcgatgatggaggtgaacagc | UHA |
| *svibla*1-DF | gctgttcacctccatcatcgcgcctccgtcgaactgctctt | DHA |
| *svibla*1-DR | tcgcgcgcggccgcggatccTCTAGAtgtccatcccggagaaccag | DHA |
| *svibla*2-vF | atgcgcacgcaccgcgccgc | Δ*svibla*2 |
| *svibla*2-vR | ccgggtgaaggtgcccaggtgttc | Δ*svibla*2 |
| *svibla*2-UF | gtaaaacgacggccagtgccAAGCTTatcctgctcacagccgggacgg | UHA |
| *svibla*2-UR | tgatccgtttcccgtcgaggaccccgctgcggtgggagaacaggtc | UHA |
| *svibla*2-DF | gacctgttctcccaccgcagcggggtcctcgacgggaaacggatca | DHA |
| *svibla*2-DR | tcgcgcgcggccgcggatccTCTAGAccaggtactcgatgcggaccttgc | DHA |
| *svibla*3-vF | gtgcctccaggggctgcgcac | Δ*svibla*3 |
| *svibla*3-vR | ctccgcccggccctgtttgacca | Δ*svibla*3 |
| *svibla*3-UF | gtaaaacgacggccagtgccAAGCTTctctcaccgatgtcgcagaactgat | UHA |
| *svibla*3-UR | caggtgcaggcagacgtggccggggtggtggtgggtgatg | UHA |
| *svibla*3-DF | catcacccaccaccaccccggccacgtctgcctgcacctg | DHA |
| *svibla*3-DR | tcgcgcgcggccgcggatccTCTAGAcgccgcagatgggcttccgcttcc | DHA |
| *svibla*4-vF | gacggagttgaccgcccgagcc | Δ*svibla*4 |
| *svibla*4-vR | ggcactcgcctggctgttcctc | Δ*svibla*4 |
| *svibla*4-UF | gtaaaacgacggccagtgccAAGCTTgctggtgttcctgggctgcgtc | UHA |
| *svibla*4-UR | tgggtgttggtggtgttggcgaggccgtgctcccgttccatgtcc | UHA |
| *svibla*4-DF | ggacatggaacgggagcacggcctcgccaacaccaccaacaccca | DHA |
| *svibla*4-DR | tcgcgcgcggccgcggatccTCTAGAttgtacccgcccctgtggctgat | DHA |
| ***E. coli* JM109 (DE3)** | | |
| *lacZ*-F | tacccaacttaatcgccttgcagcaca | *lacZ* |
| *lacZ*-R | ccgtcgatattcagccatgtgccttctt | *lacZ* |
| *lacZ*-vF | cgtttcatctgtggtgcaac | Δ*lacZ* |
| *lacZ*-vR | catcagttgctgttgactgtag | Δ*lacZ* |
| *lacZ*-UF | ggctgcaggtcgactctagaGGATCCatgagcgtggtggttatgcc | UHA |
| *lacZ*-UR | acgaagccgccctgtaaacccatgccgtgggtttcaata | UHA |
| *lacZ*-DF | tattgaaacccacggcatgggtttacagggcggcttcgt | DHA |
| *lacZ*-DR | ctagagtcgacctgcagcccAAGCTTatgcgggtcgcttcacttac | DHA |
| *lacZ*-*cat*-UF | cgcGGATCCcgatgtcggtttccgcgag | UHA |
| *lacZ*-*cat*-DR | ccccccAAGCTTcgacatccagaggcacttcacc | DHA |
| *lacZ*-U*cat*-F | attgaaacccacggcatggagacgttgatcggcacgt | UHA-*cat* |
| *lacZ*-U*cat*-R | acgtgccgatcaacgtctccatgccgtgggtttcaat | UHA-*cat* |
| *lacZ*-D*cat*-F | aaatgaagttcctattccgaagttccgggactgggtggatcagtcg | DHA-*cat* |
| *lacZ*-D*cat*-R | cgactgatccacccagtcccggaacttcggaataggaacttcattt | DHA-*cat* |
| *lacZ*-*cat*-F | gagctggtgatatgggatagtgttcaccct | t-Δ*lacZ::cat* |
| *cas*7-5-3-4-1-2-F | cggagctcgaattcggatccgtggtcgccggtgccccg | *cas*7-5-3-4-1-2 |
| *cas*7-5-3-4-1-2-R | gtttaactttaagaaggagatataccatggtcacaggatgtcactgggggcggg | *cas*7-5-3-4-1-2 |
| pCas-*cas*7-5-3-4-1-2-F | ttttaggaggcaaaagtggtcgccggtgcc | linearization of pCas without *cas*9 |
| pCas-*cas*7-5-3-4-1-2-R | catctaaaatatacttcacaggatgtcactggggg | linearization of pCas without *cas*9 |
| *cas*6-7-5-3-F | actttttattttaggaggcaaaattggtgctgacggcgcatccg | *cas*6-7-5-3 |
| *cas*6-7-5-3-R | aaataatcttcatctaaaatatacttcacaagacctccccggcgcggta | *cas*6-7-5-3 |
| pCas-*cas*6-7-5-3-F | ttttaggaggcaaaatcacaagacctccccggc | linearization of pCas without *cas*9 |
| pCas-*cas*6-7-5-3-R | catctaaaatatactttggtgctgacggcgcat | linearization of pCas without *cas*9 |
| *cas*7-5-3-F | ttattttaggaggcaaaagtggtcgccggtgccccgaac | *cas*7-5-3 |
| *cas*7-5-3-R | tcttcatctaaaatatacttcacaagacctccccggcgc | *cas*7-5-3 |
| pCas-*cas*7-5-3-F | ggaggtcttgtgaagtatattttagatgaagattatttc | linearization of pCas without *cas*9 |
| pCas-*cas*7-5-3-R | ggggcaccggcgaccacttttgcctcctaaaataaaaag | linearization of pCas without *cas*9 |
| *cas*3-F | actttttattttaggaggcaaaagtgggccgtctggacgcggtgga | *cas*3 |
| *cas*3-R | aaataatcttcatctaaaatatactttcacaagacctccccggcgcggta | *cas*3 |
| pCas-*cas*3-F | taccgcgccggggaggtcttgtgaaagtatattttagatgaagattattt | linearization of pCas without *cas*9 |
| pCas-*cas*3-R | tccaccgcgtccagacggcccacttttgcctcctaaaataaaaagt | linearization of pCas without *cas*9 |
| *Ec*-*lacZ*-ot-F | cgatcc+TTC+CCGCCCGG (xN+PAM+seed: 6+3+8=17 nt) | number of potential off-target site: 7 |
| *Ec*-*lacZ*-ot01-F/R | cgatccttcccgcccgg / cgcgcagggcgacaattttg | off-target analysis |
| *Ec*-*lacZ*-ot02-F/R | cgatccttcccgcccgg / ccgtcagcgctggatgcg | off-target analysis |
| *Ec*-*lacZ*-ot03-F/R | gaagtcaccggcgaaccg / gacgcaggatcatcaatgg | off-target analysis |
| *Ec*-*lacZ*-ot04-F/R | gcagagagttcctgttcggaag / cacacttgcggatcgcatgatg | off-target analysis |
| ***C. glutamicum* ATCC 13032** | | |
| *ldh*-F | agtgggatcgaaaatgaaagaaaccgt | *ldh* |
| *ldh*-R | actaggcgccaaagatttagaagaact | *ldh* |
| *ldh*-vF | gcctttgtgagttactcttcccag | Δ*ldh-egfp* |
| *ldh*-vR | cgatgaaacttatccacggcg | Δ*ldh-egfp* |
| *ldh*-UF | gcTCTAGAttaatactgtgctgggttaattcgc | UHA |
| *ldh*-U*egfp*-R | gctcaccattttcgatcccacttcctgatttc | UHA-*egfp* |
| *ldh*-U*egfp*-F | gtgggatcgaaaatggtgagcaagggcgag | UHA-*egfp* |
| *ldh*-D*egfp*-R | ttccaggagggttaGAATTCcttgtacagctcgtcc | DHA-*egfp* |
| *ldh*-D*egfp*-F | caagGAATTCtaaccctcctggaacacaactac | DHA-*egfp* |
| *ldh*-DR | cccAAGCTTgcagcctactctttcgttgg | DHA |
| *ldh*-*egfp*-vF | tcgtccttgaagaagatggtgc | *egfp* |
| pEC-XK99E-F | gaggtcttgtgatcgacaaggagcaagcttggctg | linearization of pEC-XK99E |
| pEC-XK99E-R | cgaattccttggtctgtttcctgtgtg | linearization of pEC-XK99E |
| pEC-XK99E-*cas*3-F | aggaattcggtgggccgtctggacgcggtg | linearization of pEC-XK99E-*cas*3-F |
| pEC-XK99E-*cas*3-R | ctccttgtcgatcacaagacctccccggcg | linearization of pEC-XK99E-*cas*3-R |
| pXMJ19-vF | tcagcttggctgttttggcggat | pXMJ19 |
| pXMJ19-vR | ttgagatcctttttttctgcgcgtaatctg | pXMJ19 |
| pEC-XK99E-*cas*3-vF | acagaccatggaattcggtgggcc | *cas*3 |
| pEC-XK99E-*cas*3-vR | gccaagcttgctccttgtcgatca | *cas*3 |
| *Cg*-*ldh*-ot-F | cacgcgg+TTC+CATTCCA (xN+PAM+seed: 7+3+7=17 nt ) | number of potential off-target site: 18 |
| *Cg*-*ldh*-ot01-F/R | aggaaggcaagttcgctgac / cttcaaagcgcacaccgatg | off-target analysis |
| *Cg*-*ldh*-ot02-F/R | ctaagaagaatgcgaagccgag / catctagttgggcgagtctg | off-target analysis |
| *Cg*-*ldh*-ot03-F/R | gcacaaccacatcagacggatc / gaagaggttaacgtccagctg | off-target analysis |
| *Cg*-*ldh*-ot04-F/R | ttgtccagcattcggctaag /cttggtctgaccacattttc | off-target analysis |
| *ldh*-UF$_a$ | gcTCTAGAatgaaagaaaccgtcggtaacaagattg | UHA |
| *ldh*-U*egfp*-R$_a$ | ctcctcgcccttgctcaccattgggttggacgccacgaggaag | UHA-*egfp* |
| *ldh*-U*egfp*-F$_a$ | cttcctcgtggcgtccaacccaatggtgagcaagggcgaggag | UHA-*egfp* |
| *ld*h-D*egfp*-R$_a$ | gtgttccaggagggttaGAATTCcttgtacagctcgtcc | DHA-*egfp* |
| *ldh*-D*egfp*-F$_a$ | caaggaattctaaccctcctggaacacaactacg | DHA-*egfp* |
| *ldh*-*egfp*-vF$_a$ | cgatgcccttcagctcgatg | *egfp* |
| *ldh*-vR$_a$ | ggatccccggaactagctc | Δ*ldh*-*egfp* |
| pEC-XK99E-*cas*3-F$_a$ | cggcatgcatttacgttgcagcctactctttcgttggg | linearization of pEC-XK99E-*cas*3 |
| pEC-XK99E-*cas*3-R$_a$ | ccattcgatggtgtcatgaaagaaaccgtcggtaacaag | linearization of pEC-XK99E-*cas*3 |
| **g-DNA** | | |
| g-*sviipe* (182+52=234 bp)<br>pJ23119 promoter: 40 bp<br>*S. pyogenes* terminator: 42 bp | ggctgcaggtcgactctagaGGATCC gatccttgacagctagctcagtcctag<br>gtataatactagt**gtcctcatcgccccttcgaggggtcgcaac***gggaacctgcac*<br>*aaatgggggtacgccccctccggcagcg***gtcctcatcgccccttcgaggggtc**<br>**gcaac**ttatcaacttgaaaaagtggcaccgagtcggtgctttttttgGAATTC g<br>taatcatgtcatagctgtt | PAM: ttc<br>R: 30+30 bp<br>S: 40 bp |
| g-*svipam*1 (182+52=234 bp)<br>pJ23119 promoter: 40 bp<br>*S. pyogenes* terminator: 42 bp | ggctgcaggtcgactctagaGGATCC gatccttgacagctagctcagtcctag<br>gtataatactagt**gtcctcatcgccccttcgaggggtcgcaac***cgagctcggccc*<br>*ccgaccgcggcaccggctacgccgtcgc***gtcctcatcgccccttcgaggggtc**<br>**gcaac**ttatcaacttgaaaaagtggcaccgagtcggtgctttttttgGAATTC g<br>taatcatgtcatagctgtt | PAM: tcc<br>R: 30+30 bp<br>S: 40 bp |
| g-*svipam*2 (182+52=234 bp)<br>pJ23119 promoter: 40 bp<br>*S. pyogenes* terminator: 42 bp | ggctgcaggtcgactctagaGGATCC gatccttgacagctagctcagtcctag<br>gtataatactagt**gtcctcatcgccccttcgaggggtcgcaac***ggtgtacaggtc*<br>*ctggtagtcggccatggcgttggtgatg***gtcctcatcgccccttcgaggggtcgc**<br>**aac**ttatcaacttgaaaaagtggcaccgagtcggtgctttttttgGAATTC gta<br>atcatgtcatagctgtt | PAM: ttc<br>R: 30+30 bp<br>S: 40 bp |
| g-*svibla*1 (182+52=234 bp)<br>pJ23119 promoter: 40 bp<br>*S. pyogenes* terminator: 42 bp | ggctgcaggtcgactctagaGGATCC gatccttgacagctagctcagtcctag<br>gtataatactagt**gtcctcatcgccccttcgaggggtcgcaac***taccaggagtcc*<br>*acccgtgaggggcagttgaggctgctgt***gtcctcatcgccccttcgaggggtcg**<br>**caac**ttatcaacttgaaaaagtggcaccgagtcggtgctttttttgGAATTC gt<br>aatcatgtcatagctgtt | PAM: ttc<br>R: 30+30 bp<br>S: 40 bp |
| g-*svibla*2 (182+52=234 bp)<br>pJ23119 promoter: 40 bp<br>*S. pyogenes* terminator: 42 bp | ggctgcaggtcgactctagaGGATCC gatccttgacagctagctcagtcctag<br>gtataatactagt**gtcctcatcgccccttcgaggggtcgcaac***aagcccgccgg*<br>*gatgacgcacaccagcaccgagttctccg***gtcctcatcgccccttcgaggggtc**<br>**gcaac**ttatcaacttgaaaaagtggcaccgagtcggtgctttttttgGAATTC g<br>taatcatgtcatagctgtt | PAM: ttc<br>R: 30+30 bp<br>S: 40 bp |
| g-*svibla*3 (182+52=234 bp)<br>pJ23119 promoter: 40 bp<br>*S. pyogenes* terminator: 42 bp | ggctgcaggtcgactctagaGGATCC gatccttgacagctagctcagtcctag<br>gtataatactagt**gtcctcatcgccccttcgaggggtcgcaac***gactacatgagc*<br>*gagaaactgaccgccgccggagcccccg***gtcctcatcgccccttcgaggggt**<br>**cgcaac**ttatcaacttgaaaaagtggcaccgagtcggtgctttttttgGAATTC<br>gtaatcatgtcatagctgtt | PAM: ttc<br>R: 30+30 bp<br>S: 40 bp |
| g-*svibla*4 (182+52=234 bp)<br>pJ23119 promoter: 40 bp<br>*S. pyogenes* terminator: 42 bp | ggctgcaggtcgactctagaGGATCCgatccttgacagctagctcagtcctaggtataatactagt**gtcctcatcgccccttcgaggggtcgcaac***ctggcccgccgcgtccactacaccgcggccgaggtcgacg***gtcctcatcgccccttcgaggggtcgcaac**ttatcaacttgaaaaagtggcaccgagtcggtgctttttttgGAATTCgtaatcatgtcatagctgtt | PAM: ttc<br>R: 30+30 bp<br>S: 40 bp |
| g-*lacZ* (222 bp)<br>T7 promoter-19 bp<br>rrnB_T1 terminator-44 bp | ggctgcaggtcgactctagaGGATCCtaatacgactcactatagggaatatt**gtcctcatcgccccttcgaggggtcgcaac***ccgcccggtgcagtatgaaggcggcggagccgacaccacg***gtcctcatcgccccttcgaggggtcgcaac**ataaaacgaaaggctcagtcgaaagactgggcctttcgttttatGAATTCgtaatcatgtcatagctgtt | PAM: ttc<br>R: 30+30 bp<br>S: 40 bp |
| g-*ldh* (366 bp)<br>ptuf promoter-179 bp<br>rrnB T1 terminator-87 bp | ccacagggtagctggtagtttgaaaatcaacgccgttgcccttaggattcagtaactggcacattttgtaatgcgctagatctgtgtgctcagtcttccaggctgcttatcacagtgaaagcaaaaccaattcgtggctgcgaaagtcgtagccaccacgaagtccaggaggacataca**gtcctcatcgccccttcgaggggtcgcaac***cattccaagccggagaatttccacactgcgtaggtcagga***gtcctcatcgccccttcgaggggtcgcaac**caaataaaacgaaaggctcagtcgaaagactgggcctttcgttttatctgttgtttgtcggtgaacgctctcctgagtaggacaaat | PAM: ttc<br>R: 30+30 bp<br>S: 40 bp |
| **t-DNA** | | |
| t-Δ*sviipe*::*cat*-1399 bp<br>(*sviipe*=**1143** nt)<br>(*sviipe* / KY243071) | Gcacctcaaccacggctcctacggggcggtcccgctgcccgtccaggaagcccaggaggccctgcgcttccaggcccacgccgaccccgacgccttcttcaacggggtctgcgaacggctcaccgccgcccgggggccggatcgcggcgcgcctgggagcggacccggacggcctcgcgttcatctccaacgccaccgaggg<br>Agacgttgatcggcacgtaagaggttccaactttcaccataatgaaataagatcactaccgggcgtatttttttgagttgtcgagattttcaggagctaaggaagctaaaatggagaaaaaaatcactggatataccaccgttgatatatcccaatggcatcgtaaagaacattttgaggcatttcagtcagttgctcaatgtacctataaccagaccgttcagctggatattacggcctttttaaagaccgtaaagaaaaataagcacaagttttatccggcctttattcacattcttgcccgcctgatgaatgctcatccggaattccgtatggcaatgaaagacggtgagctggtgatatgggatagtgttcacccttgttacaccgttttccatgagcaaactgaaacgttttcatcgctctggagtgaataccacgacgatttccggcagtttctacacata | UHA: 201 bp<br>*cat*: 807 bp<br>DHA: 391 bp<br>Del (**1141**-692 nt): 449 bp |
| | tattcgcaagatgtggcgtgttacggtgaaaacctggcctatttccctaaagggtttat<br>tgagaatatgtttttcgtctcagccaatccctgggtgagtttcaccagttttgatttaaac<br>gtggccaatatggacaacttcttcgcccccgttttcaccatgggcaaatattatacgc<br>aaggcgacaaggtgctgatgccgctggcgattcaggttcatcatgccgtctgtgat<br>ggcttccatgtcggcagaatgcttaatgaattacaacagtactgcgatgagtggcag<br>ggcggggcgtaa<u>ggcgcgccatttaaatgaagttcctattccgaagttcc</u><br>Catgggaggaccaccagggcttcccgctctccgtcgagtaccgcgccaccttcga<br>ctacaccggctggctggccgcccccgagggactcgacctcctggagcggctcgg<br>cgccgaccgggtccgggagcacaacagcgcgctggccggctacggcgccggg<br>ctgctcgcccggatccccggcctcaccccgctgccgcacaccccgggcctggcc<br>ctgcgctccctgttgctgccgcccggcgtcgccgagacccgcgaggcggccacc<br>gccctgcgcgaggaactcgccgcgaagctccgcatccgggtcctggtctggccc<br>cgcgagggcggcggcgggctgcgggtctgcgggcaggtctacaaccggcccg<br>aggagtacgaacgcct | |
| t-Δ*svipam*1-2047 bp<br>(*svipam*1=**<u>2901</u>** nt)<br>(*svipam*1 / KY243072) | Cgagaccgacgccttcctgcgcaccctcggctggcgtcaggtcgcgcaggagg<br>agtacgacaagaagctcacgcccgagacgaagaagtacctccaggcctacgccg<br>acggggtcaacgcgtggctgaaggggaagtccggcaaggacctctccgtcgaac<br>acgccgccctgaagatcaccgacggctacaagcccgagaagtggacgccggtc<br>gactcggtggcctggctcaaggcgatggcctgggacctgcgcggcaacatgcag<br>gacgagatcgaccgctcgctgatgtcgaacaagctcacgcaggcgcagatcgac<br>gagctctacccggcgtacccccttcgaccgcaacaagccgatcgtcgagggcggc<br>accgtcgcgggcggcaagtacgccccccagggcaccgcgggcgccggcaccg<br>gctcctccccgacgggcgccaccggcacgggcacgggcaccggctccaccacg<br>ggcgccgccgccggtacgggtacgggcgccggtacgggcaccggctcgcagg<br>ccacccccaacggcggcaccaaccccgccaccggcctcgccgacaacgcgag<br>cgcccagggcgcgaccgtcggcctgcgcacccagctgaccgccctgtccaaga<br>ccctggacgacatcccggccatcctcggccccaacggcagcggcatcggctcga<br>actcctgggtcatctcgggcaagtacacgaccaccggcaagccgctgctcgcgaa<br>cgacccgcacctctccccgcagctgccctcggtctggtaccagatgggcctgcact<br>gccgcgccgtctcggcccggtgccagtacgacgtcgccggcttcaccttctccgg | UHA:945 bp<br>DHA:1102 bp<br>Del (**<u>2901</u>**-2574): 327 bp |
| | catgcccggcgtggtcatcggccacaacgccgacatcgcctggggcatgaccaa<br>cctcggcgccgacgtgaccgacctctacctgga<br>Ggtcccgtcgcagaacctcatctacgccgacaacaagggccccaacggcaacat<br>cggctaccaggccccgggccgcatcccggtccgcggccagggcgacggccgc<br>atgccggcccccggctgggacccgaagtacgcctggaagggcggccgcgacg<br>gcaacgccggctacatcccgcagaacgagctgccctgggagctcaacccgcagc<br>gcggctacatcgtcaccgccaaccaggccgtcaccgagtccggcaccggcgcg<br>ggcaagtacccgtacgtgctgaccaccgactggggctacggcgcccgcagccag<br>cggatcaacgacctcatcgaggcgaagatcaaggacagcggcaagatctccacc<br>gacgacatgcgcaccatgcagatggacaacagcagcgagatcgccgcgctgctg<br>accccgatgctggccaagatcccggtctccgacccggacgtgcgctcggcgcag<br>aagctgctggagggctggaactacacccaggagaacgactcggcggcggccgc<br>gtacttcaacgcggtctggcgcaacatcctcaagctggcgttcggcgacaagatgc<br>cgaaggagctgcgcgtcgagggcagctgcatgaacgtcgtcgaggagagcacc<br>ggccccaacgacgacctggccaagaccgtccgcgagtgcggcacgcgtggtcc<br>cgactccgcgcagccggacggcggcgaccgctggttcgaggtggtccgccgcct<br>ggtcaaggacgagaagtcgccgtggtggaccgccaccccgaagaccgtccacg<br>acaagcccatcaagacccgcgacgagctcttcgcccgggcgatggaggacgcg<br>cgctgggagctgaccgccaagctcggcaaggaccagtcgacctggagctgggg<br>ccggctgcaccagctgacgctgaagaaccagaccatcggcaaggagggcccgg<br>gcttcatgcagtggctcctcaaccgcggcccgtggaacctgggcggcggcgagg<br>ccaccgtcaacgccaccggctggaacgcctccagcgggtacgacgtcacctggg<br>tgccgtcgatgcggatggtcgtcaacctcaac | |
| t-Δ*svipam*2-1030 bp<br>(*svipam*2=**<u>2049</u>** nt)<br>(*svipam*2 / KY243073) | Accaccgaggcctacgaggtctaccgcgacagctggggcatcccgcacctgcg<br>cgcgtccgacccgctccgcctctcctacgcccagggccggaccaccgcccggga<br>ccgggcctggcagctggaggtggagcggcaccgcgcccagggctccagcgcct<br>ccttcctcggggccgactgcgtgccctgggaccgcttcgcccgccaggcccgcct<br>cgacgacaccgcccgccggtgcttcgaggccctggacgcggacaccgccgcct<br>gggtcaccgcctacgtcgacggggtcaacgcgggcctggccgagggcgcggcc<br>cgcgacgaccgcttcgcggccgcgggccacacccccgcccccctggcagccgtg<br>ggtcccgctctccatctggatcggcacccacatcctcttc | UHA: 414 bp<br>DHA: 616 bp<br>Del (**<u>2049</u>**-1602 nt):447 bp |
| | Gtcgaggtcgagatcctggagaccgcccggggcccggtgatcgtcaacgacgg<br>gtccgccaccctctccctgcgctacccgccccgcgtccgctccgacctcggcttcg<br>ccgccctgcccgccctgctgcgcgcccgcaccgtcgccgacgtcgaccgcgccc<br>tcgacggctgggccgagcccgtcaacgtcgtgcacgccgccgactccgagggc<br>ggcctgctgcaccgtgtcgcgggcgccgtcccgctgcgcgaggccgcgaaccg<br>gctgcgtcccgtgcccgcctgggacgcccgccacgcctggcggggctgggccg<br>agacccccgccgagcccgtgcggggcgtcgccgtgatggcgaacgcgcgcgg<br>gatcgcgtccccccctcggcgtggagttcgcgccgccacaccgcgccgaccggat<br>ccgtgccctgctcggcggttcggccgactggtccccgaaggccatggccgaggt<br>ccaccgcgacacccacctggcctctgccgcgccgctgctcgccctgctgcccgg<br>gctcgacgccctctcccccgcggccgccgcgctccgcgaccggctgctggcctg<br>ggaccggcacatggaggccgacagca | |
| t-Δ*svibla*1-541 bp<br>(*svibla*1=**<u>2082</u>** nt)<br>(*svibla*1 / KY243074) | Ctcccccgaacacccctactacgccggggccgtggtcctcgccgggcgcggcc<br>gcacggtcgccctgcaccgggcgatgggggacgccgtccggtacgcggactac<br>gacgggcgcaccgaccggccccgggagttcccggccgccgaccggatcgccat<br>ggccgaggacaccgtcttcgacctggcctccctcagcaagctgttcacctccatcat<br>cgc<br>Gcctccgtcgaactgctcttcaccgactacaacaccgccttcccgggcgacgacc<br>acggcctcggcttcgagctctaccagcactggtacatgggcgccctggccacccc<br>ccactcggccggccacaccggcttcaccgggacctccctggtcctcgaccccctcc<br>accgactccttcctcgtcctcctcggcaactccgtgcaccccgtccgcacctggcg<br>cgccggcagcgcaccccgtgtggccaccggcaaccgcttcgcccgggccgtac<br>ccgtccgtacgaagcacggcggcccggcctggttctccgggatggaca | UHA: 219 bp<br>DHA: 322 bp<br>Del (**<u>2082</u>**-1505 nt): 577 bp |
| t-Δ*svibla*2-1066 bp<br>(*svibla*2=**<u>1542</u>** nt)<br>(*svibla*2 /KY243075) | Atcctgctcacagccgggacggccgcggcggagccgaccccgccccgcccc<br>ccagaccccgcatcagcgacgccgccgtcgccacggcggtgagccgcctcgac<br>gcggacgtcgcggacgtcatgcgccgcaccggcatccccggtgtggcggtggc<br>cgtggtccaccgcgacaaggtgctctacctcaagggcttcggcctccgccgcacg<br>ggcgagagcgcgaaggtggacccggacaccgtcttccagctggcctccgtctcg<br>aagccgctgtcctccaccgtcgtcgccggcgccctgaccgaccccgccgcctgg<br>gacgagcccctcggctcctccctgcccggcttcgcgctgaaggacccctgggtga<br>cctcccacgtcacccccgccgacctgttctcccaccgcagcg | UHA: 418 bp<br>DHA: 648 bp<br>Del (**<u>1542</u>**-1140 nt): 402 bp |
| | Gggtcctcgacgggaaacggatcatccccgccgaccacctgaaccgcacccgc<br>ctccccgagatcgtctcgcaggcccagaccttccgcggcgtcccgcagttctacg<br>ggctcggctggaacgtcagctacgacgacgacggccggctgcgcctgggccact<br>cggggggcttcgaactcggcgccaacaccaacgtcaccatgctcccgctcgaac<br>agctcggcatcgtcgtcctgaccaacgcggccccggtcggccaggccgacgcga<br>tcgccctggacttcttcgacaccgccgaacacggcaagcggaccgccgactggct<br>ggccctgaccggcgcgctctacgcacaggagctcaacgagggccaccgcccgg<br>ccaccgactacgcccacccgccggccggcgccaagcccgcccgggggggccgg<br>cacctacaccggtacctacgacaaccccttctacggccccctcaccgtcaccgccg<br>acgccaagggcgccctcaccctctccctcggccccaagccccagcgcttcccccct<br>cacccactacgacggggacacgttcagcttcgtgaccgccggcgagaacgccgt<br>cggccgcaccggggtgaccttctccgacggcaaggtccgcatcgagtacctgg | |
| t-Δ*svibla*3-838 bp<br>(*svibla*3=**1245** nt)<br>(*svibla*3 /KY243076) | Ctctcaccgatgtcgcagaactgataccagttctagttcgaggagtccacggtccg<br>gacgagggtgccggaggatccgatggacggtcaggactgttgggggacggtcg<br>acttgggctactactggaacgcgttctcgtcagagcggatcgagtcggccaggga<br>gcccccatgtcagaggtcatgtcacaggtcaccgaccacggcgggggtgtgtgg<br>ggcatcaaggtccccatccccgacaacccccctgggacacaccctcgtccacgtcc<br>tcgacacggacgcaggccccgtcctcgtcgacaccggctgggacgaccccgcgt<br>cctgggacgccctcaccggcggtctcggcgcactcggcatcgccgtggacgacg<br>tgcacggcgtggtcatcacccaccaccaccccg<br>Gccacgtctgcctgcacctggaggaggcccaccccgcgaacctgcccggcaac<br>ggccggctcttctccggcgaccacctgctgcccgggatctccccgcacatcggcc<br>tgtacgaggccccggacgacacccgggtcaccgaccccctgggcgactacctcg<br>actccctcgaacgcgtcggccgcctggagcccgccgaggtgctcccggcccacc<br>agcacgccttcaccgacgcccccgcccgcgtacgggagctcctcgcgcaccacg<br>aggagcggctggccgggctgtggcggctgctgctggccgagccgctgaccccgt<br>gggggctcgcggagcggatggagtggaaccggccctgggagcagatcccgtac<br>ggctcccggaacatcgccgtctcggaagcggaagcccatctgcggcg | UHA: 414 bp<br>DHA: 424 bp<br>Del (**1245**-927 nt): 318 bp |
| t-Δ*svibla*4-1559 bp<br>(*svibla*4=**921** nt)<br>(*svibla*4 / KY243077) | Gctggtgttcctgggctgcgtcctccacgacctggggttgtcggaacagggcaac<br>ggccaccagcggttcgaggtggacggtgcggacctggcggcccggttcctgcgc<br>gaacagggcgccggcgaggaggccgtgacggtggtgtgggacgccgtcgccct | UHA: 727 bp<br>DHA: 832 bp<br>Del (**921**-408 nt): 513 bp |
| | gcacacgtcggagggcatcgcctcccgcaagggcccggagatcgcactggccc<br>acgccggtatctcggtggacgtcctggggcgcggcaaggagctcctgcccgagg<br>ggttcgccgaccgggtgcacgccgcgttcccccgcgccgacctcggctacgcca<br>tcacggacgtgatcgtccggcagatgctggacaaccccgccaaggccggcccgc<br>tcagcttccccgggcagctgctgcgcgcccacgtgccggccggcacgctgcccg<br>actggtccgacctgatcgacgcggccggctggggagaccggccggcgggcgg<br>gccggtgcaaccgtcctgagacggttcgcgtgtttcctcccgtagtgatgatcaacg<br>atgaggagacgatgtgaagcgcgtggcaagccgtccgtcccggcgtacggtcct<br>gaccctcgcgaccgggggcctggccgccctcggcctcggcgccgccgctccgg<br>cgtccgcgtccgcggacacgcggccgggcccgggccggggcgccccgcgcc<br>ggctgcgggacatggaacgggagcacggc<br>Ctcgccaacaccaccaacacccatcgcttcaaggagggcctgcccggctggacc<br>ctggccgacaagaccggcggaggcgacgactacggcgtcgccaacgacgtgg<br>gcgtggtgtggagcccggccggtgtgccggtggtgatggcggtcctgacgacca<br>ggccgcaggccccgcacggcccgaacgacagcgagctcgtcgccgagacggc<br>ccgcctgctgggcgaaaccgtcctggactgacccgcaccggcccgcaccggcc<br>ccatcccgctcgcaccgacccgcacgggccggctcctcatggcgctcgccgcgc<br>gggcgtacggcggtccgtgggcgggaggggcgcgatgccggtgaaggggcac<br>cgggcgctcccgggtgcgggggtaccggccgaagcggccggtgctgcgacgg<br>gacgggggcggaggtcgtgcggatggtcggtgatgtcgagggccagcttgtggg<br>gttggaggatggtgcgggtgctgaagacgggcgtggtgcccggggtgcggttga<br>tctggtagctgctggtgagggtggcggtggtcacggtcatcgcgacggtggcgaa<br>gtgggttcccaggcaggtgcgcggtccgccgccgaagggcatgaacgcgtactt<br>gggcagcggatcggcgggggagcggccggtccagcggtcggggaggaaggc<br>gtccggctggtggtagtggcgggggtcgcgctggaggacgaaggggctgacgg<br>cgacgcgctgcccggggcgcagggggaagccgccgagttcggtgggttcttcga<br>cggtcctttcgatcagccacaggggcgggtacaa | |
| t-Δ*lacZ*-1144 bp<br>(*lacZ*=**<u>3075</u>** nt)<br>(*Escherichia coli* str. K-12 substr. MG1655, ~4.6 Mbp) | Atgagcgtggtggttatgccgatcgcgtcacactacgtctgaacgtcgaaaaccc<br>gaaactgtggagcgccgaaatcccgaatctctatcgtgcggtggttgaactgcaca<br>ccgccgacggcacgctgattgaagcagaagcctgcgatgtcggtttccgcgaggt<br>gcggattgaaaatggtctgctgctgctgaacggcaagccgttgctgattcgaggcg | UHA: 423 bp<br>DHA: 721 bp<br>Del (**<u>3075</u>**-2655 nt): 420 bp |
| | ttaaccgtcacgagcatcatcctctgcatggtcaggtcatggatgagcagacgatg<br>gtgcaggatatcctgctgatgaagcagaacaactttaacgccgtgcgctgttcgcat<br>tatccgaaccatccgctgtggtacacgctgtgcgaccgctacggcctgtatgtggtg<br>gatgaagccaatattgaaacccacggcatgg<br>Gtttacagggcggcttcgtctgggactgggtggatcagtcgctgattaaatatgatg<br>aaaacggcaacccgtggtcggcttacggcggtgattttggcgatacgccgaacga<br>tcgccagttctgtatgaacggtctggtctttgccgaccgcacgccgcatccagcgct<br>gacggaagcaaaacaccagcagcagtttttccagttccgtttatccgggcaaaccat<br>cgaagtgaccagcgaatacctgttccgtcatagcgataacgagctcctgcactgga<br>tggtggcgctggatggtaagccgctggcaagcggtgaagtgcctctggatgtcgct<br>ccacaaggtaaacagttgattgaactgcctgaactaccgcagccggagagcgccg<br>ggcaactctggctcacagtacgcgtagtgcaaccgaacgcgaccgcatggtcaga<br>agccgggcacatcagcgcctggcagcagtggcgtctggcggaaaacctcagtgt<br>gacgctccccgccgcgtcccacgccatcccgcatctgaccaccagcgaaatggat<br>ttttgcatcgagctgggtaataagcgttggcaatttaaccgccagtcaggctttctttc<br>acagatgtggattggcgataaaaaacaactgctgacgccgctgcgcgatcagttca<br>cccgtgcaccgctggataacgacattggcgtaagtgaagcgacccgcat | |
| t-Δ*lacZ*::*cat*-1400 bp<br>(*lacZ*=**<u>3075</u>** nt)<br>(*Escherichia coli* str. K-12 substr. MG1655, ~4.6 Mbp) | Cgatgtcggtttccgcgaggtgcggattgaaaatggtctgctgctgctgaacggca<br>agccgttgctgattcgaggcgttaaccgtcacgagcatcatcctctgcatggtcagg<br>tcatggatgagcagacgatggtgcaggatatcctgctgatgaagcagaacaacttt<br>aacgccgtgcgctgttcgcattatccgaaccatccgctgtggtacacgctgtgcgac<br>cgctacggcctgtatgtggtggatgaagccaatattgaaacccacggcatgg<br>Agacgttgatcggcacgtaagaggttccaactttcaccataatgaaataagatcact<br>accgggcgtatttttttgagttgtcgagattttcaggagctaaggaagctaaaatggag<br>aaaaaaatcactggatataccaccgttgatatatcccaatggcatcgtaaagaacatt<br>ttgaggcatttcagtcagttgctcaatgtacctataaccagaccgttcagctggatatt<br>acggcctttttaaagaccgtaaagaaaaataagcacaagttttatccggcctttattca<br>cattcttgcccgcctgatgaatgctcatccggaattccgtatggcaatgaaagacgg<br>tgagctggtgatatgggatagtgttcacccttgttacaccgttttccatgagcaaactg<br>aaacgttttcatcgctctggagtgaataccacgacgatttccggcagtttctacacata<br>tattcgcaagatgtggcgtgttacggtgaaaacctggcctatttccctaaagggtttat | UHA: 278 bp<br>*cat*: 807 bp<br>DHA: 315 bp<br>Del (**<u>3075</u>**-2797 nt): 441 bp |
| | tgagaatatgtttttcgtctcagccaatccctgggtgagtttcaccagttttgatttaaac<br>gtggccaatatggacaacttcttcgcccccgttttcaccatgggcaaatattatacgc<br>aaggcgacaaggtgctgatgccgctggcgattcaggttcatcatgccgtctgtgat<br>ggcttccatgtcggcagaatgcttaatgaattacaacagtactgcgatgagtggcag<br>ggcggggcgtaaggcgcgccatttaaatgaagttcctattccgaagttcc<br>Gggactgggtggatcagtcgctgattaaatatgatgaaaacggcaacccgtggtc<br>ggcttacggcggtgattttggcgatacgccgaacgatcgccagttctgtatgaacg<br>gtctggtctttgccgaccgcacgccgcatccagcgctgacggaagcaaaacacca<br>gcagcagtttttccagttccgtttatccgggcaaaccatcgaagtgaccagcgaata<br>cctgttccgtcatagcgataacgagctcctgcactggatggtggcgctggatggtaa<br>gccgctggcaagcggtgaagtgcctctggatgtcg | |
| t-Δ*ldh*::*egfp*-2402 bp (*ldh*=**<u>945</u>** nt) (*Corynebacterium glutamicum* ATCC 13032, ~3.3 Mbp) | Ttaatactgtgctgggttaattcgccggtgatcagcagcgcgccgtaccccaaggt<br>gccgacactaatgcccgcgatcgtctccttcggtccaaaattcttctgcccaatcagc<br>cggatttgggtgcgatgcctgatcaatcccacaaccgtggtggtcaacgtgatggc<br>accagttgcgatgtgggtggcgttgtaaattttcctggatacccgccggttggttctg<br>gggaggatcgagtggattcccgtcgctgccgcatgccccaccgcttgtaaaacag<br>ccaggttagcagccgtaacccaccacggtttcggcaacaatgacggcgagagag<br>cccaccacattgcgatttccgctccgataaagccagcgcccatatttgcagggagg<br>attcgcctgcggtttggcgacattcggatccccggaactagctctgcaatgacctgc<br>gcgccgagggaggcgaggtgggtggcaggtttagtgcgggtttaagcgttgcca<br>ggcgagtggtgagcagagacgctagtctggggagcgaaaccatattgagtcatctt<br>ggcagagcatgcacaattctgcagggcataggttggttttgctcgatttacaatgtga<br>ttttttcaacaaaaataacacttggtctgaccacattttcggacataatcgggcataatt<br>aaaggtgtaacaaaggaatccgggcacaagctcttgctgattttctgagctgctttgt<br>gggttgtccggttagggaaatcaggaagtgggatcgaaa<br>Atggtgagcaagggcgaggagctgttcaccggggtggtgcccatcctggtcgag<br>ctggacggcgacgtaaacggccacaagttcagcgtgtccggcgagggcgaggg<br>cgatgccacctacggcaagctgaccctgaagttcatctgcaccaccggcaagctg<br>cccgtgccctggcccaccctcgtgaccaccctgacctacggcgtgcagtgcttcag<br>ccgctaccccgaccacatgaagcagcacgacttcttcaagtccgccatgcccgaa<br>ggctacgtccaggagcgcaccatcttcttcaaggacgacggcaactacaagaccc | UHA: 776 bp<br>*egfp*: 717+9 bp<br>DHA: 900 bp<br>Del (**<u>945</u>**+1053 nt): 1998 bp |
| | gcgccgaggtgaagttcgagggcgacaccctggtgaaccgcatcgagctgaagg<br>gcatcgacttcaaggaggacggcaacatcctggggcacaagctggagtacaacta<br>caacagccacaacgtctatatcatggccgacaagcagaagaacggcatcaaggtg<br>aacttcaagatccgccacaacatcgaggacggcagcgtgcagctcgccgaccact<br>accagcagaacacccccatcggcgacggccccgtgctgctgcccgacaaccact<br>acctgagcacccagtccgccctgagcaaagaccccaacgagaagcgcgatcaca<br>tggtcctgctggagttcgtgaccgccgccgggatcactctcggcatggacgagctg<br>tacaaggaattctaa<br>Ccctcctggaacacaactacgaccgctcccgggtctacggcatccccgccgtagt<br>tcagcgcatcaacctcaaagtcggcgaccgcctcatccttaccgacgaagaactca<br>cctacgatccatccctcggatccggccgcacaccacgcatcagctgcacccttcca<br>caagcagtcgatgcaattaaagtcgggcaccgcgtgcttttcgacgacggagccat<br>cgccgcagtctgcatcgacaagacctccactgccgacggccacaacgacgtaga<br>attggaagtcacccacgcccgcccacaaggcgtaaacctggccgcatacaaggg<br>aatcaacctcccagactccgaacttccactcccaagcctcactgaagaagacctcc<br>aacacctgcgctttgtcgtcaaatacgccgacatcgcagccatctccttcatccgaa<br>acgtcgccgacgtggaatacctcctccaagcactcgccgacatcggagatccagt<br>agccgtcgaacgccttggcctcgtccttaaaatcgagaccatcccaggctacgaag<br>gcctcgcccaaatcctcctgaccggcatgcgccacgaaaacttcggcatcatgatc<br>gcccgcggagacctcgccgtcgaactcggcttcgaccgcatggcagaagtcccc<br>caactgatcatggcccttgccgaagccgcccacgtcccaaccatcttggccaccca<br>agtcctggaaaacatggccaaaaacggactcccatctcgcgcagaaatcaccgac<br>gcagcaatggcacttcgcgctgaatgcgtcatgctgaacaagggaccacacatca<br>acgacgccatcaaggtcctcaccgaaatgagccgcaaacttggtgcatcccaacg<br>aaagagtaggctgc | |
| t-Δ*ldh*::*egfp*-2001 bp<br>(*ldh*=**<u>945</u>** nt)<br>(*Corynebacterium glutamicum* ATCC 13032) | Atgaaagaaaccgtcggtaacaagattgtcctcattggcgcaggagatgttggagt<br>tgcatacgcatacgcactgatcaaccagggcatggcagatcaccttgcgatcatcg<br>acatcgatgaaaagaaactcgaaggcaacgtcatggacttaaaccatggtgttgtgt<br>gggccgattcccgcacccgcgtcaccaagggcacctacgctgactgcgaagacg<br>cagccatggttgtcatttgtgccggcgcagcccaaaagccaggcgagacccgcct | UHA: 375 bp<br>*egfp*: 717+9 bp<br>DHA: 900 bp<br>Del (**<u>945</u>**-375+1053 nt): 1623 bp |
| | ccagctggtggacaaaaacgtcaagattatgaaatccatcgtcggcgatgtcatgg<br>acagcggattcgacggcatcttcctcgtggcgtccaaccca<br>Atggtgagcaagggcgaggagctgttcaccggggtggtgcccatcctggtcgag<br>ctggacggcgacgtaaacggccacaagttcagcgtgtccggcgagggcgaggg<br>cgatgccacctacggcaagctgaccctgaagttcatctgcaccaccggcaagctg<br>cccgtgccctggcccaccctcgtgaccaccctgacctacggcgtgcagtgcttcag<br>ccgctaccccgaccacatgaagcagcacgacttcttcaagtccgccatgcccgaa<br>ggctacgtccaggagcgcaccatcttcttcaaggacgacggcaactacaagaccc<br>gcgccgaggtgaagttcgagggcgacaccctggtgaaccgcatcgagctgaagg<br>gcatcgacttcaaggaggacggcaacatcctggggcacaagctggagtacaacta<br>caacagccacaacgtctatatcatggccgacaagcagaagaacggcatcaaggtg<br>aacttcaagatccgccacaacatcgaggacggcagcgtgcagctcgccgaccact<br>accagcagaacacccccatcggcgacggccccgtgctgctgcccgacaaccact<br>acctgagcacccagtccgccctgagcaaagaccccaacgagaagcgcgatcaca<br>tggtcctgctggagttcgtgaccgccgccgggatcactctcggcatggacgagctg<br>tacaaggaattctaa<br>Ccctcctggaacacaactacgaccgctcccgggtctacggcatccccgccgtagt<br>tcagcgcatcaacctcaaagtcggcgaccgcctcatccttaccgacgaagaactca<br>cctacgatccatccctcggatccggccgcacaccacgcatcagctgcacccttcca<br>caagcagtcgatgcaattaaagtcgggcaccgcgtgcttttcgacgacggagccat<br>cgccgcagtctgcatcgacaagacctccactgccgacggccacaacgacgtaga<br>attggaagtcacccacgcccgcccacaaggcgtaaacctggccgcatacaaggg<br>aatcaacctcccagactccgaacttccactcccaagcctcactgaagaagacctcc<br>aacacctgcgctttgtcgtcaaatacgccgacatcgcagccatctccttcatccgaa<br>acgtcgccgacgtggaatacctcctccaagcactcgccgacatcggagatccagt<br>agccgtcgaacgccttggcctcgtccttaaaatcgagaccatcccaggctacgaag<br>gcctcgcccaaatcctcctgaccggcatgcgccacgaaaacttcggcatcatgatc<br>gcccgcggagacctcgccgtcgaactcggcttcgaccgcatggcagaagtcccc<br>caactgatcatggcccttgccgaagccgcccacgtcccaaccatcttggccaccca<br>agtcctggaaaacatggccaaaaacggactcccatctcgcgcagaaatcaccgac<br>gcagcaatggcacttcgcgctgaatgcgtcatgctgaacaagggaccacacatca | |
| | acgacgccatcaaggtcctcaccgaaatgagccgcaaacttggtgcatcccaacg aaagagtaggctgc | |

*Capital letters represent the site for restriction enzyme digestion; letters with dash-line represent complementary region with plasmid; letters with single underline represent promoter; letters with double underline represent terminator; boldface letters represent direct repeat; italic letters represent spacer; UHA: upstream homologous arm; DHA: downstream homologous arm; *cat*: a gene acronym for chloramphenicol acetyl transferase; UF / UR: primers used for amplification of UHA by PCR; DF / DR: primers used for amplification of DHA by PCR; vF / vR: primers used for verification of edited sequences (two fragment sequences in a genome normally locate at both sides of an edited sequence but not in t-DNA); NCBI database: KY243071- KY243077.

**Table S3.** Main biochemical properties of *Svi*Cas proteins in the subtype I-B-*Svi* Cas system* (http://web.expasy.org/protparam) [11]

| Name | ID in NCBI | Size (AA) | MW(Da) | C/M | pI | I-ind | AI | GRAVY | Functions |
|---|---|---|---|---|---|---|---|---|---|
| *Svi*Cas1 | KY176008 | 326 | 36722.99 | 2/6 | 9.59 | 48.45 | 99.69 | -0.196 | A metal-dependent DNAase |
| *Svi*Cas2 | KY176009 | 87 | 9976.31 | 1/0 | 6.19 | 46.59 | 99.66 | -0.164 | A small endo-RNAase |
| *Svi*Cas3 | KY176010 | 771 | 84352.40 | 2/4 | 5.78 | 35.49 | 95.91 | -0.153 | Both Hel-D (cas3') and HD (cas3'') |
| *Svi*Cas4 | KY176011 | 165 | 18758.08 | 5/0 | 8.65 | 58.89 | 77.58 | -0.557 | A RecB-like nuclease |
| *Svi*Cas5 | KY176012 | 220 | 23514.74 | 1/2 | 9.72 | 56.70 | 86.00 | -0.165 | A RNAase generating crRNAs |
| *Svi*Cas7 | KY176013 | 332 | 35946.55 | 0/2 | 6.18 | 40.91 | 90.78 | -0.315 | A subunit of Cascade positioning the RNA guide for DNA binding |
| GVGL007770 / *Svi*Cas6 | KY243078 | 469 | 51276.86 | 6/4 | 6.18 | 43.09 | 88.83 | -0.272 | Potential Cas6 generating crRNAs and forming a hairpin structure on the 3'flank from pre-crRNA |

*:C/M: cysteine/methionine; pI: theoretical isoelectric point; I-ind: instability index (smaller than 40 is predicted as stable ); AI: aliphatic index (regarded as a positive factor for the increase of thermostability of globular proteins); GRAVY: grand average of hydropathicity (+: hydrophobic, -: hydrophilic); red letters with dark blue background represent the highest value; red letters with green background represent the lowest value.

**Table S4.** Genome editing efficiency of microbial and mammalian cells based on the *Svi*Cas system*

| Conditions / Cells | HCN | MCN | NCGEC | TE (%) | HRE (%) | GEE (%) |
|---|---|---|---|---|---|---|
| *S. virginiae* IBL14 | | | | 1.1 | 92.0 | 1.0 |
| *Svi*$^{p}$ (×10$^{5}$/ml) | 4.13±0.24 | NT | NT | | | |
| *Svi*-M (×10$^{3}$/ml) | NT | 4.43±0.89 | NT | | | |
| *S. virginiae* IBL14-Δ*gene* | NT | NT | 46 | | | |
| *E. coli* JM109 (DE3) | | | | 3.28×10$^{-4}$ | 27.3 | 9.0×10$^{-5}$ |
| *Ec*$^{c}$ (×10$^{6}$/ml) | 7.5±1.12 | NT | NT | | | |
| *Ec*-BW (×10$^{3}$/ml) | NT | 2.46±1.60 | NT | | | |
| *Ec*-W (×10$^{3}$/ml) | NT | 1.29±0.95 | NT | | | |
| *E. coli* JM109 (DE3)-Δ*lacZ* | NT | NT | 26 | | | |
| *C. glutamicum* ATCC 13032 | | | | 1.8×10$^{-6}$ | 48.0 | 8.6×10$^{-7}$ |
| *Cg*-pEK$^{c}$ (×10$^{7}$/ml) | 2.59±0.25 | NT | NT | | | |
| *Cg*$^{m}$ (×10/ml) | NT | 4.7±0.19 | NT | | | |
| *C. glutamicum* ATCC 13032-Δ*ldh::egfp* | NT | NT | 24 | | | |

*Concentration and number of cells are expressed as mean ± SEM of n ≥ 3; NT: not tested.

*Svi*$^{p}$: *S. virginiae* IBL14 protoplasts growing on MR2YE plates without antibiotics; *Svi*-M: mutants of *S. virginiae* IBL14 transformed by pKC1139-t/g-*gene abbreviation* growing on MR2YE plates with apramycin; EC$^{c}$: *E. coli* JM109 (DE3) competent cells growing on LBPET plates without antibiotics; *Ec*-BW: blue, white and blue-white colonies of *E. coli* JM109 (DE3) transformed by pCas-*cas*3 and pKC1139-t/g-Δ*lacZ* growing on LBPET plates with kanamycin and apramycin; *Ec*-W: white and blue-white colonies of *E. coli* JM109 (DE3) transformed by pCas-*cas*3 and pKC1139-t/g-Δ*lacZ* growing on LBPET plates with kanamycin and apramycin (about one half of the tested *Ec*-W colonies are correctly gene-edited mutants); *Cg*-pEK$^{c}$: competent cells of *C. glutamicum* ATCC 13032-pEC-XK99E-*cas*3 growing on LBHIS plates with kanamycin; *Cg*$^{m}$: mutants of *C. glutamicum* ATCC 13032 transformed by pXMJ19-t/g-Δ*ldh::egfp* and pEC-XK99E-*cas*3 growing on LBHIS plates with chloramphenicol and kanamycin. HCN: host cell number; MCN: mutant cell number; NCGEC: number of correctly gene-edited cells identified from 50 potential gene-edited cells by PCR and/or DNA sequencing; TE: transformation efficiency; HRE: homologous recombination efficiency; GEE: genome editing efficiency.

**Table S5.** Main biochemical characteristics of the typical signature proteins in various type CRISPR-Cas systems * (http://web.expasy.org/protparam; http://www.ncbi.nlm.nih.gov; http://www.addgene.org/) [23-25]

| AA \ Protein | *Svi*Cas3 type I | *Sp*Cas9 type II | *Cj*Cas9 type II | *Ll*Cas10 type III | *As*Cas12a type V | *Ls*Cas13a type VI |
|---|---|---|---|---|---|---|
| Ala (A) | 14.5% | 5.3% | 6.5% | 7.5% | 6.0% | 2.0% |
| Arg (R) | 9.2% | 5.6% | 4.3% | 5.3% | 4.4% | 3.9% |
| Asn (N) | 1.2% | 5.1% | 6.8% | 6.7% | 5.7% | 10.7% |
| Asp (D) | 6.9% | 7.2% | 6.0% | 7.5% | 5.5% | 7.3% |
| Cys (C) | 0.3%(2) | 0.1%(2) | 0.7%(7) | 0.8%(6) | 0.6%(8) | 0.5%(7) |
| Gln (Q) | 3.2% | 3.8% | 3.0% | 3.4% | 4.5% | 1.7% |
| Glu (E) | 6.4% | 7.9% | 8.7% | 6.7% | 7.6% | 10.4% |
| Gly (G) | 7.7% | 5.0% | 3.7% | 4.6% | 4.4% | 2.7% |
| His (H) | 3.8% | 2.3% | 1.7% | 1.6% | 3.0% | 0.8% |
| Ile (I) | 3.2% | 6.8% | 6.1% | 7.4% | 7.0% | 12.7% |
| Leu (L) | 12.3% | 10.8% | 10.5% | 10.0% | 10.6% | 8.4% |
| Lys (K) | 1.3% | 11.0% | 14.7% | 7.5% | 9.0% | 15.3% |
| Met (M) | 0.5% | 1.6% | 1.0% | 2.9% | 1.4% | 1.2% |
| Phe (F) | 3.4% | 4.6% | 6.2% | 4.8% | 5.8% | 5.2% |
| Pro (P) | 5.7% | 2.6% | 1.8% | 1.2% | 3.7% | 0.7% |
| Ser (S) | 4.0% | 5.6% | 5.9% | 6.7% | 5.7% | 4.3% |
| Thr (T) | 5.4% | 4.8% | 3.3% | 5.4% | 6.0% | 3.4% |
| Trp (W) | 1.7% | 0.5% | 0.4% | 0.8% | 0.8% | 0.4% |
| Tyr (Y) | 2.2% | 4.0% | 4.4% | 5.3% | 4.3% | 5.0% |
| Val (V) | 7.1% | 5.3% | 4.3% | 3.7% | 3.9% | 3.5% |
| Total number of AA | 771 | 1368 | 984 | 757 | 1307 | 1389 |
| MW (Da) | 84352.40 | 158441.41 | 114896.12 | 87181.92 | 151206.55 | 166277.01 |
| pI | 5.78 | 8.98 | 9.37 | 5.75 | 8.01 | 8.76 |
| I-ind | 35.49 | 37.57 | 30.50 | 33.83 | 35.53 | 40.26 |
| AI | 95.91 | 89.44 | 83.49 | 86.26 | 85.99 | 94.23 |
| GRAVY | -0.153 | -0.594 | -0.678 | -0.429 | -0.492 | -0.749 |

*AA: amino acid; *Svi*Cas3: from *Streptomyces virginiae* IBL14; *Sp*Cas9: from *Streptococcus pyogenes*; *Cj*Cas9: from *Campylobacter jejuni*; *Ll*Cas10: from *Lactococcus lactis* subsp. Lactis; *As*Cas12a / Cpf1: from *Acidaminococcus sp*. BV3L6; *Ls*Cas13a / C2c2: from *Leptotrichia shahii*; pI: theoretical isoelectric point; I-ind: instability index (smaller than 40 is predicted as stable ); AI: aliphatic index (regarded as a positive factor for the increase of thermostability of globular proteins); GRAVY: grand average of hydropathicity (+: hydrophobic, -: hydrophilic); red letters with dark blue background represent the highest value; red letters with green background represent the lowest value.

## Supplementary references

1 Chylinski, K., Makarova, K. S., Charpentier, E. & Koonin, E. V. SURVEY AND SUMMARY Classification and evolution of type II CRISPR-Cas systems. *Nucleic acids research* **42**, 6091-6105 (2014).

2 Gasiunas, G., Sinkunas, T. & Siksnys, V. Molecular mechanisms of CRISPR-mediated microbial immunity. *Cell Mol Life Sci* **71**, 449-465, doi:10.1007/s00018-013-1438-6 (2014).

3 Nimkar, S. & Anand, B. Cas3/I-C mediated target DNA recognition and cleavage during CRISPR interference are independent of the composition and architecture of Cascade surveillance complex. *Nucleic acids research* **48**, 2486-2501, doi:10.1093/nar/gkz1218 (2020).

4 Zhang, X. H., Tee, L. Y., Wang, X. G., Huang, Q. S. & Yang, S. H. Off-target Effects in CRISPR/Cas9-mediated Genome Engineering. *Molecular therapy. Nucleic acids* **4**, e264, doi:10.1038/mtna.2015.37 (2015).

5 Cho, S. W. *et al.* Analysis of off-target effects of CRISPR/Cas-derived RNA-guided endonucleases and nickases. *Genome research* **24**, 132-141, doi:10.1101/gr.162339.113 (2014).

6 Kim, D. *et al.* Digenome-seq: genome-wide profiling of CRISPR-Cas9 off-target effects in human cells. *Nature methods* **12**, 237-243, 231 p following 243, doi:10.1038/nmeth.3284 (2015).

7 Wang, G., Du, M., Wang, J. & Zhu, T. F. Genetic variation may confound analysis of CRISPR-Cas9 off-target mutations. *Cell discovery* **4**, 18, doi:10.1038/s41421-018-0025-2 (2018).

8 Rutkauskas, M. *et al.* Directional R-Loop Formation by the CRISPR-Cas Surveillance Complex Cascade Provides Efficient Off-Target Site Rejection. *Cell Rep* **10**, 1534-1543, doi:10.1016/j.celrep.2015.01.067 (2015).

9 Xiao, Y. *et al.* Structure Basis for Directional R-loop Formation and Substrate Handover Mechanisms in Type I CRISPR-Cas System. *Cell* **170**, 48-60 e11, doi:10.1016/j.cell.2017.06.012 (2017).

10 Mohanraju, P. *et al.* Diverse evolutionary roots and mechanistic variations of the CRISPR-Cas systems. *Science (New York, N.Y.)* **353**, aad5147, doi:10.1126/science.aad5147 (2016).

11 Makarova, K. S. *et al.* Evolution and classification of the CRISPR-Cas systems. *Nat Rev Microbiol* **9**, 467-477, doi:10.1038/nrmicro2577 (2011).

12 Mulepati, S. & Bailey, S. Structural and biochemical analysis of nuclease domain of clustered regularly interspaced short palindromic repeat (CRISPR)-associated protein 3 (Cas3). *The Journal of biological chemistry* **286**, 31896-31903, doi:10.1074/jbc.M111.270017 (2011).

13 Dolan, A. E. *et al.* Introducing a Spectrum of Long-Range Genomic Deletions in Human Embryonic Stem Cells Using Type I CRISPR-Cas. *Molecular cell* **74**, 936-950 e935, doi:10.1016/j.molcel.2019.03.014 (2019).

14 He, L., St John James, M., Radovcic, M., Ivancic-Bace, I. & Bolt, E. L. Cas3 Protein-A Review of a Multi-Tasking Machine. *Genes* **11**, 208, doi:10.3390/genes11020208 (2020).

15 Xue, C. & Sashital, D. G. Mechanisms of Type I-E and I-F CRISPR-Cas Systems in Enterobacteriaceae. *EcoSal Plus* **8**, doi:10.1128/ecosalplus.ESP-0008-2018 (2019).

16 Osakabe, K. *et al.* Genome editing in plants using CRISPR type I-D nuclease. *Communications biology* **3**, 648, doi:10.1038/s42003-020-01366-6 (2020).

17 Xiao, Y., Luo, M., Dolan, A. E., Liao, M. & Ke, A. Structure basis for RNA-guided DNA degradation by Cascade and Cas3. *Science (New York, N.Y.)* **361**, 1-15, doi:10.1126/science.aat0839 (2018).

18 Li, X. *A construction method of recombinant Streptomyces viginiae IBL-14 used for penicillin producation* MS thesis, Anhui University, (2016).

19 Qiu, C.-H. *Primary studies on both the Type I-B-svi CRISPR-Cas system in Streptomyces virginiae IBL14 and the self-targeting gene editing to the genes involved in penicillin metabolism* MS thesis, Anhui University, (2017).

20 Zhang, Y. *Study on the Physiological Function and Steroid Transformation Mechanism of CYP105 Family Genes in Streptomyces . virginiae IBL14* PhD thesis, Anhui University, (2017).

21 Xu, X. *Functional Analyses of the sviscsn Gene Cluster in Streptomyces virginiae IBL14* MS thesis, Anhui university, (2017).

22 Cao, S.-L. *Self-targeting gene editing of O-methyltransferase genes in Streptomyces virginiae IBL14* MS thesis, Anhui University, (2018).

23 Kim, E. *et al.* In vivo genome editing with a small Cas9 orthologue derived from *Campylobacter jejuni*. *Nature communications* **8**, 14500, doi:10.1038/ncomms14500 (2017).

24 Zetsche, B. *et al.* Cpf1 is a single RNA-guided endonuclease of a class 2 CRISPR-Cas system. *Cell* **163**, 759-771, doi:10.1016/j.cell.2015.09.038 (2015).

25 Abudayyeh, O. O. *et al.* C2c2 is a single-component programmable RNA-guided RNA-targeting CRISPR effector. *Science (New York, N.Y.)* **353**, aaf5573, doi:10.1126/science.aaf5573 (2016).

# Appendix

Prof. / Dr. Wang-Yu Tong
Integrated Biotechnology Laboratory
School of Life Sciences
Anhui University
111 Jiulong Road, Hefei 230601, China
Cell: (86) 150-56036299
E-mail: tongwy@ahu.edu.cn

**May 9, 2023**

Dear editors,

We are pleased to submit our original work, **"Prokaryotic genome editing based on the subtype I-B-*Svi* CRISPR-Cas system"**, written by *Wang-Yu Tong*, De-Xiang Yong, Xin Xu, Cai-Hua Qiu, Yan Zhang, Xing-Wang Yang, Ting-Ting Xia, Qing-Yang Liu, Su-Li Cao, Yan Sun and Xue Li*, to you (arXiv) according to the "Author Guide" from Your website.

As we all know, genome-editing technology is the most important tool for creating and transforming organisms. Here, we recommend this paper for publication as a Research Article in your "arXiv" journal for the following reasons:

(1) The single compact *Svi*Cas3 can be programmed to implement strict template-based genome editing in microbial cells, breaking the view that a full Cascade is necessary for genome editing directed by type I CRISPR-Cas systems.

(2) This finding implies that the *Svi*Cas3 may have a novel enzymatic mechanism, that is, it may be an endonuclease that can recognize R-loop or D-loop (also see its sister article: "Template-based eukaryotic genome editing directed by the *Svi*Cas3"). In addition, the *Svi*Cas system lacks the Cas8 subunit of Cascade, the PAM recognition component, implying that this reaction may be PAM independent.

(3) No off-target cases and indel formation were detected in prokaryotic microbial genome editing mediated by the single *Svi*Cas3, indirectly supporting the view that for the most part, Cascade-Cas3 interaction features a conformation-capture rather than an induced-fit mechanism.

(4) In addition, the *Svi*Cas3 has the following advantages over commercial *Sp*Cas9: (i) a small molecular weight suitable for vector carrying (*Svicas*3 / 2316 bp,

*Spcas*9 / 4107 bp), (ii) simple design of g-DNA (flexible PAM and repeat sequences, as well as without tracrRNA), (iii) a bacterial protein of non-animal origin with low possibility producing antibody of the *Svi*Cas3, and (iv) better biocompatibility (e.g., the *Svi*Cas3 can be successfully applied to the genome editing of *C. glutamicum* where *Sp*Cas9 is difficult to work).

In conclusion, we believe that the *Svi*Cas3 from type I CRISPR-Cas system in *Streptomyces virginiae* IBL14 will become a major competitor to the *Sp*Cas9 from type II CRISPR-Cas system in *S. pyogenes*, becoming the most important tool for creating and modifying organisms.

Reviewers are welcome to point out academic mistakes in this research paper objectively, rather than subjectively. In particular, relevant researchers are welcome to verify the functional authenticity of the *Svi*Cas3 protein.

All data for this study are included in this Manuscript or Supplementary Information, or on the websites marked in the paper.

A patent application (CN107557373A / WO2019056848A1 / EP3556860A1 / US11286506 B2) has been filed for the content disclosed in this study.

The work was not funded by any agency.

We obey the copyright policy of your editorial office as to the copyright of the paper.

We look forward to hearing from you.

Yours sincerely,

Wang-Yu Tong